\documentclass[10pt,letterpaper]{article}
\usepackage[top=0.85in,left=2.75in,footskip=0.75in]{geometry}

\usepackage{amsmath,amssymb}
\usepackage[section]{placeins}
\usepackage{changepage}
\usepackage{float}
\usepackage{textcomp,marvosym}

\usepackage{cite}
\usepackage{tabularray, multirow}

\usepackage{nameref,hyperref}

\usepackage[table]{xcolor}

\usepackage{array}
\usepackage{comment}
\usepackage{makecell}

\newcolumntype{+}{!{\vrule width 2pt}}

\newlength\savedwidth

\newcommand\thickhline{\noalign{\global\savedwidth\arrayrulewidth\global\arrayrulewidth 2pt}%
\hline
\noalign{\global\arrayrulewidth\savedwidth}}

\raggedright
\usepackage[aboveskip=1pt,labelfont=bf,labelsep=period,justification=raggedright,singlelinecheck=off]{caption}

\makeatletter
\renewcommand{\@biblabel}[1]{\quad#1.}
\makeatother

\usepackage{lastpage,fancyhdr,graphicx}
\usepackage{epstopdf}
\fancyheadoffset[L]{2.25in}
\fancyfootoffset[L]{2.25in}
\usepackage{subcaption}
\usepackage{svg}

\begin{document}

\vspace*{0.2in}

\begin{flushleft}
{\Large
\textbf\newline{The statistical and dynamic modeling of the first part of the 2013-2014 Euromaidan protests in Ukraine: The Revolution of Dignity and preceding times.} 
}
\newline
\\
Yassin Bahid\textsuperscript{1,*},
Olga Kutsenko\textsuperscript{2},
Nancy Rodr\'iguez\textsuperscript{1},
David White\textsuperscript{3},
\\
\bigskip
\textbf{1} Department of Applied Mathematics, University of Colorado, Boulder, Colorado, USA
\\
\textbf{2} Department of Sociology, Taras Shevchenko National University of Kyiv, Ukraine; Berlin Technical University, Germany
\\
\textbf{3} Department of Mathematics, Denison University, Granville, Ohio, USA
\\
\bigskip

%
%





* Corresponding Author: yassin.bahid@colorado.edu

\end{flushleft}
\section*{Abstract}
Ukraine's tug-of-war between Russia and the West has had significant and lasting consequences for the country. In 2013, Viktor Yanukovych, the Ukrainian president aligned with Russia, opted against signing an association agreement with the European Union. This agreement aimed to facilitate trade and travel between the EU and Ukraine. This decision sparked widespread protests that coalesced in Kyiv's Maidan Square, eventually becoming known as the Euromaidan protests. In this study, we analyze the protest data from 2013, sourced from Ukraine's Center for Social and Labor Research. Despite the dataset's limitations and occasional inconsistencies, we demonstrate the extraction of valuable insights and the construction of a descriptive model from such data. Our investigation reveals a pre-existing state of self-excitation within the system even before the onset of the Euromaidan protests. This self-excitation intensified during the Euromaidan protests. A statistical analysis indicates that the government's utilization of force correlates with increased future protests, exacerbating rather than quelling the protest movement.  Furthermore, we introduce the implementation of Hawkes process models to comprehend the spatiotemporal dynamics of the protest activity. Our findings highlight that, while protest activities spread across the entire country, the driving force behind the dynamics of these protests was the level of activity in Kyiv. Furthermore, in contrast to prior research that emphasized geographical proximity as a key predictor of event propagation, our study illustrates that the political alignment among oblasts, which are the distinct municipalities comprising Ukraine, had a more profound impact than mere geographic distance. This underscores the significance of social and cultural factors in molding the trajectory of political movements.

\section*{Introduction}
Ukraine's historical and cultural ties with Russia forged over centuries and solidified by its emergence as an independent nation from the dissolution of the Soviet Union in December 1991, stand in stark contrast to its more recent inclinations toward Western nations. This dynamic has given rise to a complex and ongoing struggle for national identity\cite{ConFR}. While a significant portion of the Ukrainian population maintains enduring connections with Russia, an equally substantial contingent leans toward Western influences. In 1994, Ukraine entered into an important partnership and cooperation agreement with the European Union, a landmark accord detailed in \cite{agreement1994}. This agreement not only facilitated bilateral trade but also encompassed diverse domains such as science, technology, mining, and maritime transport, signifying the comprehensive scope of engagement between Ukraine and the EU.

During the early years of the 21st century, a pressing question emerged within Ukraine's political discourse: whether to align itself with the European Union. This deliberation was viewed by Russian President Vladimir Putin as a destabilizing development \cite{Marples15}. The extent to which individuals within Ukraine aspired to align with the West has evolved into a central issue catalyzing political polarization within the nation \cite{Marples15}.
From 2005 to 2010, the Ukrainian presidency was held by Viktor Yushchenko, who pursued closer integration with the European Union. Yushchenko's aspiration to formalize this commitment through an association agreement, aiming for a deepened engagement, was hindered by the limitations of his political influence. The outcome of the 2010 presidential election underscored the challenges he faced, as he was defeated by President Viktor Yanukovych, who maintained a more Russia-oriented stance\cite{Ind10}.
The year 2013 marked a turning point when President Yanukovych, despite broad parliamentary approval, declined to sign the aforementioned association agreement with the European Union. The interplay of domestic and external influences, including significant pressure from Russia, guided his decision to reject the agreement \cite{Aljazeera13}. This choice was the triggering event for the Euromaidan protests, known as the Revolution of Dignity, which swept across Ukraine, predominantly centered in Kyiv, before spreading throughout the nation \cite{Marples15}.

 In this study, our primary objective is to discern the underlying determinants propelling these protests. Specifically, our investigation centers on two key aspects: the potential diffusion of protest activities and the ramifications stemming from the frequency of arrests and injuries witnessed within a given protest incident, and the relationship to subsequent protest dynamics. To study these phenomena, we make use of an extensive repository of protest data amassed by the Center for Social and Labor Research in Ukraine, accessible through their \href{https://www.cslr.org.ua/en/protests/}{website}.
Our analytical approach is anchored in the deployment of both regression and ARIMA models. Through these quantitative methodologies, we aim to understand the interplay of variables. Our findings underscore the existence of self-propagation within the realm of protests, with prominence during the latter phase of 2013, coinciding with the critical Euromaidan upheaval.

The second fundamental aim of this study is the development of models of the spatial and temporal dynamics characterizing the escalation of events during the 2013 phase of the Euromaidan revolution. This analysis covers a spectrum of events, encompassing both pro-Maidan and anti-Maidan manifestations. To accomplish this, we make use of Hawkes processes, a class of point processes known for modeling self-excitation. 
These models can detail the trajectories of escalation, which extend beyond Kyiv. Notably, our investigation unveils Kyiv as the central force behind the spread of events throughout the nation. While geographical distance does not exert a discernible effect, our study discerns a significant impact emerging from the political affinity of the different regions. Moreover, we identify contagion patterns both within Kyiv and across many oblasts, which represent the nation's principal administrative subdivisions.

The Euromaidan movement has encompassed a diverse set of events, ranging from peaceful protests to more tumultuous episodes such as violent riots. Although there are substantial distinctions among these various event types, as outlined in \cite{Lemos2013}, all of these occurrences contribute to and can be regarded as components of civil unrest. While a more comprehensive discussion could delve into the nuances of each event type, this falls beyond the purview of our research. The deployment of mathematical models has gained recognition as a powerful framework for studying such events, and diverse methodological approaches have been developed.
In previous works such as \cite{Epstein2012} and \cite{Lemos2013}, an agent-based paradigm was adopted to model civil unrest, encapsulating the intricate dynamics underlying such phenomena. A distinct avenue of inquiry ventured into the use of non-linear dynamics and chaos theory, as showcased in \cite{Andreev1997}, wherein the Russian Empire's labor strikes during the period spanning 1895 to 1905 were subjected to rigorous analysis. Other pertinent dynamical approaches can be found in \cite{Berestycki2015} and \cite{alsulami2022dynamical}. Meanwhile, evolutionary game theory has found application in the works \cite{quek2009evolutionary} and \cite{Gavrilets2015}.
Epidemiological models, as exemplified in \cite{Bonnasse-Gahot2018}, \cite{Khosaeva2015}, and \cite{caroca2020anatomy}, have gained traction as tools to understand these societal events. A noteworthy framework is that of kinetic theory, which was used in \cite{dimarco2021kinetic} to model the dynamics of epidemic propagation.

In this work, we opt to use stochastic point processes known as Hawkes processes to model the spatiotemporal dynamics of the 2013 protesting events. While Hawkes processes were initially introduced to model seismic activities, their utility has extended to the modeling of events in diverse fields including finance, neuroscience, social sciences, and computer science. Some examples of the uses of Hawkes processes include but are not limited to EU instances of terrorist attacks \cite{tench2016spatio}, gun violence \cite{loeffler2018gun}, gang-related violence \cite{brantingham2021gang}, and incidents of disorder that unfolded during the COVID-19 pandemic \cite{campedelli2021temporal}. Hawkes processes are particularly well suited to model self-propagating events, a pattern observed in the dynamics of Euromaidan.

\section*{Historical context and timeline}

\subsection*{The Orange Revolution and Ukraine's pull between the West and Russia}

In the aftermath of World War I and the Russian Revolution of 1917, a substantial portion of the Ukrainian territory came under the dominion of the Soviet Union. Notably, certain enclaves of western Ukraine found themselves divided among Poland, Romania, and Czechoslovakia \cite{BritHist}. Ukraine's independence materialized in 1991, following the dissolution of the Soviet Union. 
The initial phase of Ukrainian autonomy was marked by numerous challenges as the nation grappled to navigate the implementation of both economic and political reforms. These initial stages were marked by escalating social tensions, which grew into a watershed moment characterized by large-scale demonstrations, now known as the Orange Revolution. The origins of this movement lay in the contested aftermath of the 2004 presidential elections, during which President Viktor Yanukovych proclaimed victory, an assertion contested \cite{Marples15, Reznik2016-vb}. The incendiary nature of these events led to the nullification of the election results, forcing Ukraine's Supreme Court to mandate a new election process. After the new election period, Viktor Yushchenko was named president of Ukraine.
However, the support for the protests was not uniform among the Ukrainian people. The support primarily coalesced through a collaborative effort spearheaded by individuals coming from the western and central regions of Ukraine \cite{Kuzio2010}. 

Certain scholars, including Kuzio, a distinguished professor of political science at the National University of Kyiv-Mohyla Academy, advance the perspective that Ukraine's trajectory post-independence has led to the emergence of two distinct strands of nationalism. Within this paradigm, civic nationalism takes root in western and central Ukraine, while pro-Russian sentiment gains prominence in the eastern regions \cite{Kuzio2010}. Kuzio underscores that the pronounced mobilization witnessed during the Orange Revolution was propelled by the fervor of civic nationalism.

The 2010 Viktor Yanukovych became the fourth president of Ukraine. This event in itself served as a testament to the enduring fracture in Ukraine's political landscape. A trend surfaces when studying the political allegiances of the various administrative divisions. Generally, the oblasts situated in the western and central regions align more closely with Western principles and ideologies, while their eastern and southern counterparts tend to exhibit a pro-Russian disposition. This dichotomy is illustrated in Figure \ref{fig:votemap}, which shows the support for the pro-European Union candidate across the 2004, 2010, and 2014 presidential elections. The figure underscores a gradual attenuation in pro-Western inclinations as one traverses from the western to the eastern extremities of the nation, rather than a stark demarcation between the two political inclinations.

\subsection*{Euromaidan: the revolution of dignity}

\begin{figure}[!h]
    \centering
     \subcaptionbox{ \label{fig:vote2004}}{\includegraphics[width=.45\textwidth]{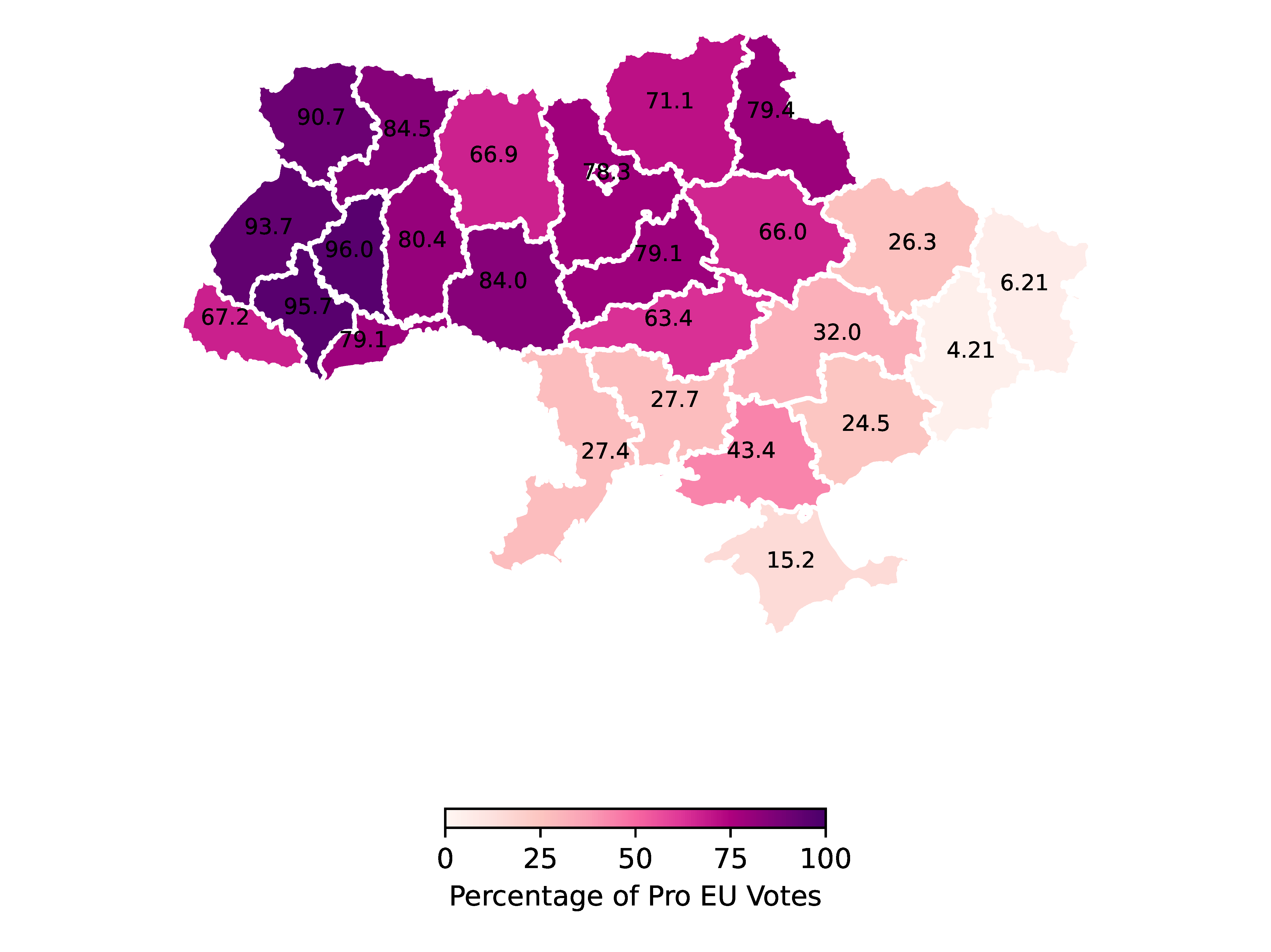}}\hspace{1em}%
     \subcaptionbox{ \label{fig:vote2010}}{\includegraphics[width=.45\textwidth]{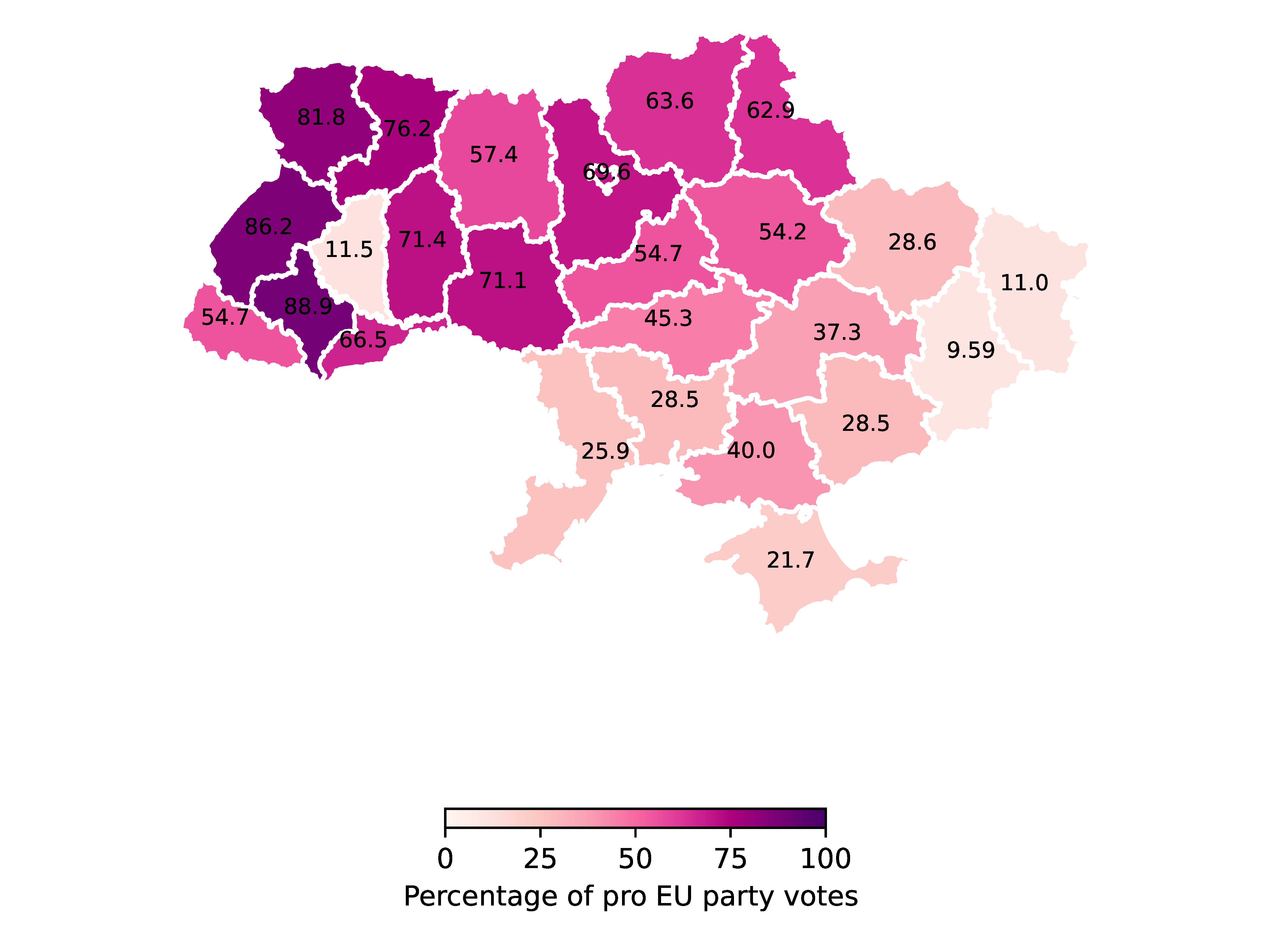}}\hspace{1em}%
     \subcaptionbox{ \label{fig:vote2014}}{\includegraphics[width=.45\textwidth]{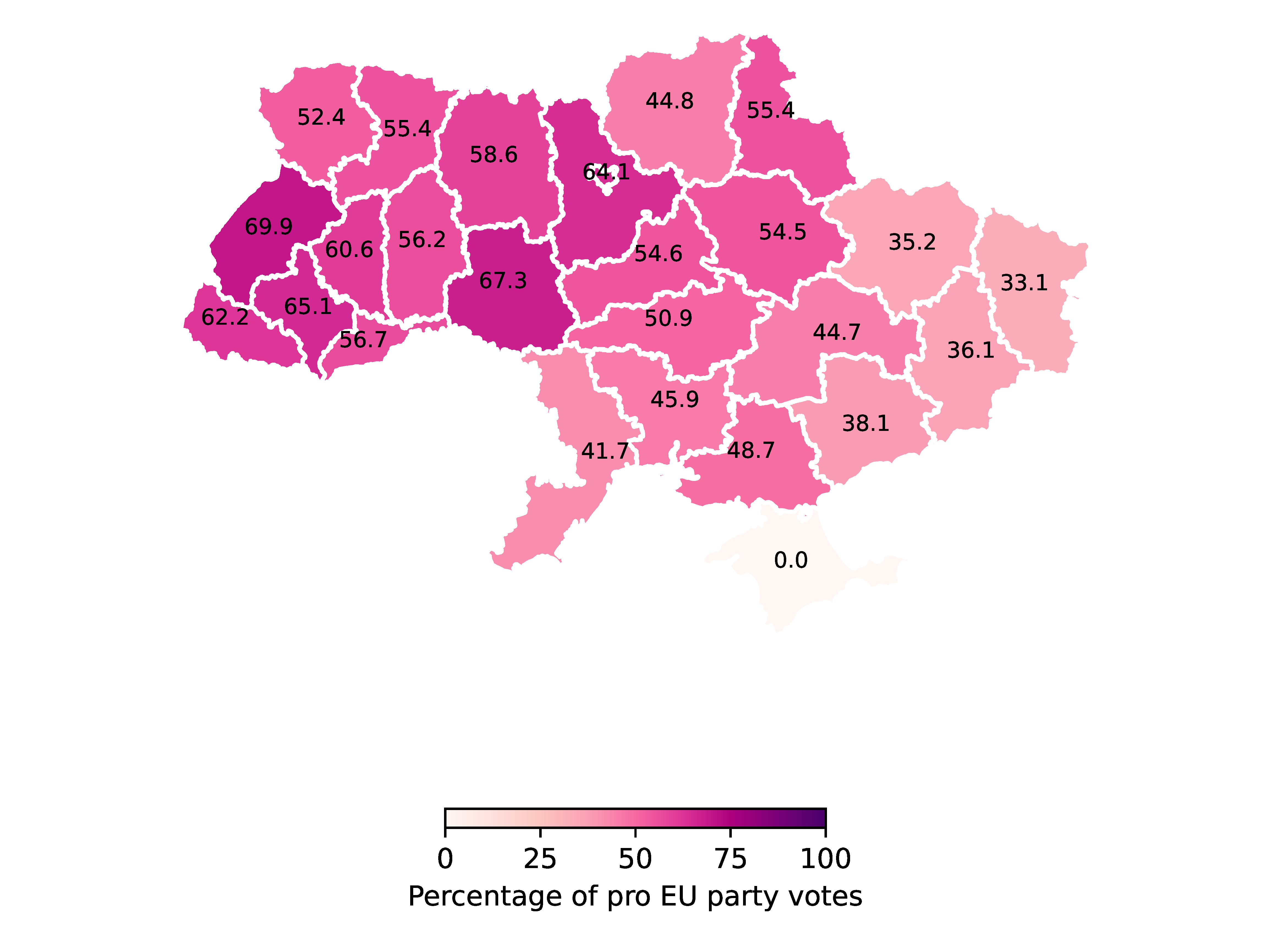}}\hspace{1em}%
    \caption{The presidential voting maps of Ukraine: \textbf{(a)} 2004 \textbf{(b)} 2010 \textbf{(c)} 2014.}
    \label{fig:votemap}
\end{figure}

In 2009 the Eastern Partnership (EaP), a joint initiative involving the European Union, its Member States, and six eastern European Partner countries: Armenia, Azerbaijan, Belarus, Georgia, the Republic of Moldova, and Ukraine, was launched \cite{EaP}.  A high point of this partnership for Ukraine was the Association Agreement with the European Union, a bilateral agreement between the EU and a third country. It served as a way to open up borders between the EU and the third-party countries for trade and travel\cite{EUtrade}. This agreement established political and economic ties between Ukraine and the West \cite{EuAssAgree}. 
The Ukrainian parliament had approved the final agreement and in late November 2013, President Yanukovych signaled his support for the agreement. However, a visit to Moskow appears to have changed his mind.  On November 21, 2013, Ukraine rejected draft laws that aimed to release jailed opposition leader Yulia Tymoshenko, a requirement imposed by the European Union for the association agreement, and suspended plans for the signing of this landmark agreement.  Moreover, Ukraine announced that it would renew dialogue with Russia \cite{Aljazeera13}. While President Yanukovych's support in the western oblasts was always small, his support in the eastern regions started to decline after the contested decisions associated with his presidency, Figure \ref{fig:vote2014}.

The Euromaidan protests began in earnest the evening of November 21, 2013, when more than 1,500 Ukrainians marched to the Maidan Square in Kyiv in disapproval of the decisions taken by their government earlier that day \cite{NYT2013}.  In the next days, the peaceful civil protests continued in Maidan Square and throughout other regions of the country, with these events achieving a very large number of protesters over the weekend \cite{Marples15}. On the night of November 30, 2013, things became violent when the government ordered the {\it Berkut} police, a special unit of the Ukrainian police within the Ministry of Internal Affairs, to disperse the square.   It was documented that the unit used force to violently disperse the Maidan Square protesters\cite{Lapatina}. The next day the protesters reoccupied the square.  That day saw multiple riots in Kyiv and a large number of journalists being injured by the police \cite{WikiDec1}.  On December 11, 2013, the Berkut special police unit and interior ministry troops descended on protesters violently in an attempt to break up the Euromaidan protests \cite{WikiDec11}.

On January 16, 2014, the Ukrainian Parliament signed new anti-protest laws, which restricted freedom of speech and freedom of assembly \cite{Guard14}.  
In reaction, on Sunday, January 19, 2014, over 200,000 Ukrainians showed up to protest in the center of Kyiv \cite{NST14}. More violence erupted on January 22, 2014, on Hrushevskoho Street in Kyiv resulting in the deaths of protesters \cite{Unian14}. A few days later, on January 25, 2014, Arsenii Yatsenyuk, the former Economy and Foreign Minister of Ukraine and leader
of the Batkivshchyna, an EU-leaning party was offered the Prime Minister's position, which he declined, citing the need for the Ukrainian citizens to decide on their future leader and not the current government they were protesting against \cite{Marples15}. 
More deadly protests occurred between February 18-20, leaving more than 100 individuals dead \cite{Econ14}.  The European Union became involved on February 21, 2014, and introduced sanctions against the Ukrainian leaders.  On February 22, 2014, President Yanukovych fled to Donetsk and then Crimea and parliament voted to remove him from office \cite{BBC14}. Afterward, the Russian army annexed Crimea, which led to the Russo-Ukrainian War. The election, first scheduled for March 2015, was held following Euromaidan. It led to the victory of President Petro Poroshenko\cite{BBCNews_2014} who led Ukraine through the integration with the EU by signing the EU Association Agreement\cite{Smith_Spark_Brumfield_Krever_2014}.
\begin{figure}[!h]
 \centering
 \includegraphics[width=\textwidth]{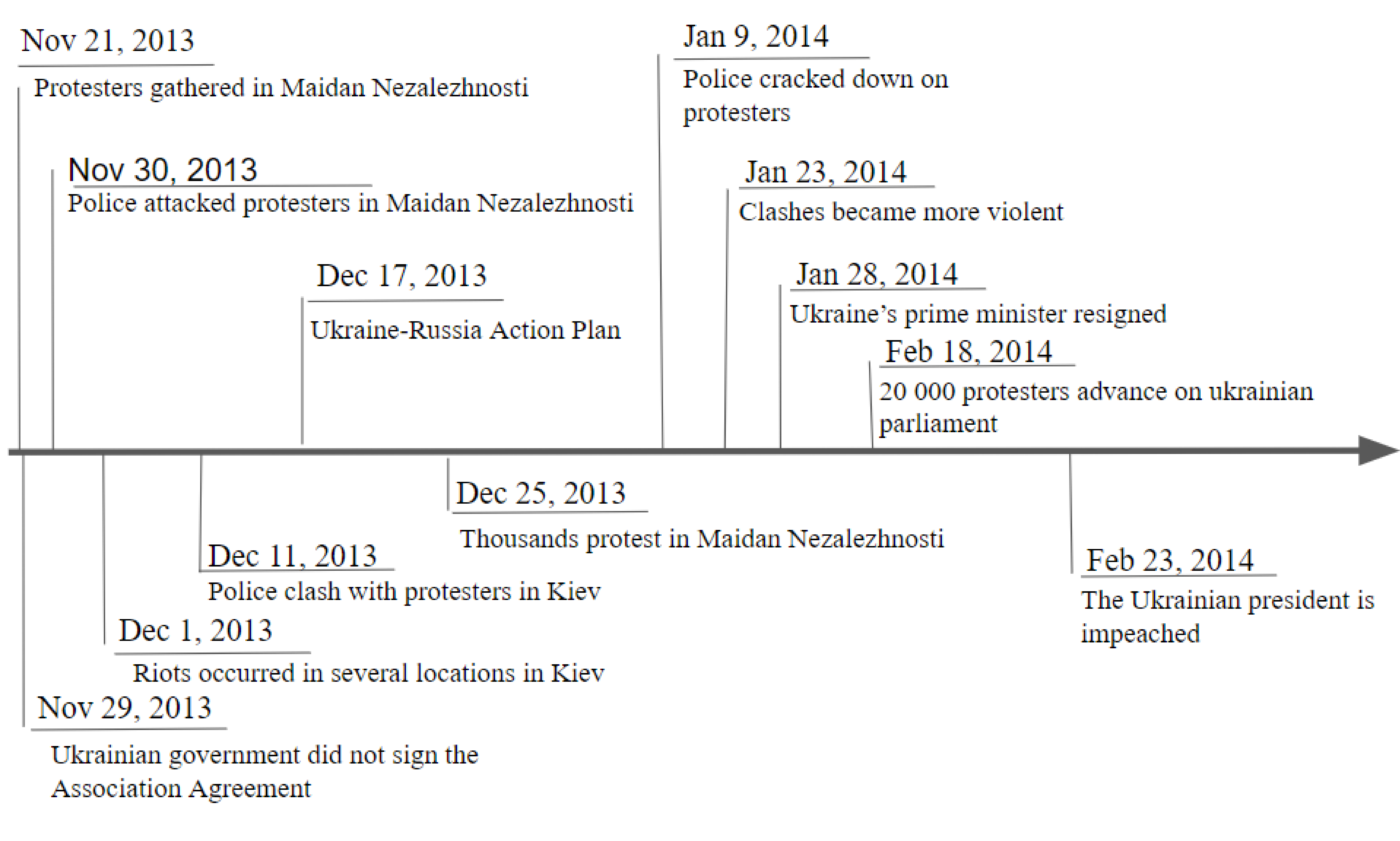}
 \caption{The timeline of Euromaidan from November 25, 2013, to February 23, 2014.}
 \label{fig:timeline}
\end{figure}

\section*{Materials and methods}
\subsection*{Data collection and limitations}

Created in 2013, the Center for Social and Labor Research is a Ukrainian non-profit organization focused on gathering and analyzing data about socioeconomic problems in Ukraine. To model the spatiotemporal dynamics observed in the Euromaidan protests, we use their protests data set\cite{data} from 2013 (although some events spill into 2014). This data set was created and curated, largely by academics without any explicit political agenda \cite{data-prov}, using information aggregated from newspaper articles found through monitoring of news feeds from more than 190 national, regional, and activist web media. The data set has been continually updated from 2013 until the present day. The data set includes protests, rallies, riots, and police crackdowns, among others, as events spanning a range of days \cite{codebook}. The main unit of the data set is a ``protest event" and therefore we use the terms ``protest" and ``event" interchangeably. The data takes into consideration the bias of certain events and their over-representation. The data curators addressed the over-representation by adhering to certain principles, e.g., including data from national, regional, and local news sources, and taking care not to double-count any event.  Potential bias is also taken into consideration, as certain events, e.g., those backed and instigated by the government, were not counted. The data set contains both large and successful events that were reported in mass media, as well as events that were considered as failed, e.g., because of violent law enforcement intervention.  In some cases, the data set details the number, demographics, and political leanings of the participants. In every case, the data set details the date(s) and the locations of the events. In some cases, e.g., an extremely large protest or a failed protest, data on the number of protesters is missing. We discuss this in more detail in Appendix \ref{sec:App2}.  While the report does give a breakdown of the level at which the events happen, e.g., neighborhoods or cities,  we choose to aggregate the events per oblast as we want to look at a holistic view of the events' spread, but also because certain events' location is only defined on the oblast level.  Note that some oblasts share the same name as certain cities in the region, however, in this work, we use these names only to refer to the oblasts. For example, when we refer to Kyiv we mean the oblast and not the city.

For every event in the data set, we know which oblast it occurred in. However, since the data comes from media reports, it is possible that events happened that were never included in the data set. For each oblast, the data curators drew from eight or nine news sources to determine which events happened in that oblast. It is important to recognize that an event only appears in the data set if it was reported in the news, and this presents the potential of bias in the data. We did find evidence of events that occurred but were not reported in the media, especially in eastern oblasts. We did not find evidence that this phenomenon was widespread, but we nevertheless stress the caveat that, if the data set exhibited significant bias, this would affect our models. We encourage future researchers to replicate our findings on data sets with less bias. Related research (replicating the appropriateness of Hawkes process models for protest dynamics, and the relationship between injuries and the number of protests) on a data set of protests in the United States has been carried out in \cite{Rodriguez-White}.

The statistical models below demonstrate that events exhibit self-exciting behavior, e.g., more events today are associated with even more tomorrow. The statistical model is based on the entire data set and does not omit any oblasts. 
There is a variable, ``Action type", that takes values such as ``protest", ``positive response", or ``negative response." The ``negative response" events are ones where law enforcement was involved in a way perceived as negative. More precisely, the term ``negative response" refers to events featuring repression, suppression, or obstruction of the protesters, either physical or legal \cite{codebook}. An event is coded as a ``negative response" if it features arrests, attacks, beatings, a blockade, a confrontation with police, a conviction, deportation, employees being fired (or students being expelled) for participating in a protest, a fight or gunfight, harassment, hacking, interrogation, imprisonment, lockout in response to the protest, martial law, a criminal case, law enforcement preemptive obstruction, police searches of participants, or a shooting. There is also a variable, ``Event series", that gives an idea of what the event is about. When this variable is valued as ``Euromaidan", the event was about the Euromaidan movement.

Our best model includes covariates detailing the number of ``Euromaidan" events per day, the number of ``negative response" events per day, and the number of civilian injuries per day. Of the 6627 events, 3220 have the string ``Euromaidan" occurring in the ``Event series" variable, which denotes that these events were associated with the Euromaidan event. This includes both pro and anti-Euromaidan events, in any oblast in the country. Similarly, there are 1102 events with ``negative response" and 405 that are both ``Euromaidan" and ``negative response."

\subsubsection*{Pre-Euromaidan}
\begin{figure}[!h]
    \centering
    \includegraphics[width=0.6\textwidth]{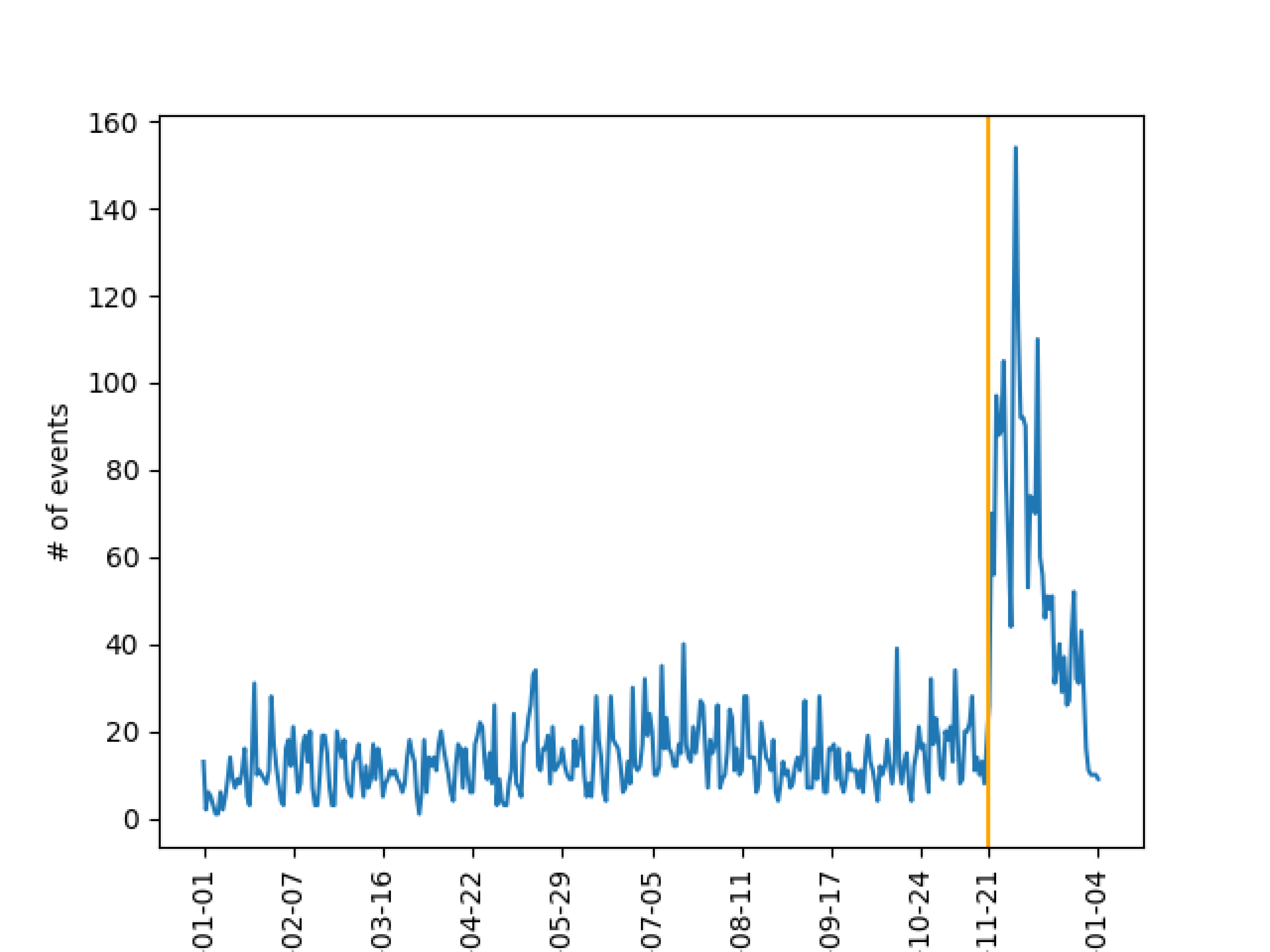}
    \caption{The number of events including protests, riots, and negative police response in Ukraine from January 1, 2013, to January 04, 2014.}
    \label{fig:year_roun}
\end{figure}

Figure \ref{fig:year_roun} depicts the number of events per day from January 1 to December 31, 2013. We observed a relatively small activity before Euromaidan, which lasted between November 21, 2013, and February 23, 2014. The average number of events before the beginning of the Euromaidan protests is very low, less than one per day per oblast. We shall use each oblast's average number of events before self-excitation as a base number of protests in our model. Most of the events presented during this time were small-scale protests with few protesters.

\subsubsection*{Euromaidan}
The impact of the EU trade agreements on Ukraine's relationship with Russia, Ukraine's biggest trade partner at the time, pushed President Yanukovych to delay signing the agreement. In turn, this act triggered the protests constituting Euromaidan. As seen in Figure \ref{fig:timeline}, protesters started gathering in the Maidan square on November 21, 2013. After the failure of the EU trade agreement on November 29, there were major police crackdowns on protesters, which in turn led to riots in Kyiv. The protests continued for multiple weeks afterward. 

\begin{figure}[!h]
    \centering
    \subcaptionbox{\label{fig:spt0}}{\includegraphics[width=.29\textwidth]{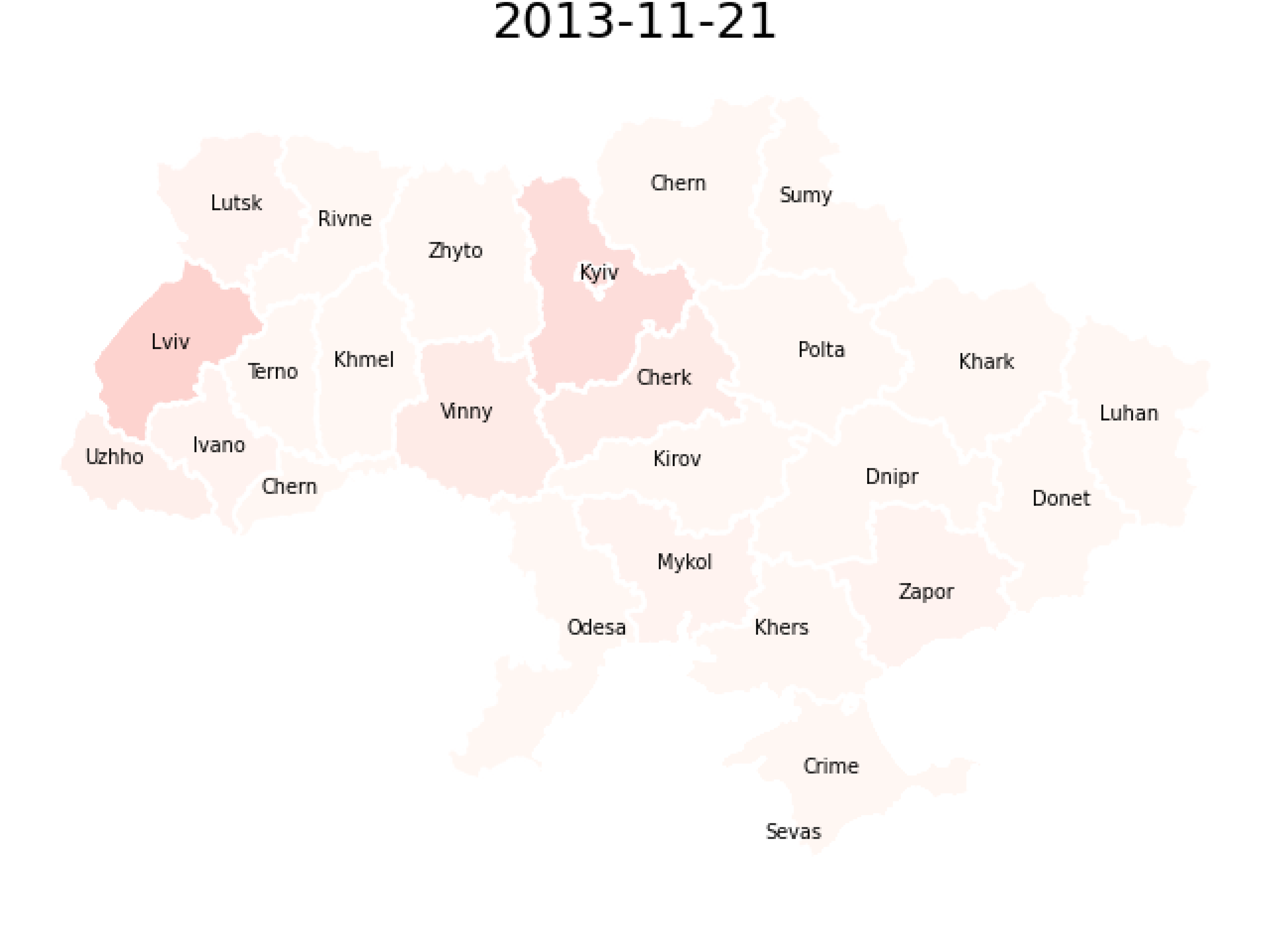}}\hspace{1em}%
    \subcaptionbox{\label{fig:spt1}}{\includegraphics[width=.29\textwidth]{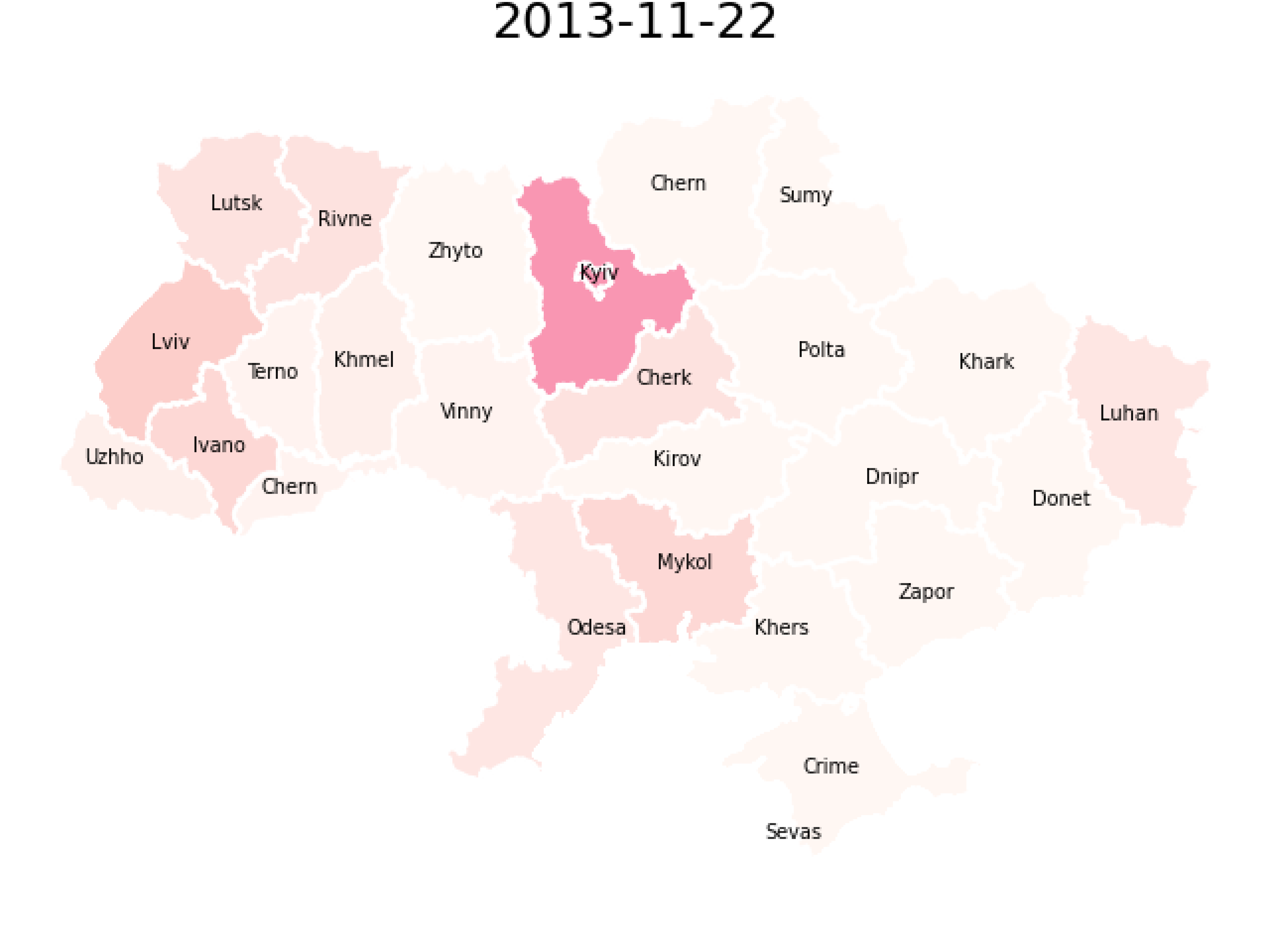}}\hspace{1em}%
    \subcaptionbox{\label{fig:spt2}}{\includegraphics[width=.32\textwidth]{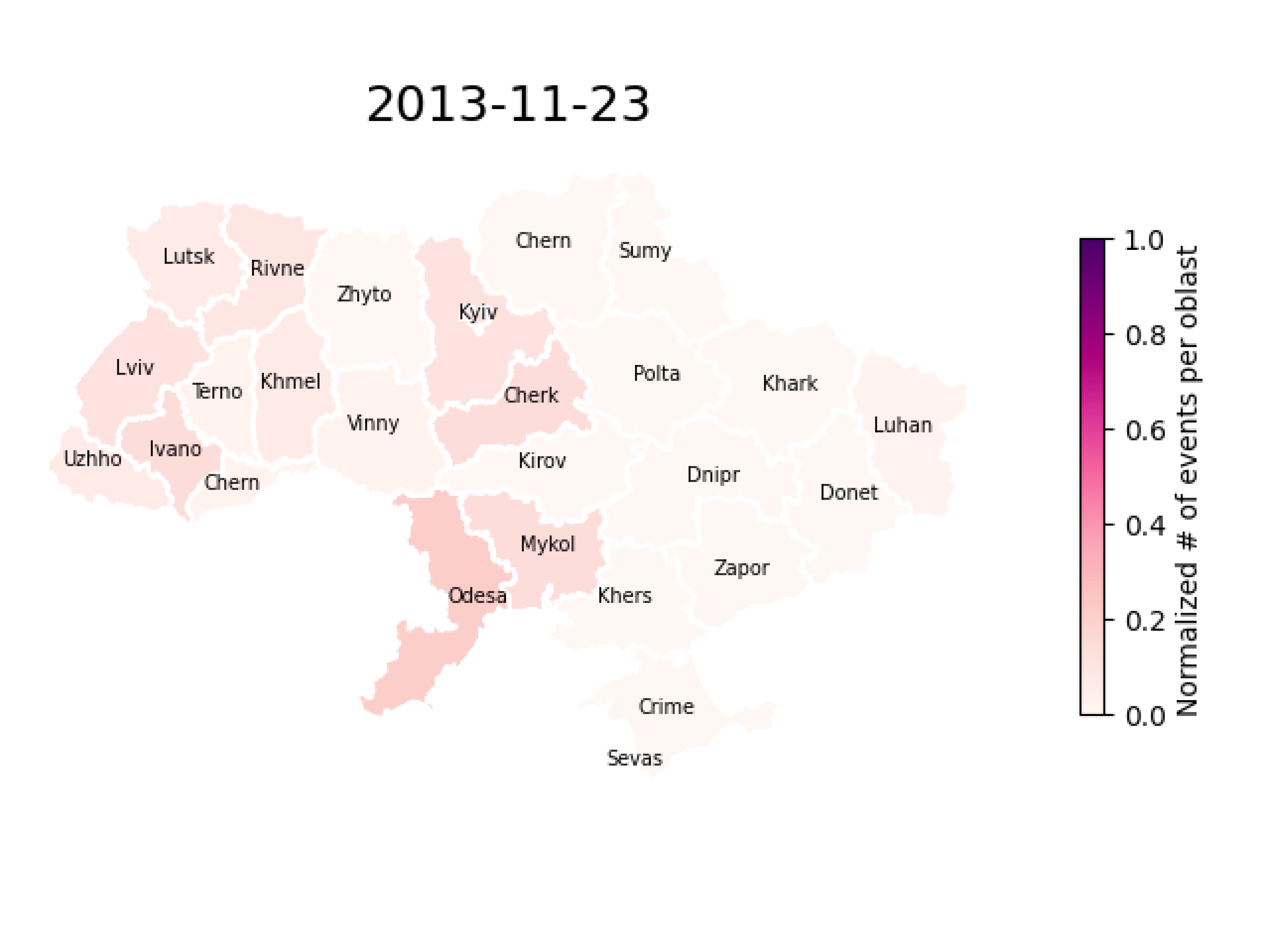}}\hspace{1em}%
    \\
    \subcaptionbox{\label{fig:spt3}}{\includegraphics[width=.29\textwidth]{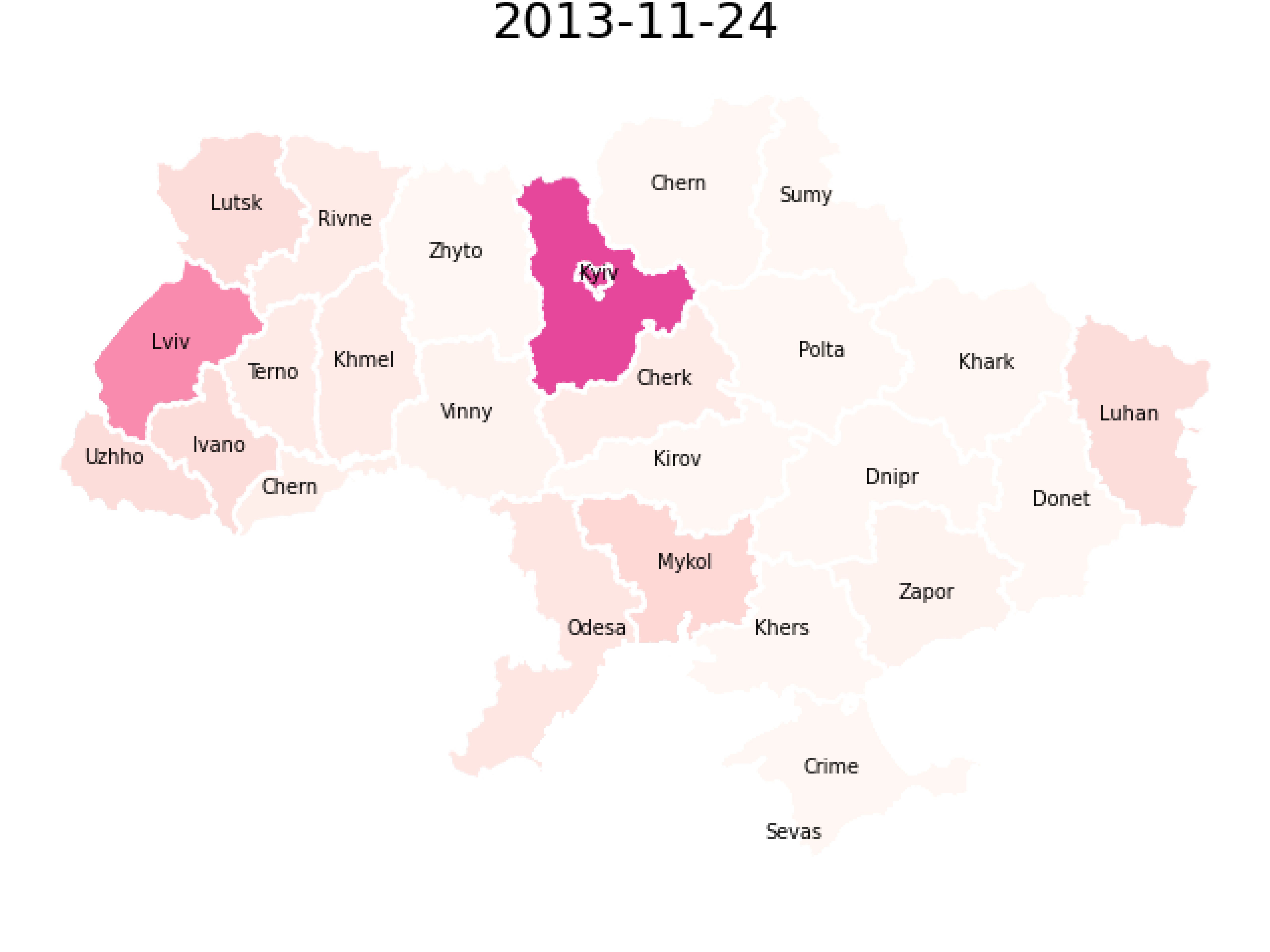}}\hspace{1em}%
    \subcaptionbox{\label{fig:spt4}}{\includegraphics[width=.29\textwidth]{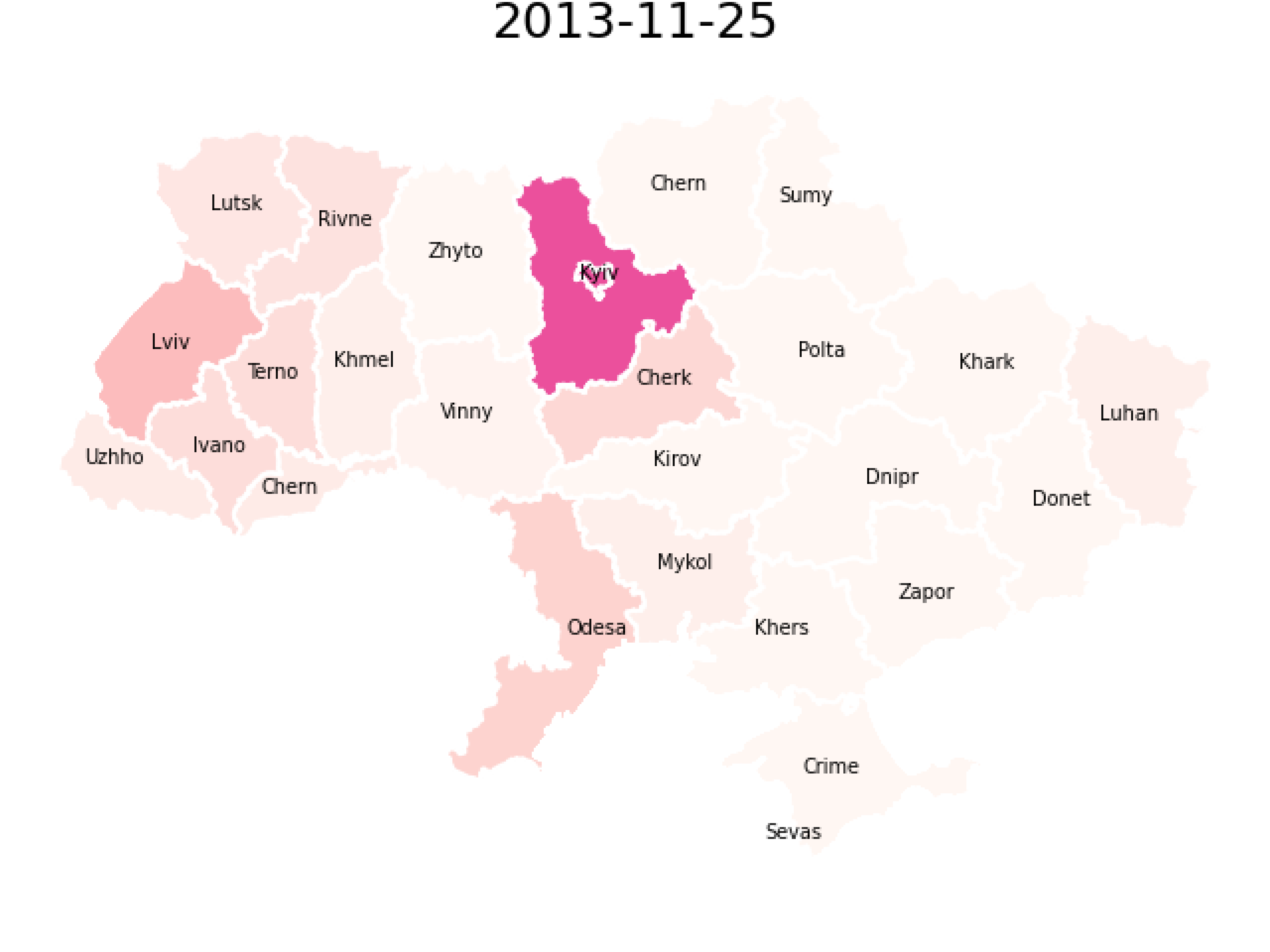}}\hspace{1em}%
    \subcaptionbox{\label{fig:spt5}}{\includegraphics[width=.32\textwidth]{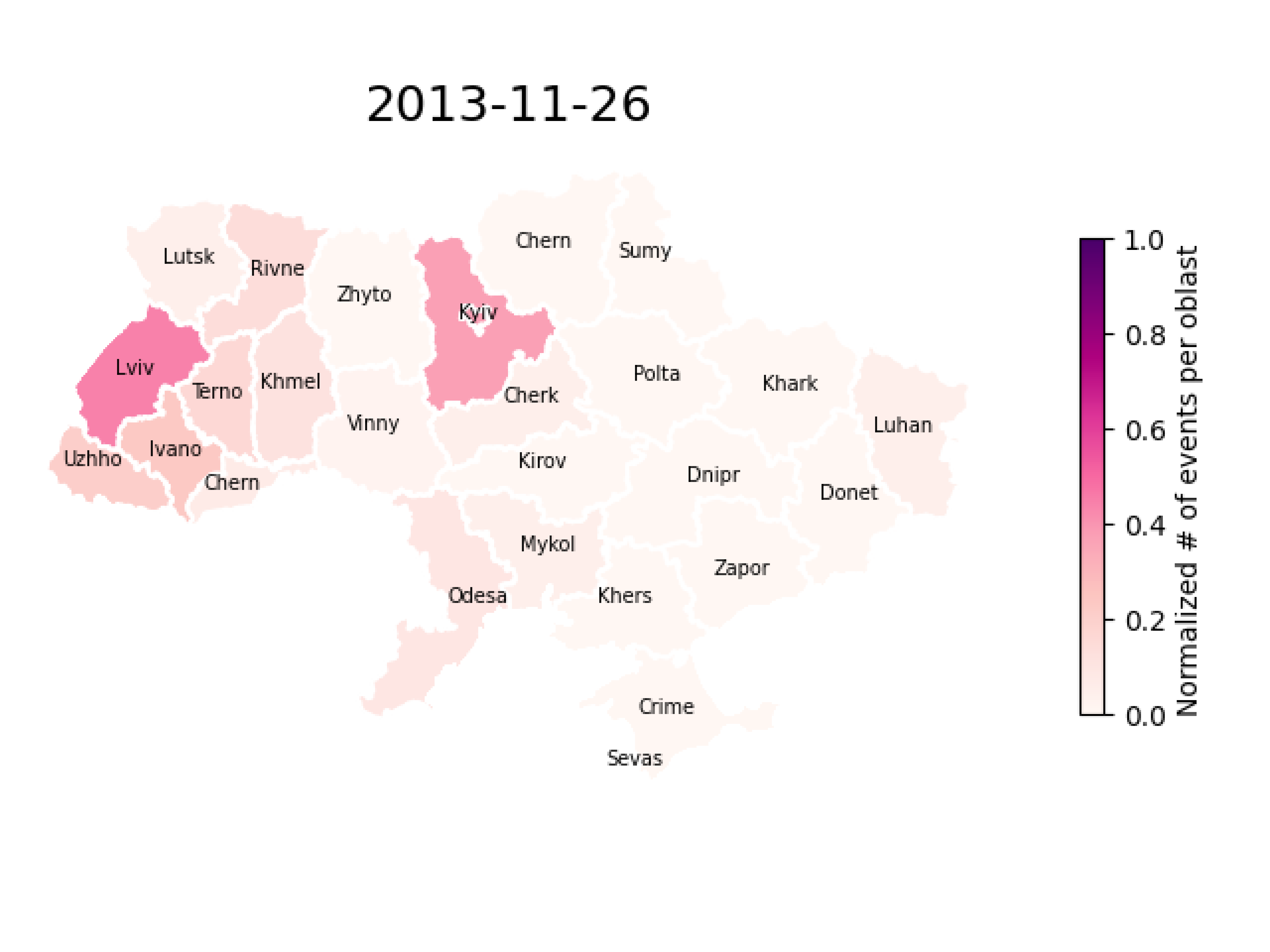}}\hspace{1em}%
    \\
    \subcaptionbox{\label{fig:spt6}}{\includegraphics[width=.29\textwidth]{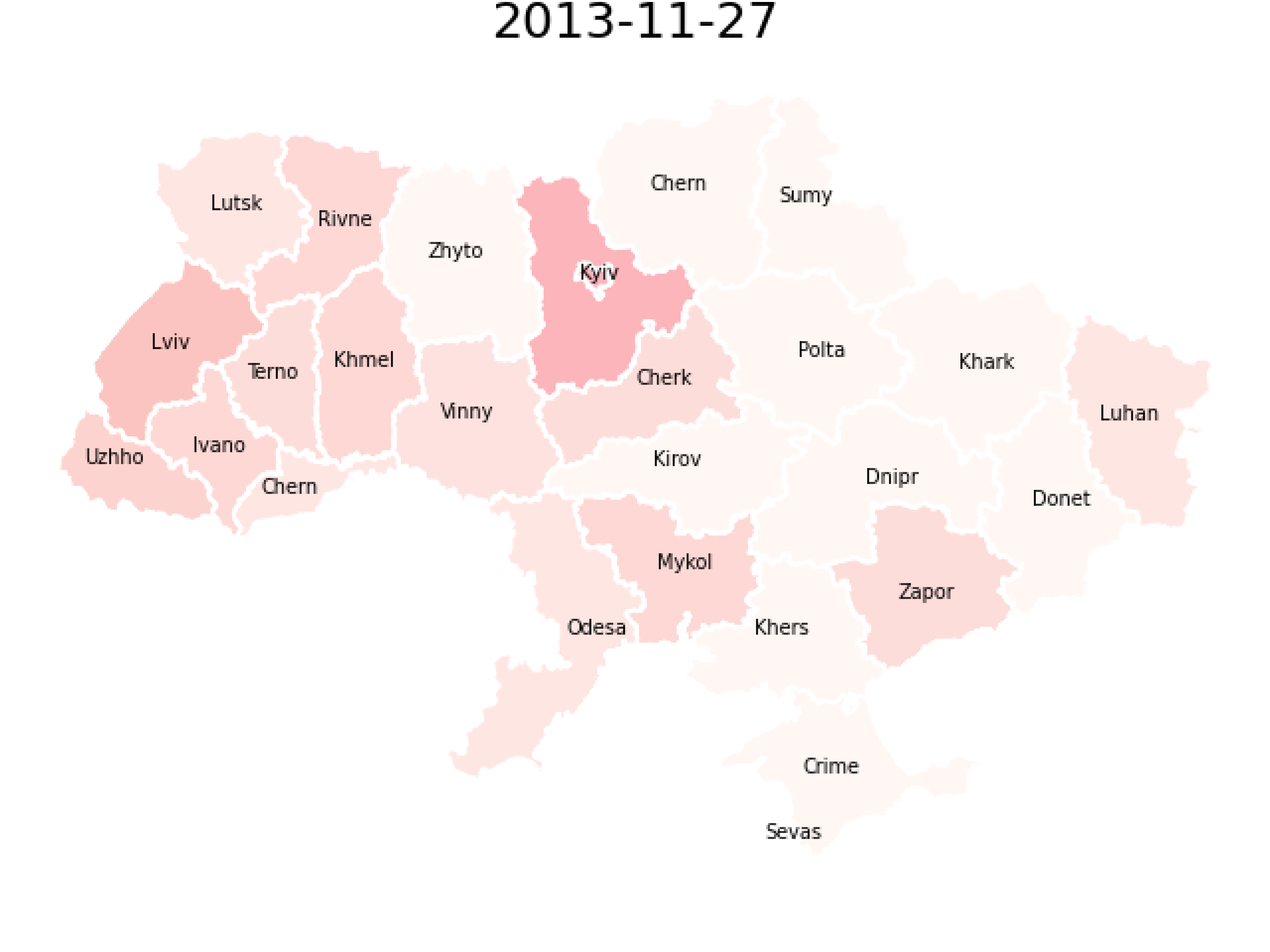}}\hspace{1em}%
    \subcaptionbox{\label{fig:spt7}}{\includegraphics[width=.29\textwidth]{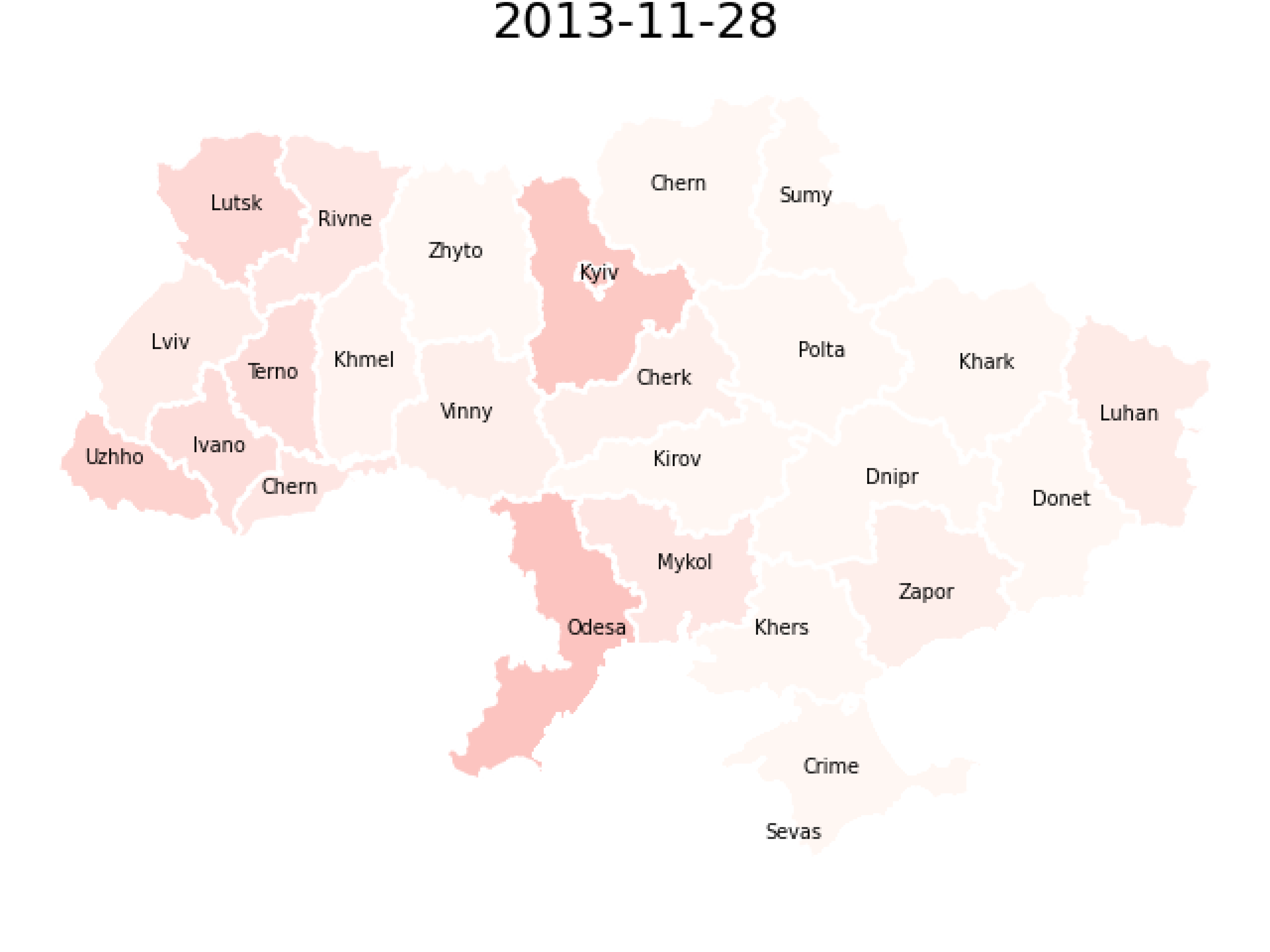}}\hspace{1em}%
    \subcaptionbox{\label{fig:spt8}}{\includegraphics[width=.32\textwidth]{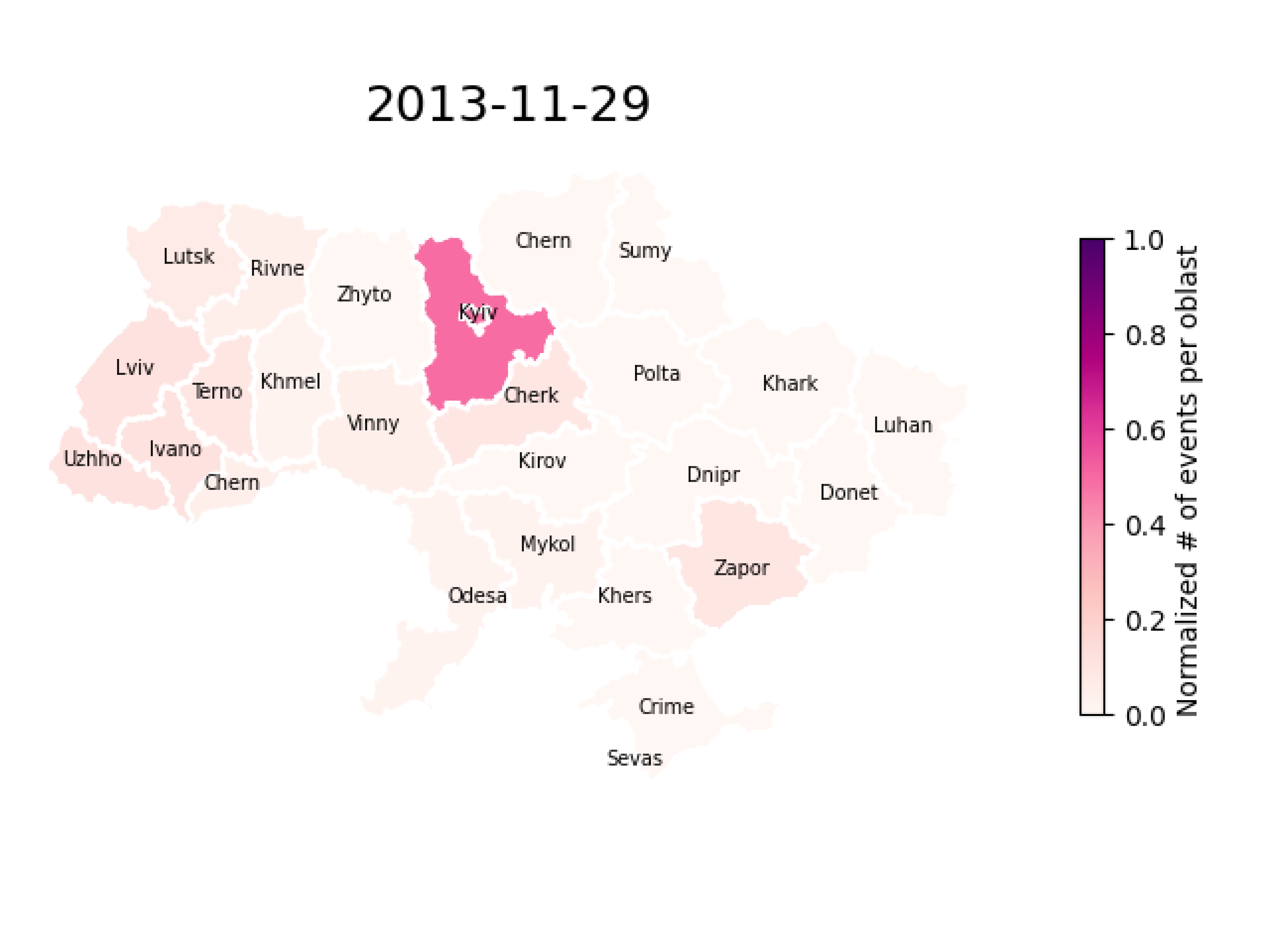}}\hspace{1em}%
    \\
    \subcaptionbox{\label{fig:spt9}}{\includegraphics[width=.29\textwidth]{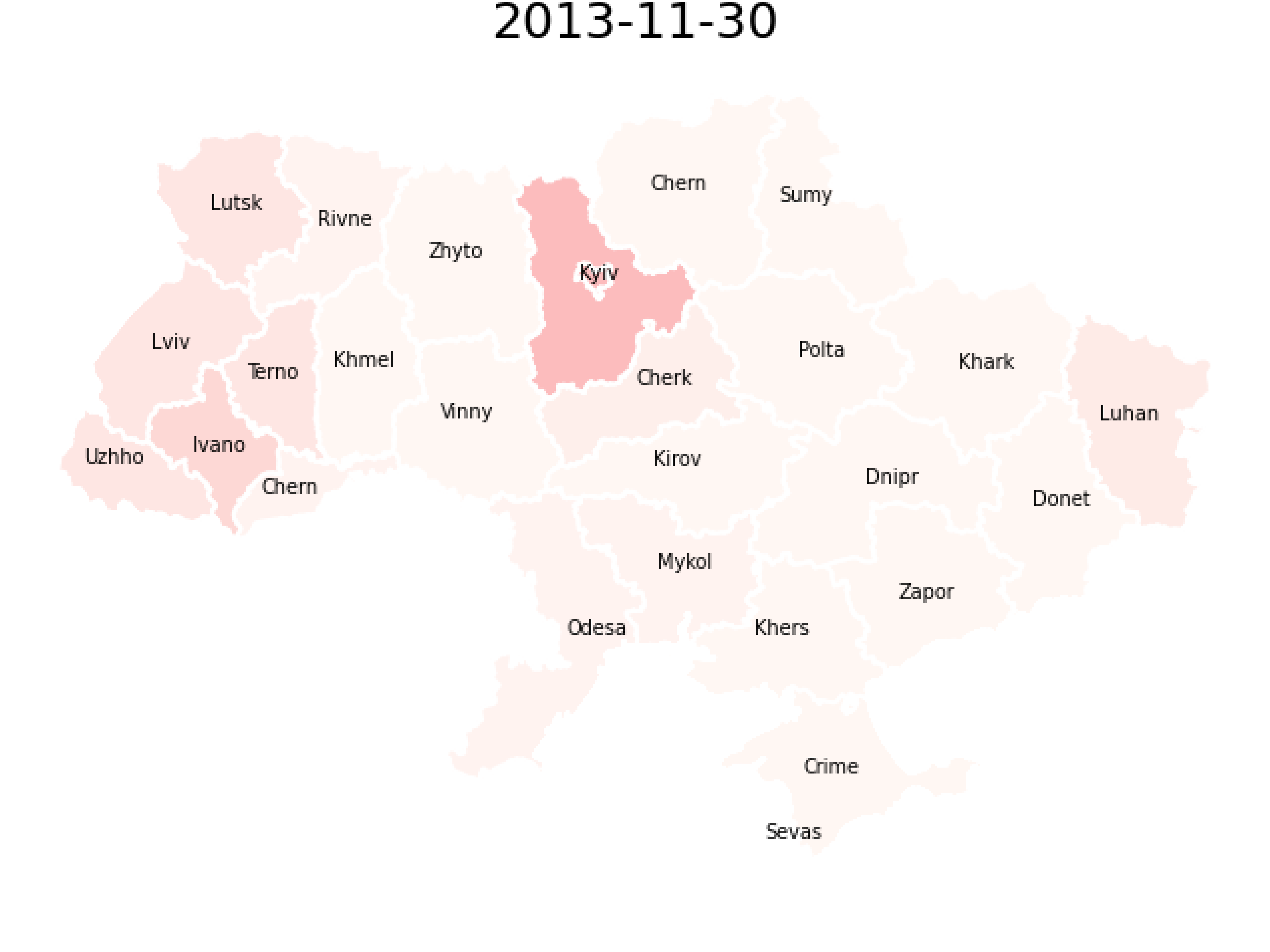}}\hspace{1em}%
    \subcaptionbox{\label{fig:spt10}}{\includegraphics[width=.29\textwidth]{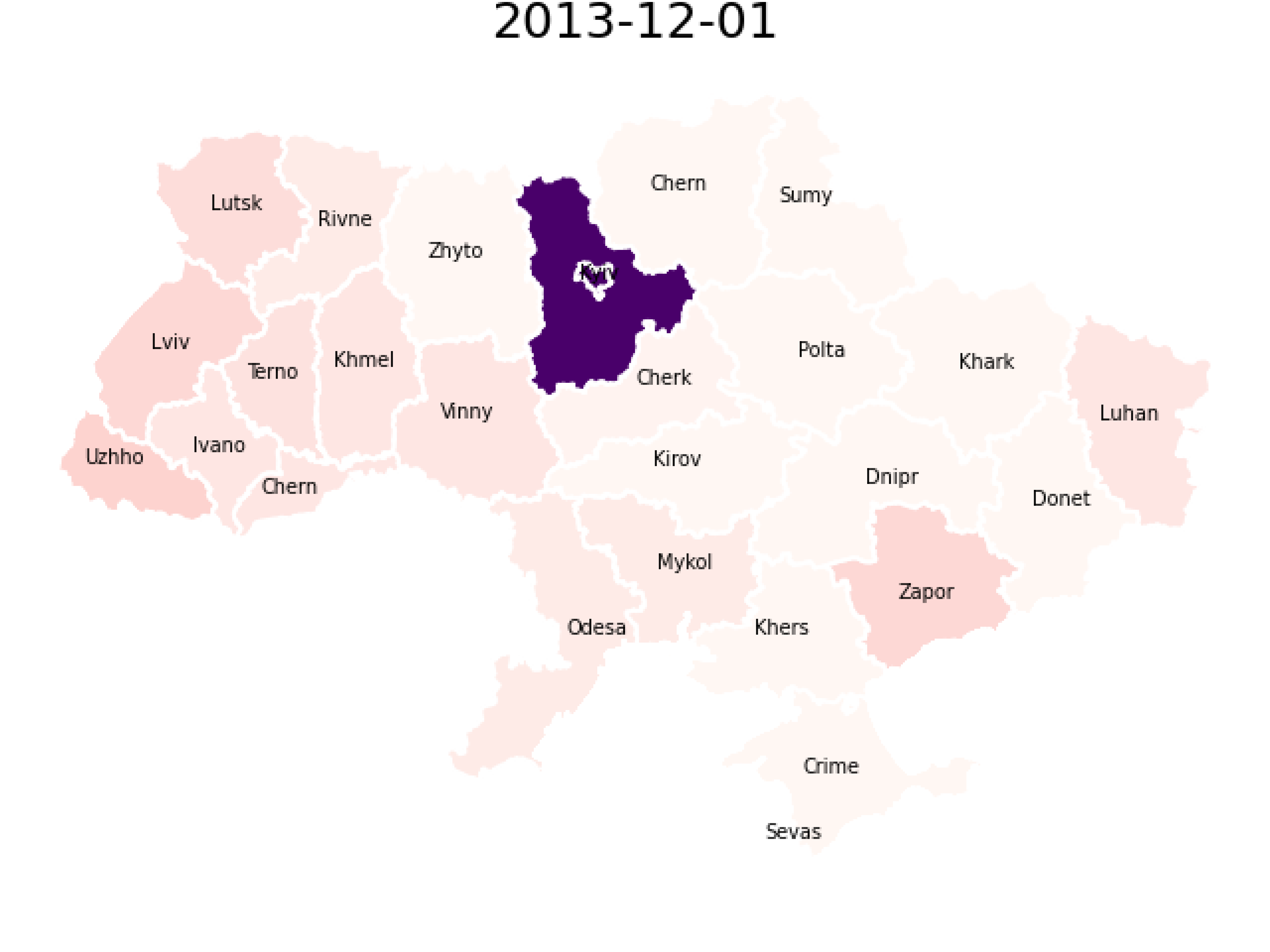}}\hspace{1em}%
    \subcaptionbox{\label{fig:spt11}}{\includegraphics[width=.32\textwidth]{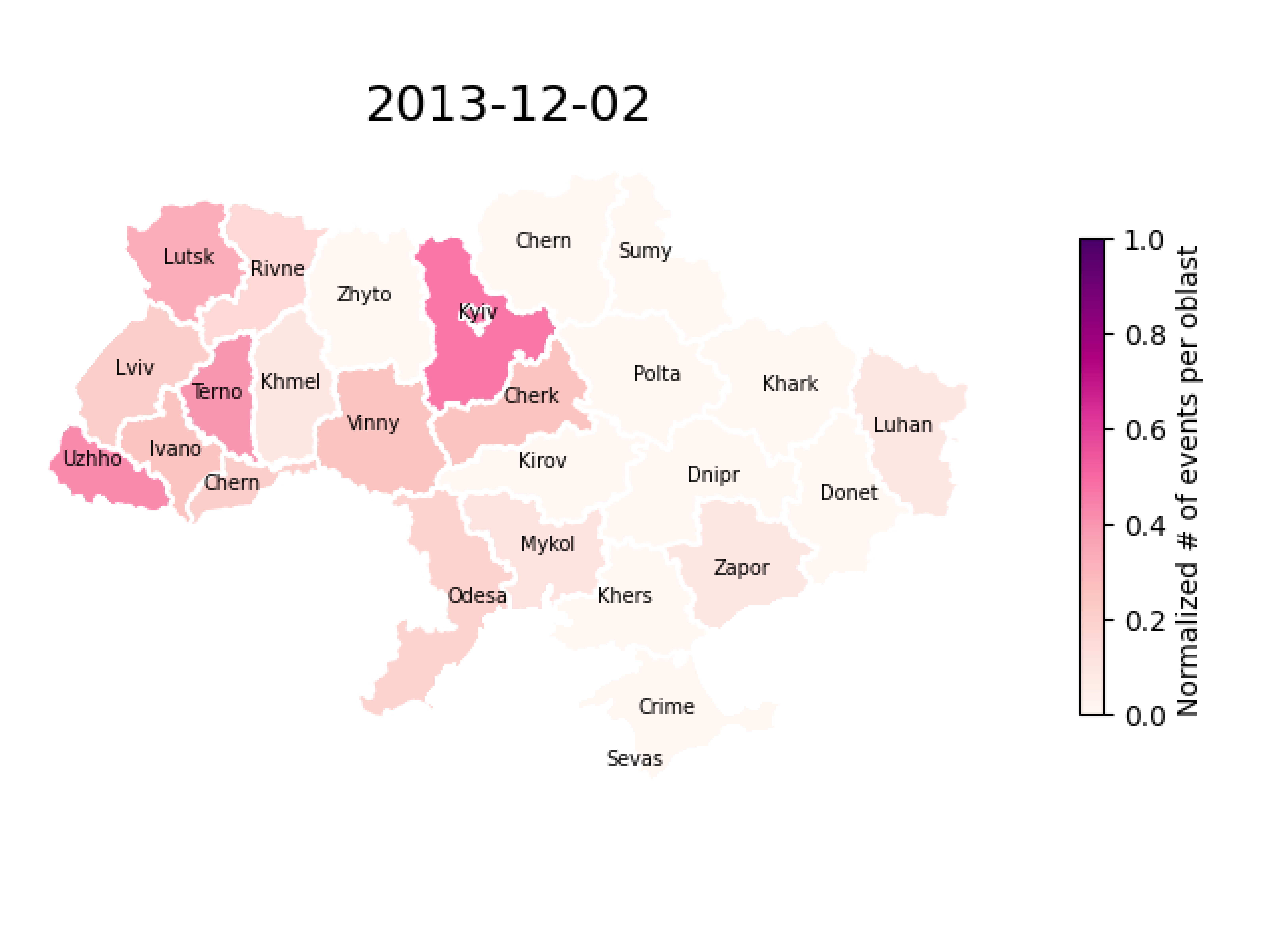}}\hspace{1em}%
    \\
    \subcaptionbox{\label{fig:spt12}}{\includegraphics[width=.29\textwidth]{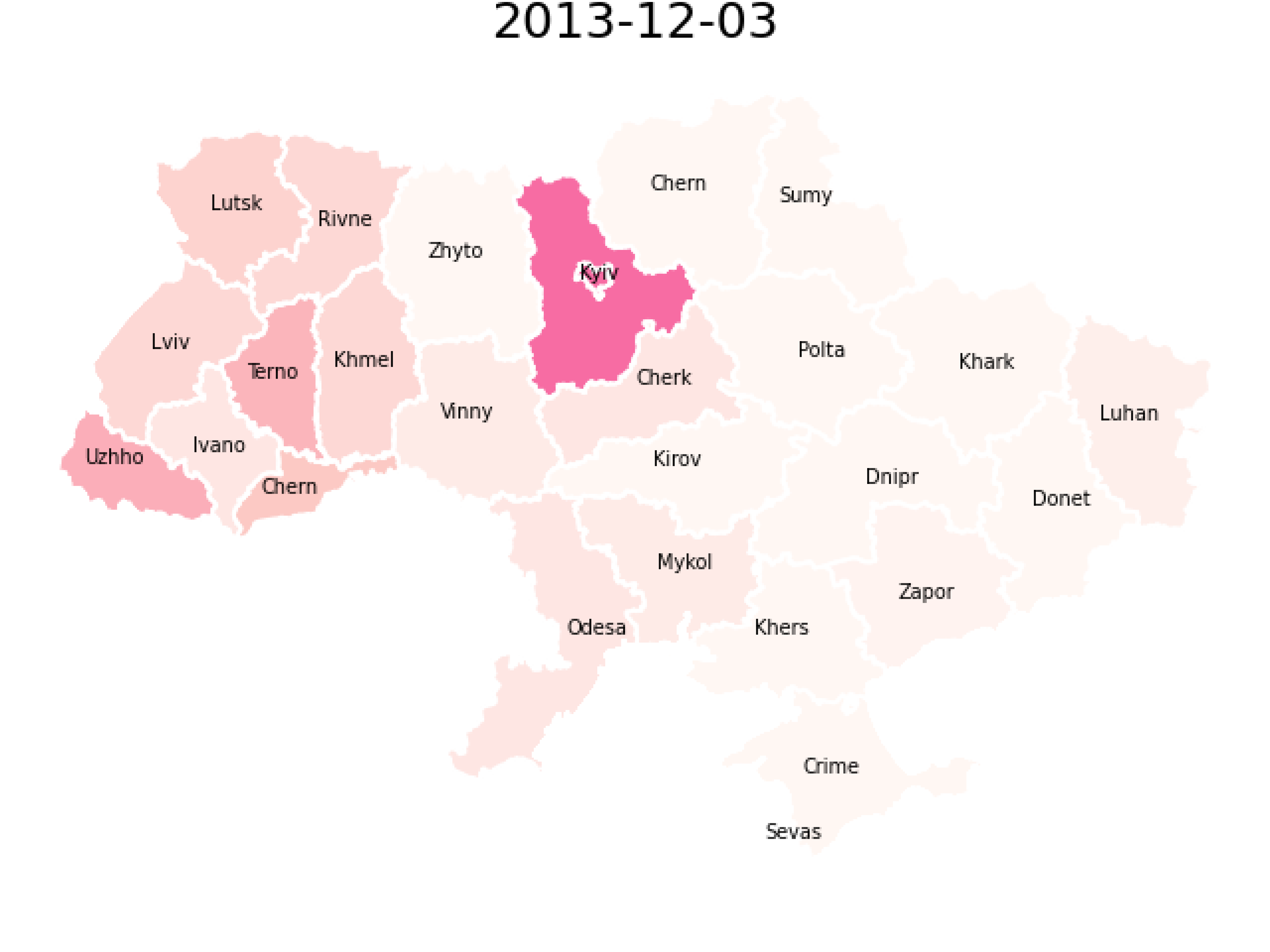}}\hspace{1em}%
    \subcaptionbox{\label{fig:spt13}}{\includegraphics[width=.29\textwidth]{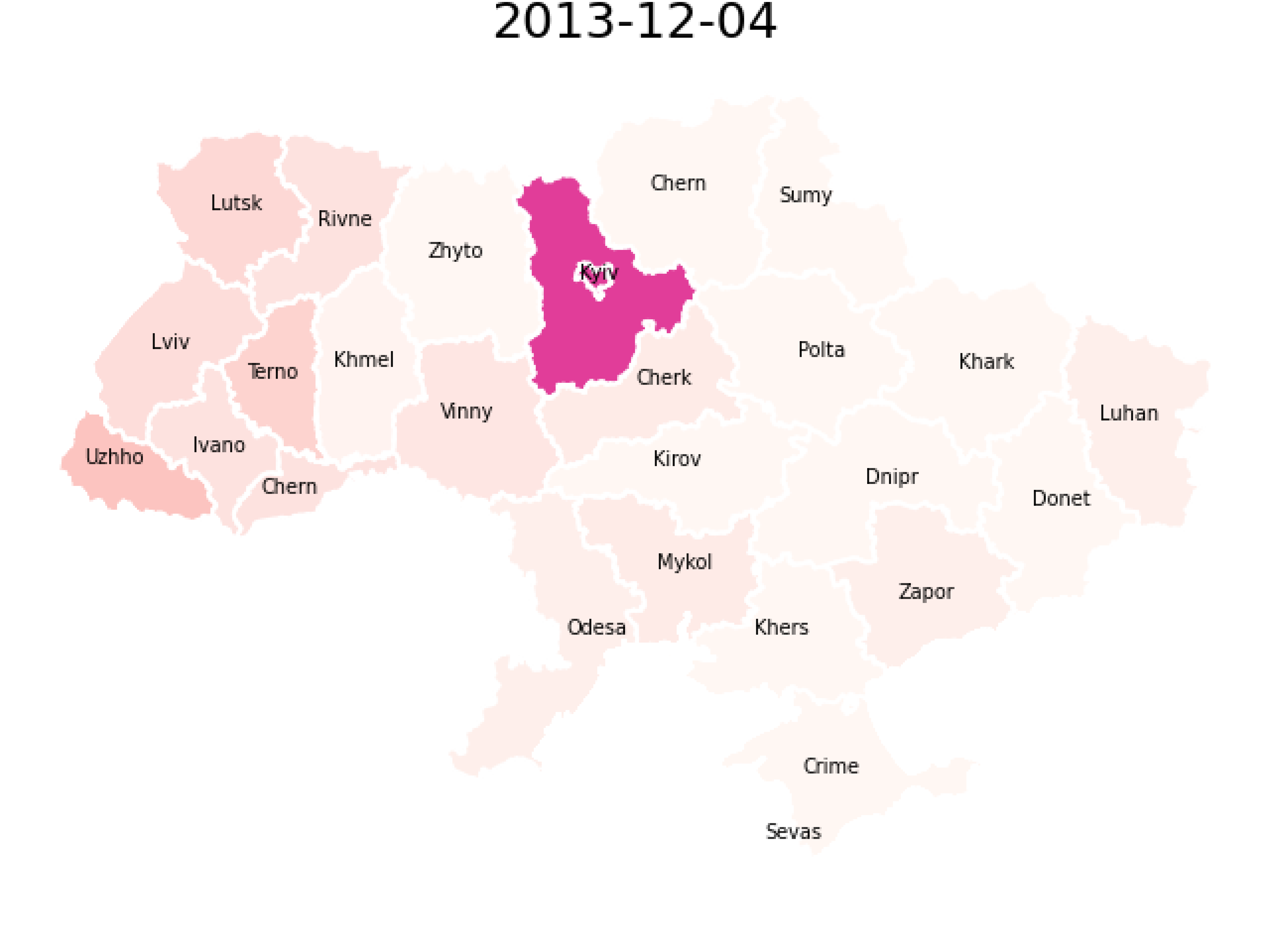}}\hspace{1em}%
    \subcaptionbox{\label{fig:spt14}}{\includegraphics[width=.32\textwidth]{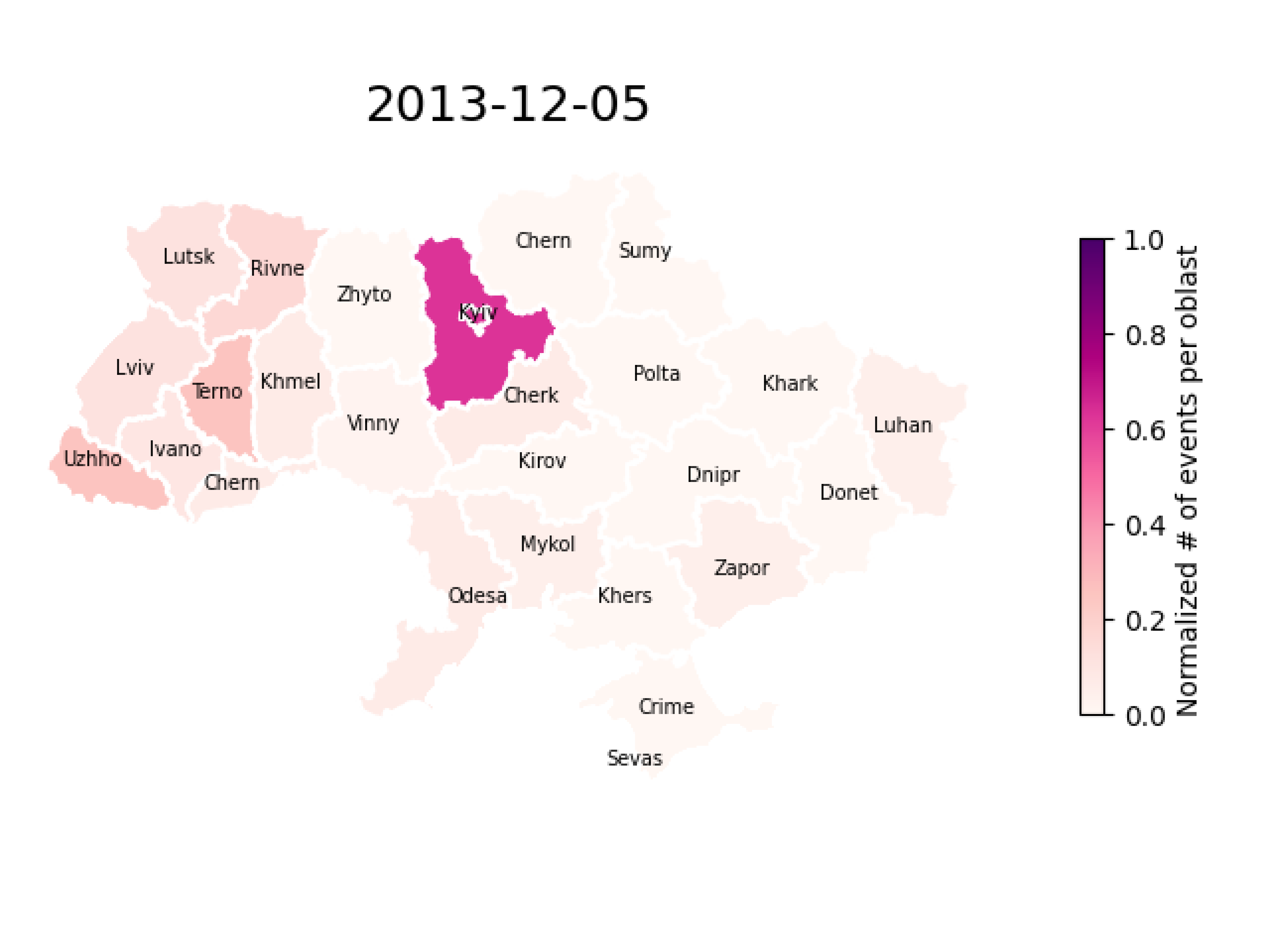}}\hspace{1em}%

    \caption{A spatiotemporal illustration of the number of total events per oblast from November 21, 2013, to December 05, 2013.  These events include protests, rallies, riots, and police crackdowns.}
    \label{fig:sptp1}
\end{figure}

Figure \ref{fig:sptp1} offers a geographic visual of the spread of events for the first six days of protests. The protests were heavily localized in the Kyiv oblast, and more specifically in the Maidan square located in the capital. Once these events reach their peak, we observe the spread of activity throughout the western and central parts of the country. These are the regions that are most pro-EU \cite{young2015} as one can see in Figure \ref{fig:votemap} which shows the votes for candidate Petro Poroshenko, the pro-EU candidate, in the 2014 elections following Euromaidan. We shall use these votes as an indicator of the political leanings of each oblast in our model later on. We believe that the 2014 votes are most reflective of the Ukrainian political map as they are from the election that happened in response to the protests and reflect the public opinion's shift from the 2010 election where President Yanukovych won by a thin margin.

\begin{figure}[!h]
    \centering
    \subcaptionbox{\label{fig:obev}}{\includegraphics[width=.31\textwidth]{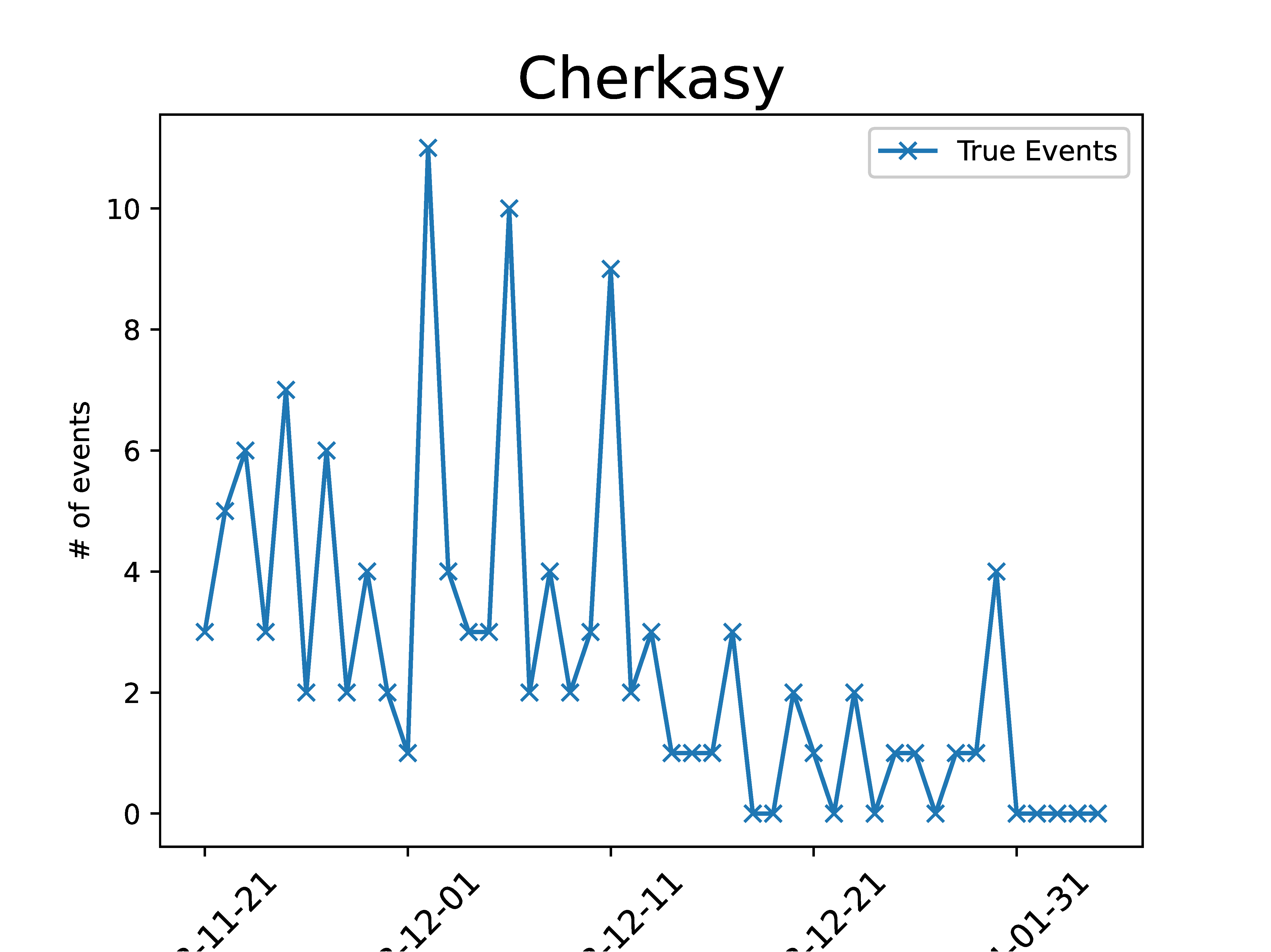}}\hspace{1em}%
    \subcaptionbox{\label{fig:obev}}{\includegraphics[width=.31\textwidth]{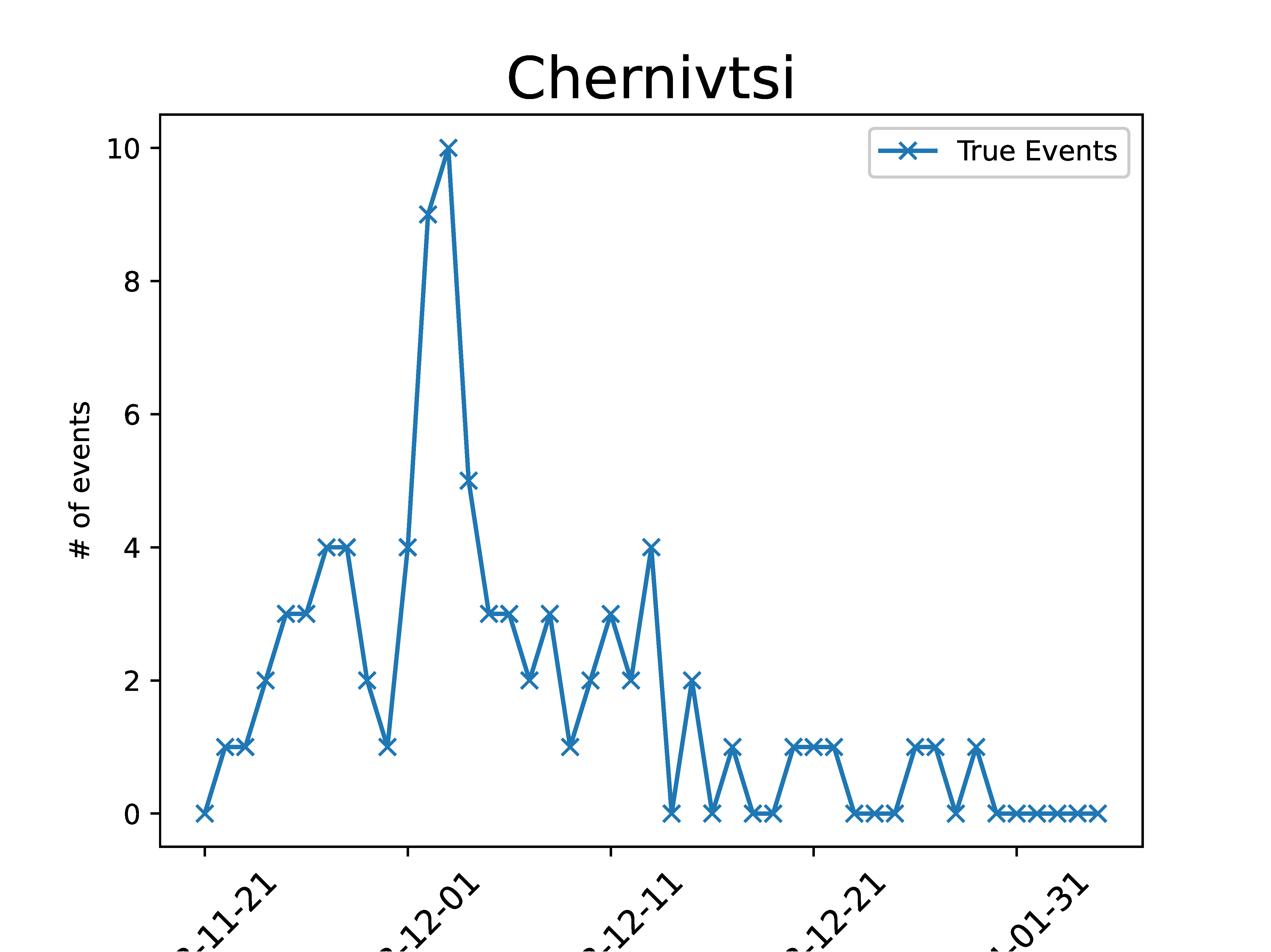}}\hspace{1em}%
    \subcaptionbox{\label{fig:obev}}{\includegraphics[width=.31\textwidth]{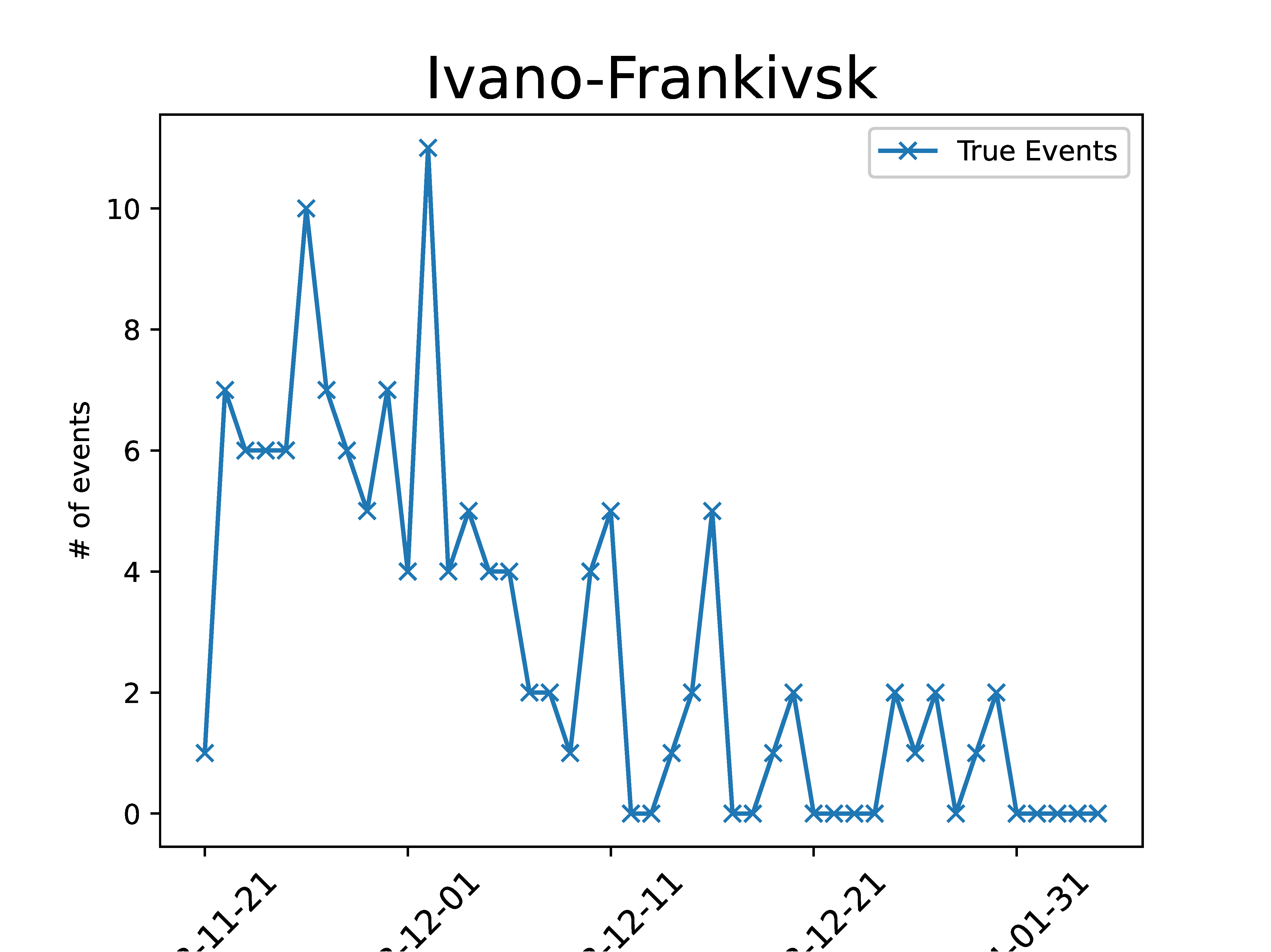}}\hspace{1em}%
    \\
    \subcaptionbox{\label{fig:obev}}{\includegraphics[width=.31\textwidth]{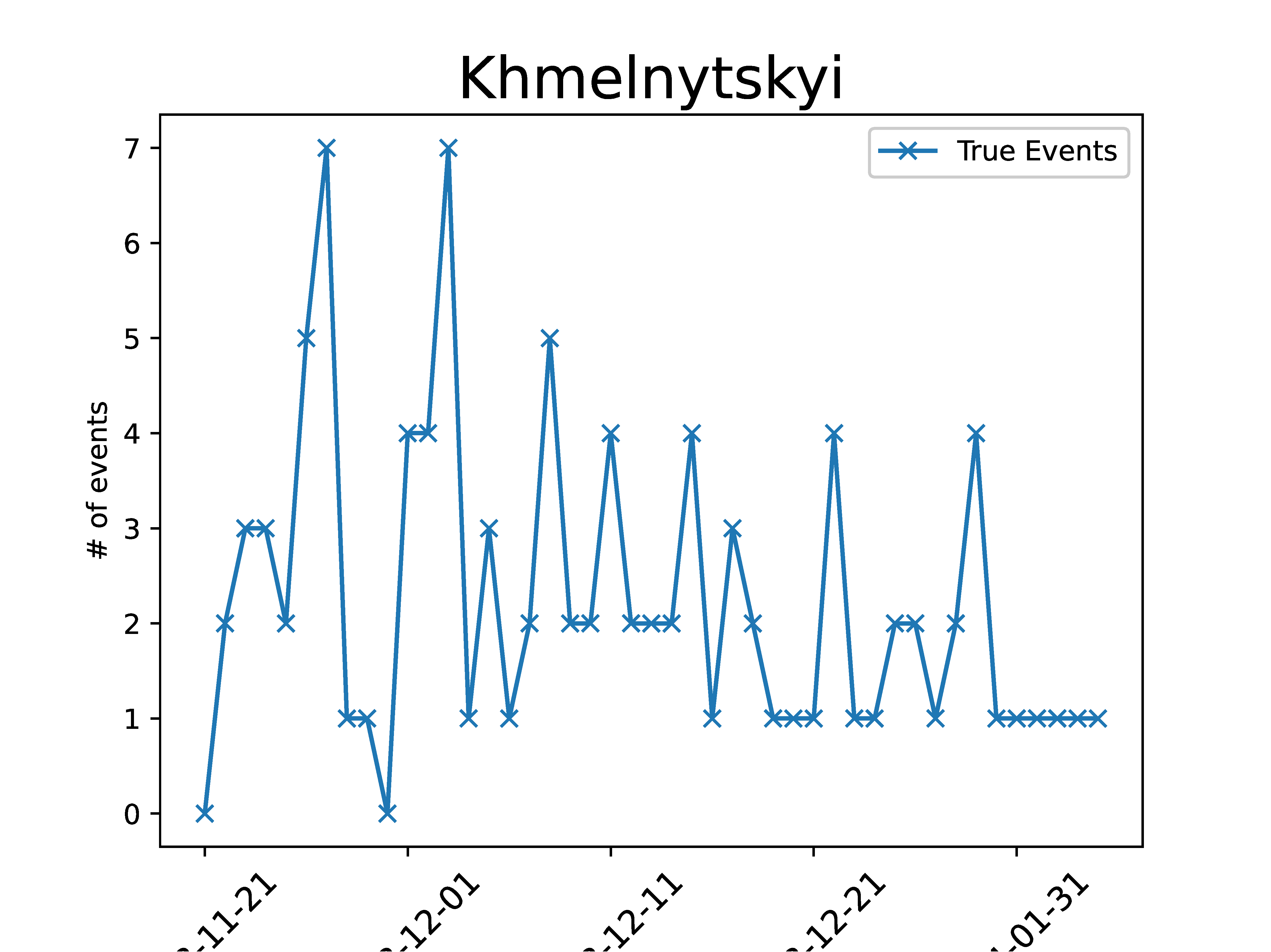}}\hspace{1em}%
    \subcaptionbox{\label{fig:obev}}{\includegraphics[width=.31\textwidth]{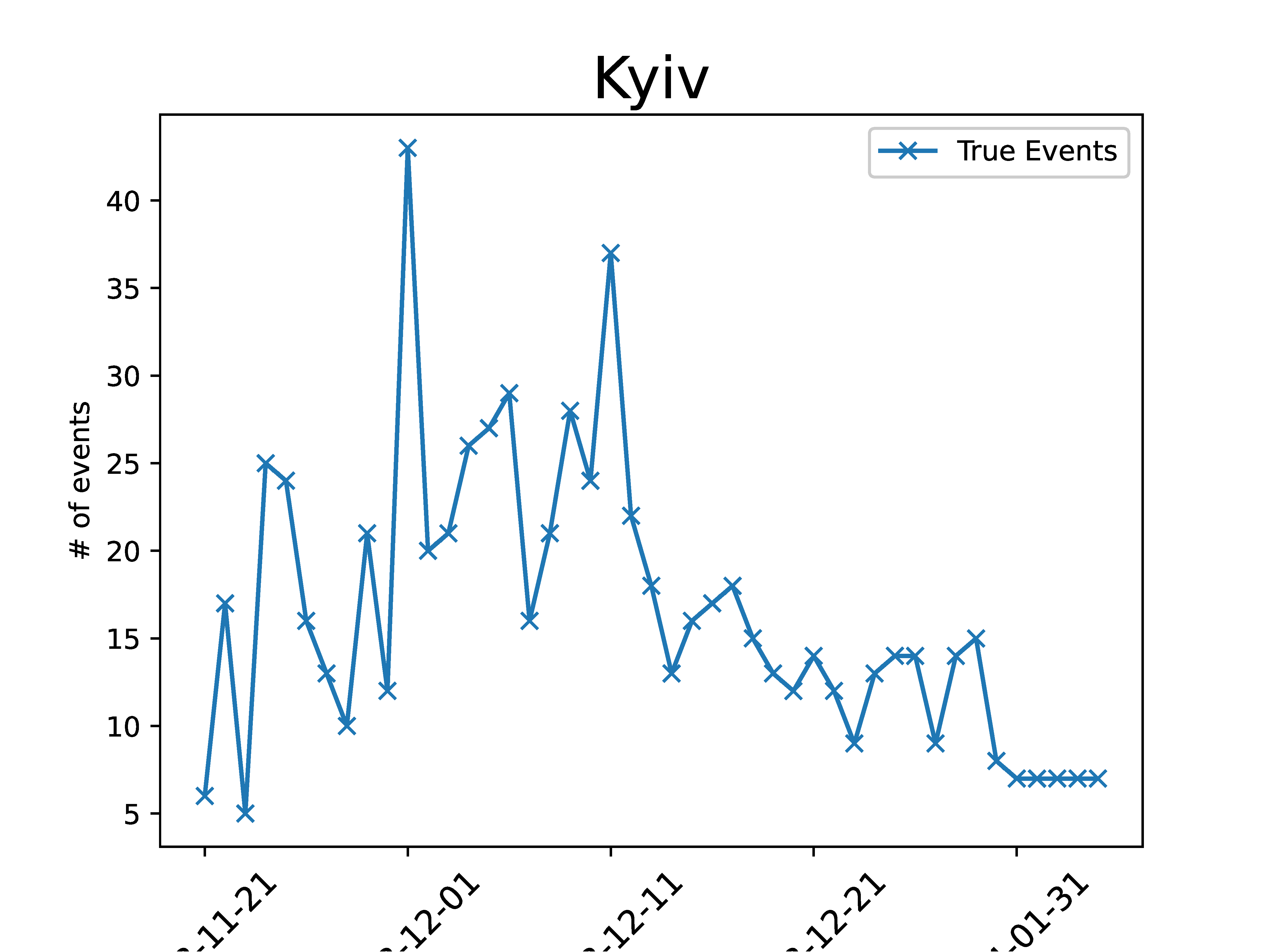}}\hspace{1em}%
    \subcaptionbox{\label{fig:obev}}{\includegraphics[width=.31\textwidth]{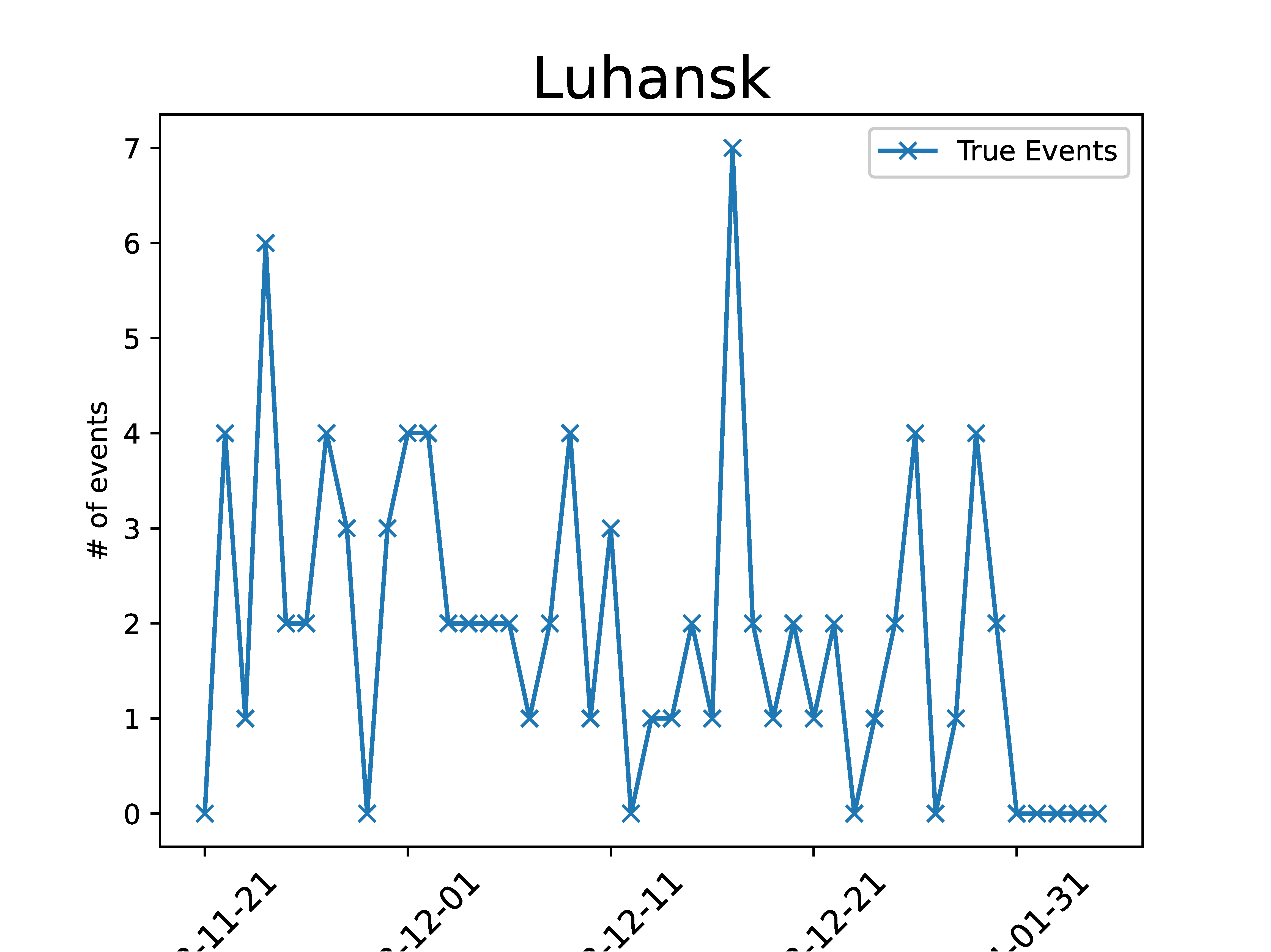}}\hspace{1em}%
    \\
    \subcaptionbox{\label{fig:obev}}{\includegraphics[width=.31\textwidth]{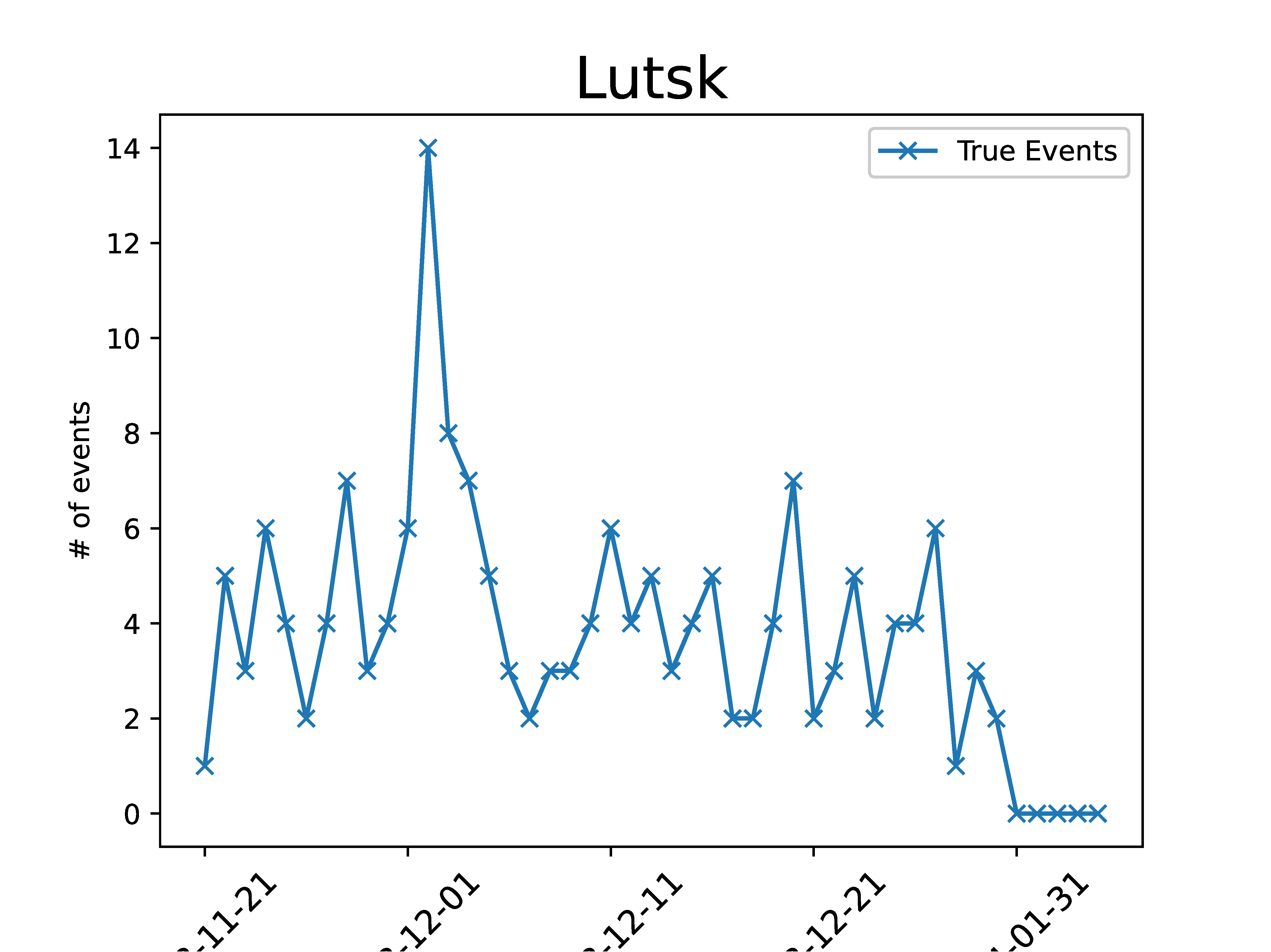}}\hspace{1em}%
    \subcaptionbox{\label{fig:obev}}{\includegraphics[width=.31\textwidth]{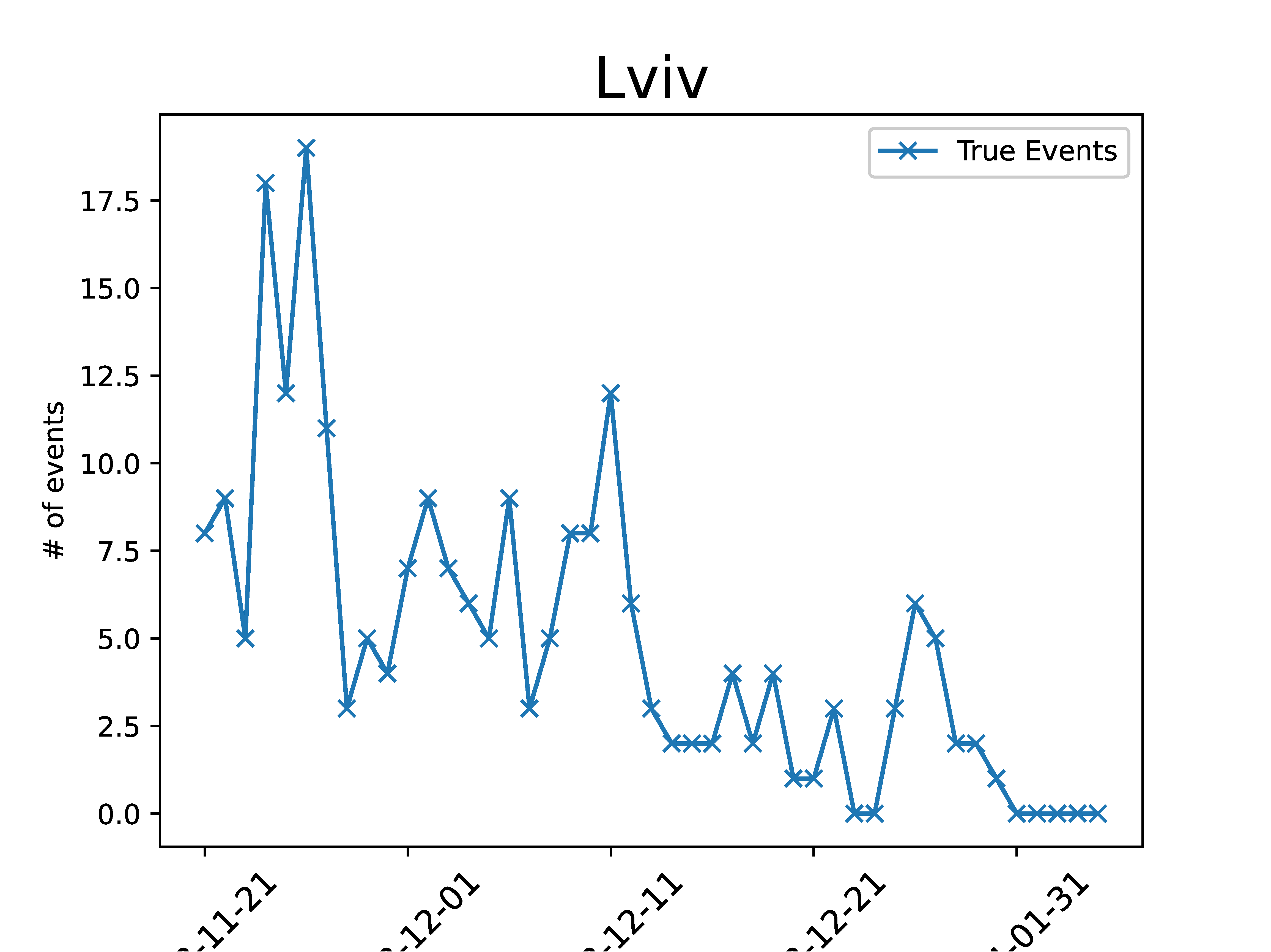}}\hspace{1em}%
    \subcaptionbox{\label{fig:obev}}{\includegraphics[width=.31\textwidth]{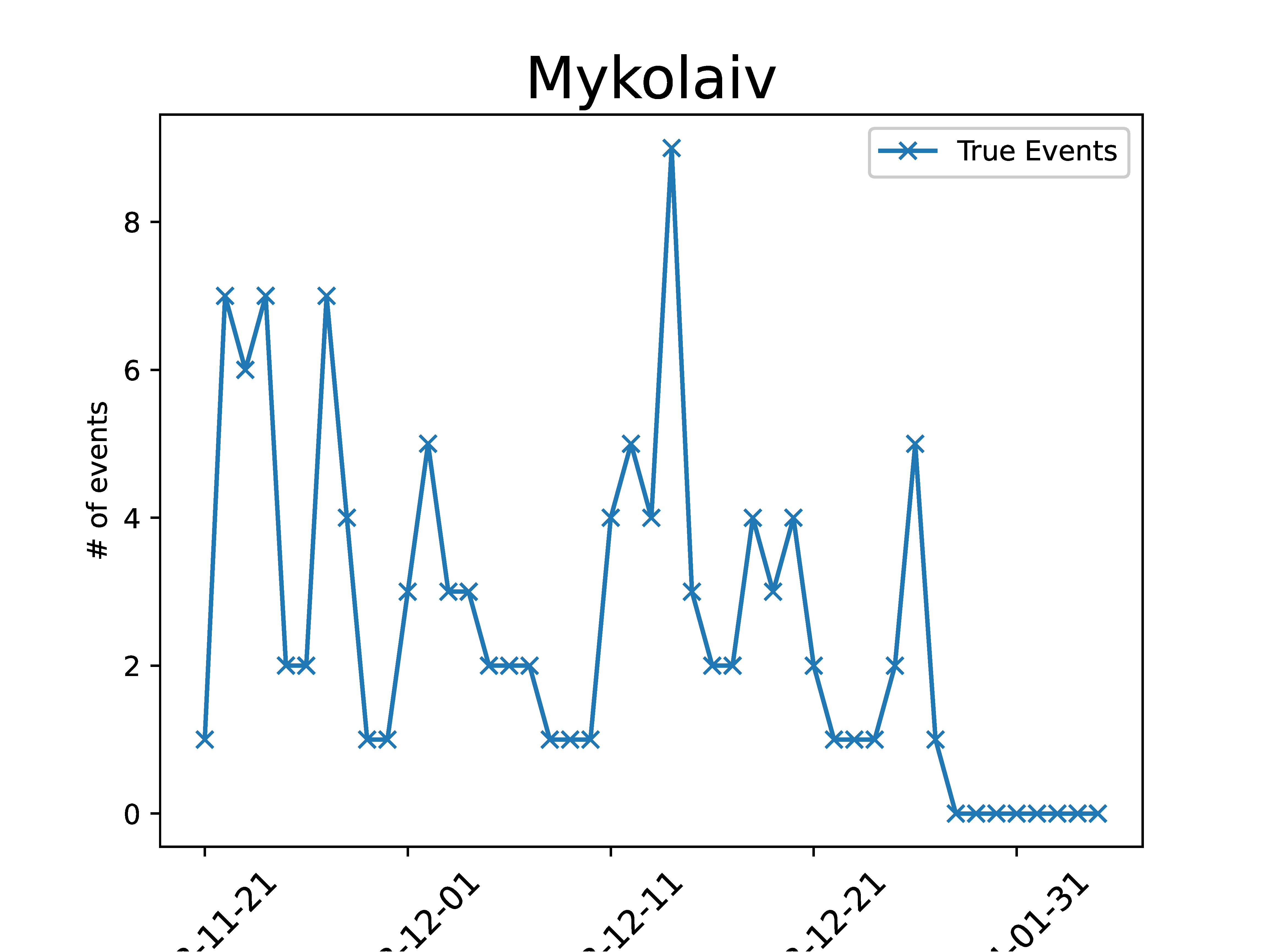}}\hspace{1em}%
    \\
    \subcaptionbox{\label{fig:obev}}{\includegraphics[width=.31\textwidth]{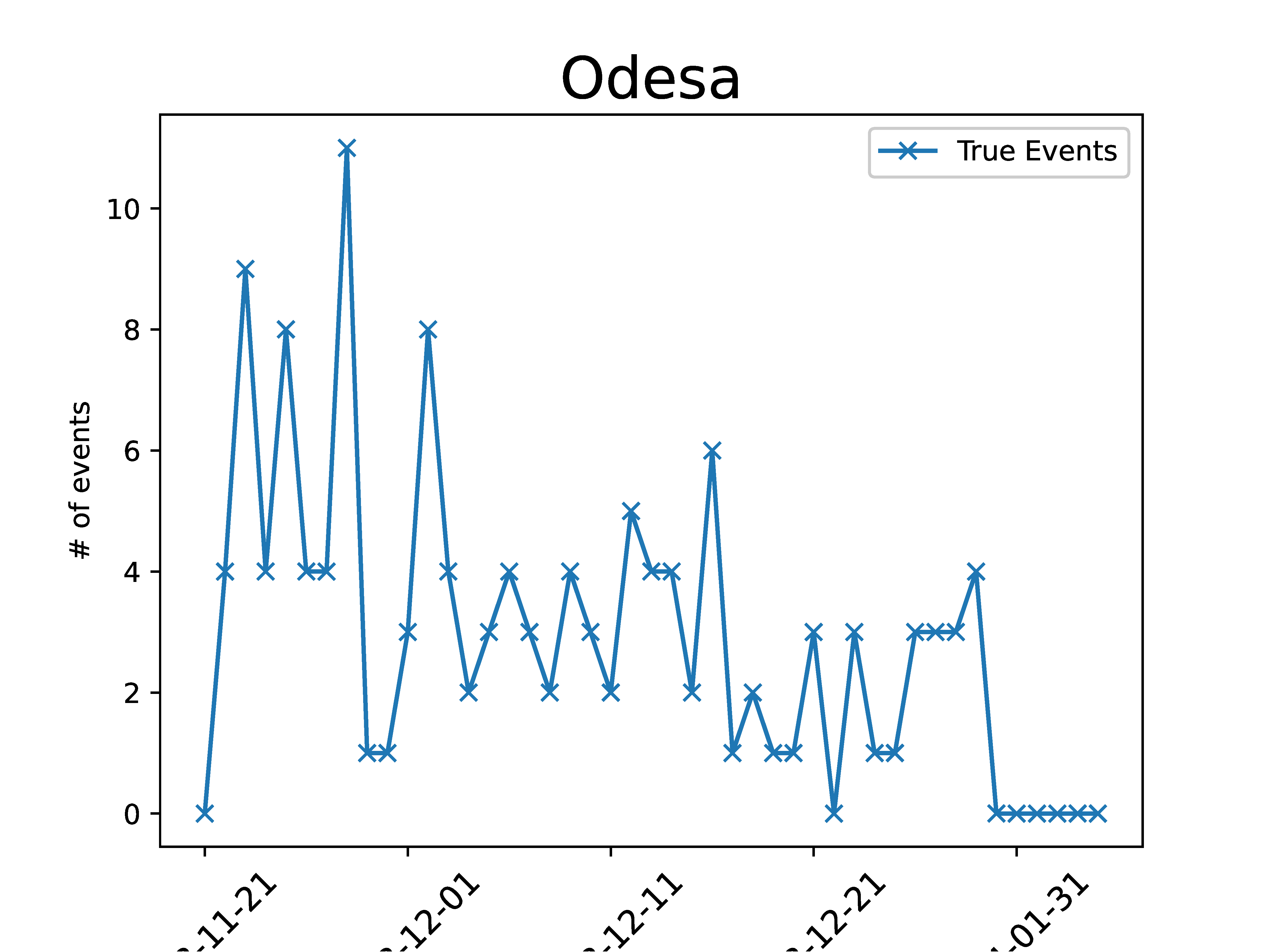}}\hspace{1em}%
    \subcaptionbox{\label{fig:obev}}{\includegraphics[width=.31\textwidth]{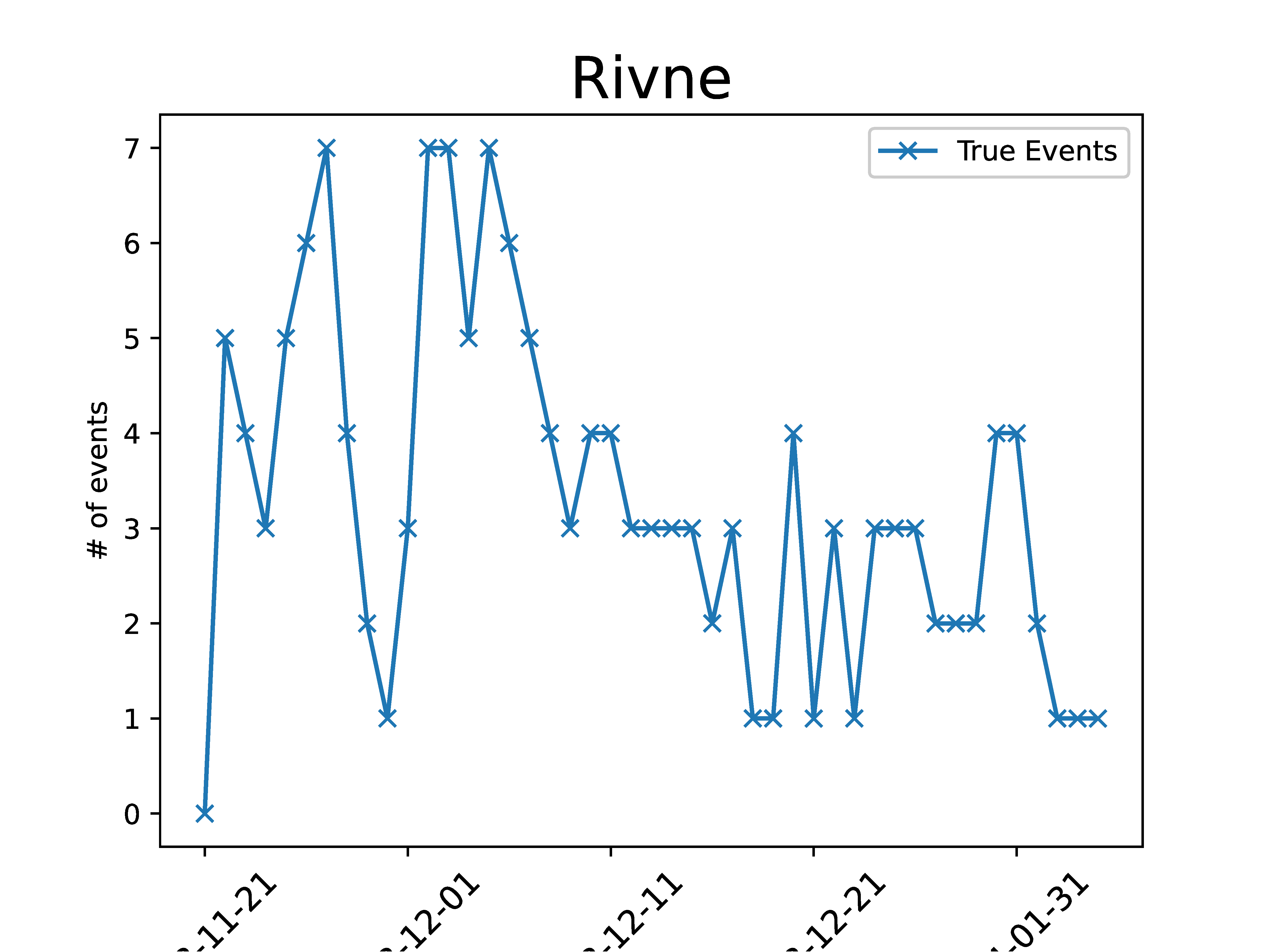}}\hspace{1em}%
    \subcaptionbox{\label{fig:obev}}{\includegraphics[width=.31\textwidth]{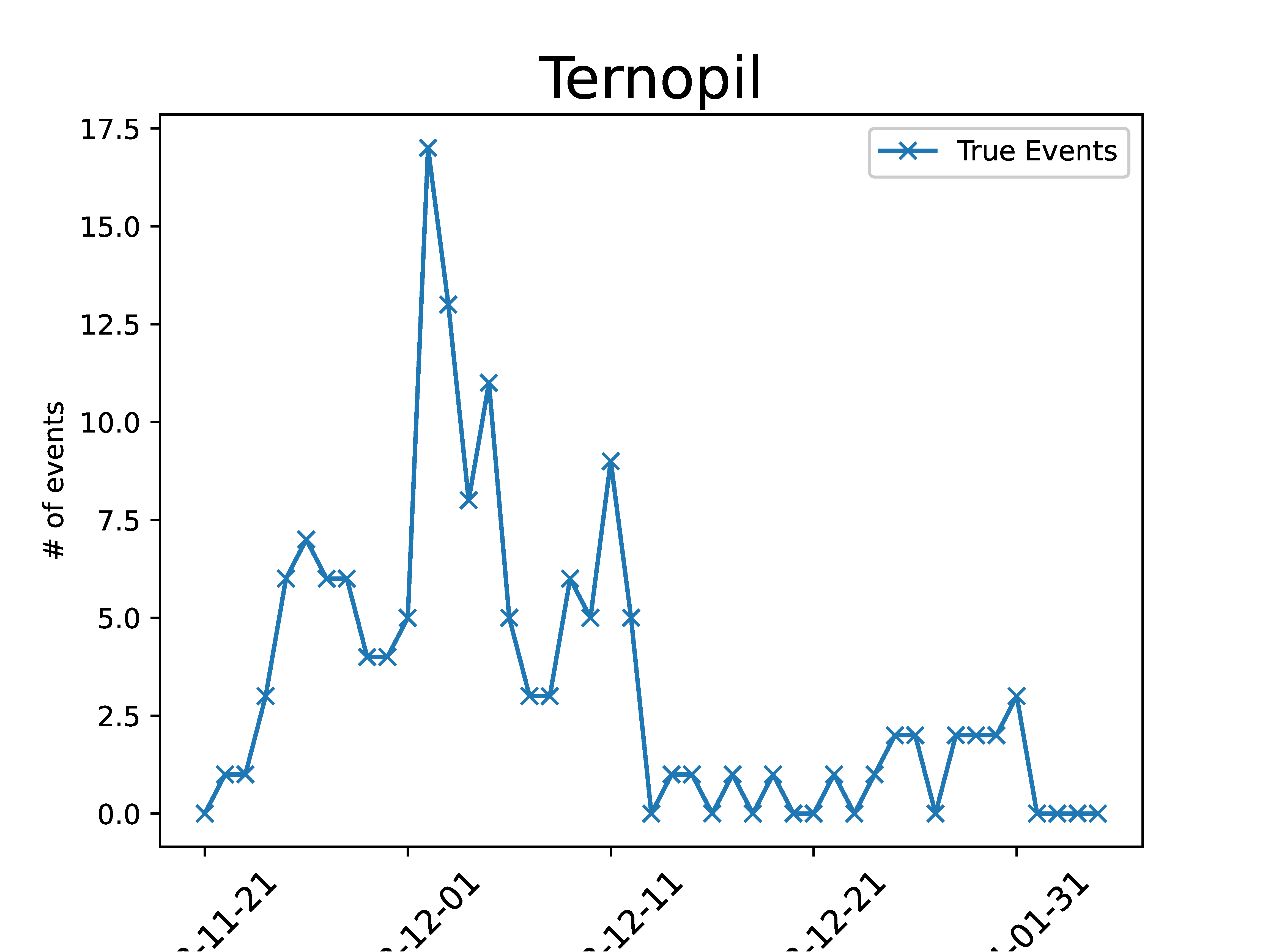}}\hspace{1em}%
    \\
    \subcaptionbox{\label{fig:obev}}{\includegraphics[width=.31\textwidth]{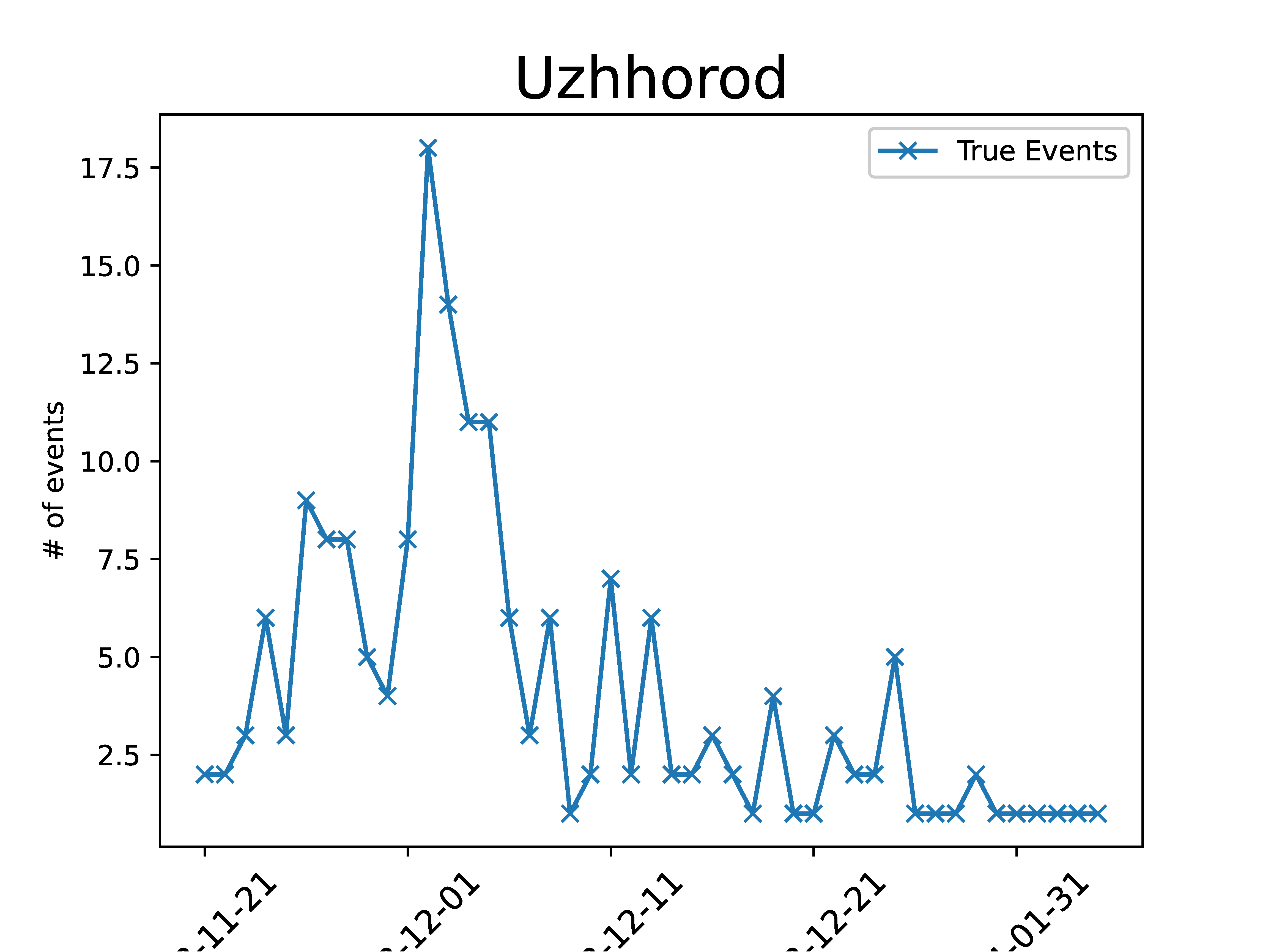}}\hspace{1em}%
    \subcaptionbox{\label{fig:obev}}{\includegraphics[width=.31\textwidth]{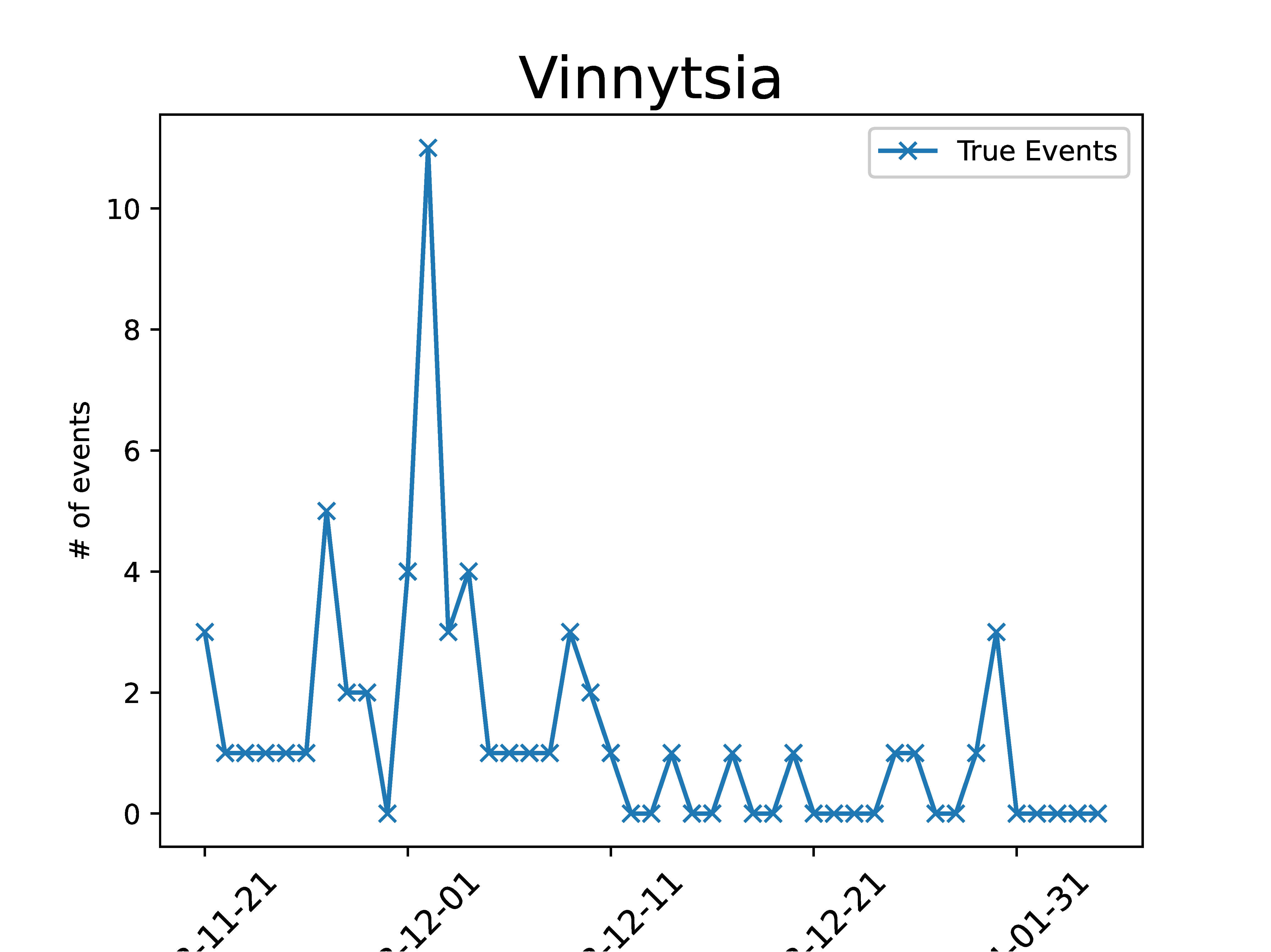}}\hspace{1em}%
    \subcaptionbox{\label{fig:obev}}{\includegraphics[width=.31\textwidth]{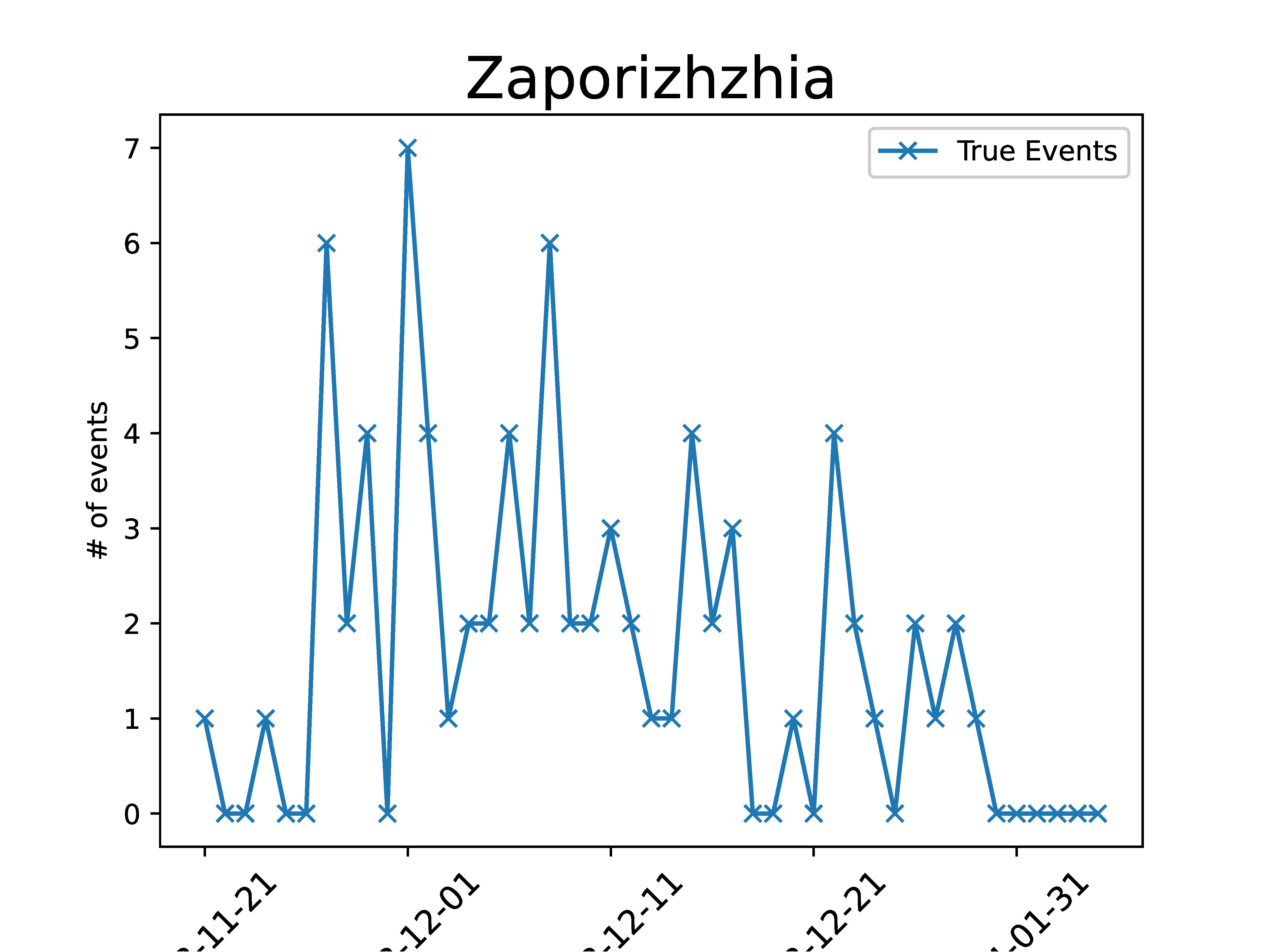}}\hspace{1em}%
  \caption{The number of events, such as protests, throughout different oblasts in Ukraine from November 25, 2013 to February 20, 2014.}
  \label{fig:oblasttmp}
\end{figure}

\begin{figure}[!h]
    \centering
     \subcaptionbox{\label{fig:sfig22}}{\includegraphics[width=.45\textwidth]{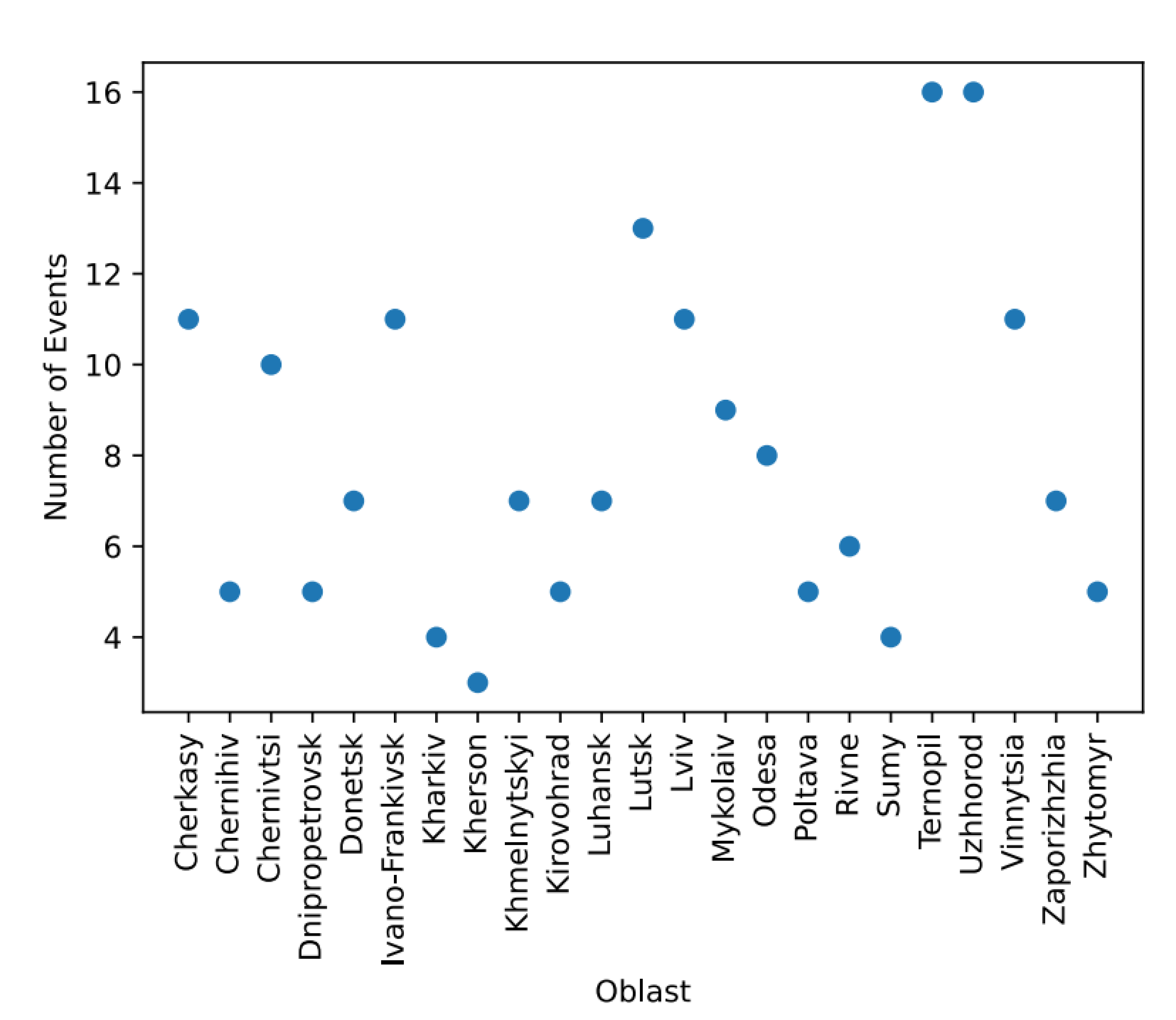}}\hspace{1em}%
     \subcaptionbox{\label{fig:sfig23}}{\includegraphics[width=.45\textwidth]{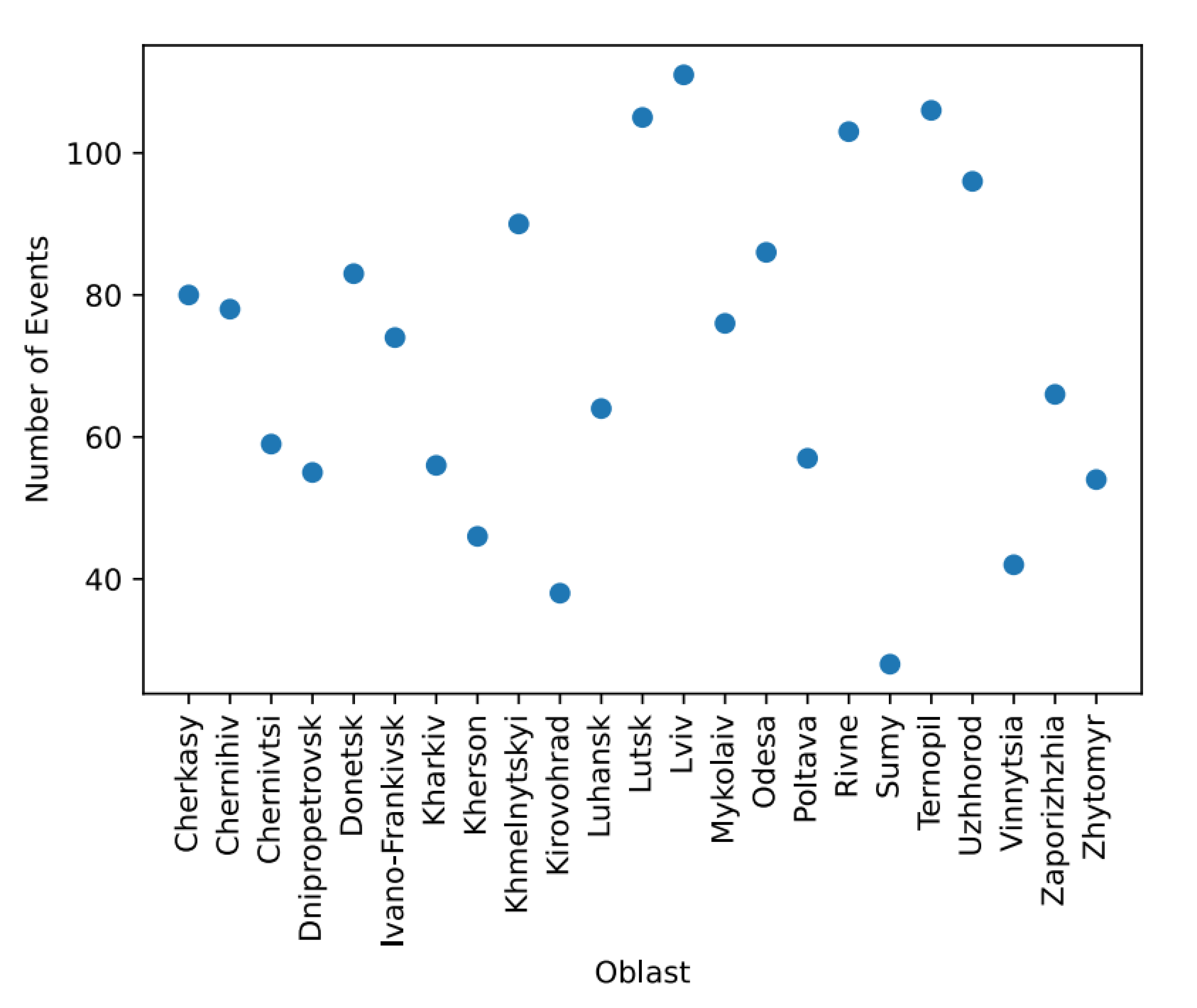}}\hspace{1em}%
\caption{(a) The maximum daily number of events in all oblasts excluding Kyiv. (b) The total number of events from November 11, 2013, to February 02, 2014, in all oblasts excluding Kyiv.}
\label{fig:maxtot1}
\end{figure}

Figure \ref{fig:oblasttmp} shows the wide difference between the magnitude of events in Kyiv and all other oblasts. Figure \ref{fig:maxtot1} details the maximum number of events per day and total number of events per oblast. These graphs purposefully omit Kyiv oblast, as the total number of events and the maximum number of events far exceeds any other oblast (a total of more than 800 events over the entire period with a maximum of 47 events for one day). Moreover, one can see that some regions register little activity, namely: Chernihiv, Donetsk, Dnipropetrovsk, Kharkiv, Kherson, Kirovohrad, Poltava, Sumy, and Zhytomyr. The geopolitical situation, i.e., their proximity to Russia and their political leanings, might be the direct cause of these discrepancies. As there is no represented activity in such oblasts, we choose to omit them from our model. The lack of activity is mainly due to the lack of representation of such oblasts' activity in the media.  There were regular pro-EU protests and later anti-Maidan ones in these oblasts. However, they were not very crowded, rarely manifested violent events, and were not sensational. Thus, they did not regularly make the pages of popular media. 

A careful look at the data, unfortunately, does show some discrepancies. First, the number of protesters is not reported for all events. Second, the number of deceased individuals is heavily underrepresented. For example, the Kharkiv Human Rights Protection Group, one of Ukraine's oldest and most respected human rights organizations,  reports the death of Pavlo Mazurenko, on the 24th of December, 2013,  in a Kyiv hospital after a violent police interaction \cite{khpg2013}. However, this tragic event is missing from our data set. Third, and most important is the fact that the data set is missing important activity from the first two months of 2014. The only events reported in 2014 are those that began in 2013 and spilled over into 2014.  This period has seen a resurgence in protests and clashes between the pro-Euromaidan protesters and police \cite{fisher_2021}\cite{UA_2014} \cite{kyivpost2014}. The turmoil that Ukraine plunged into after the Euromaidan revolution, namely the Russo-Ukrainian war, made it difficult to collect data or local news reports in the country. The international news unfortunately does not offer the level of granularity needed for this type of analysis we are conducting.

\begin{figure}[!h]
 \centering
 \includegraphics[width = 0.6\textwidth]{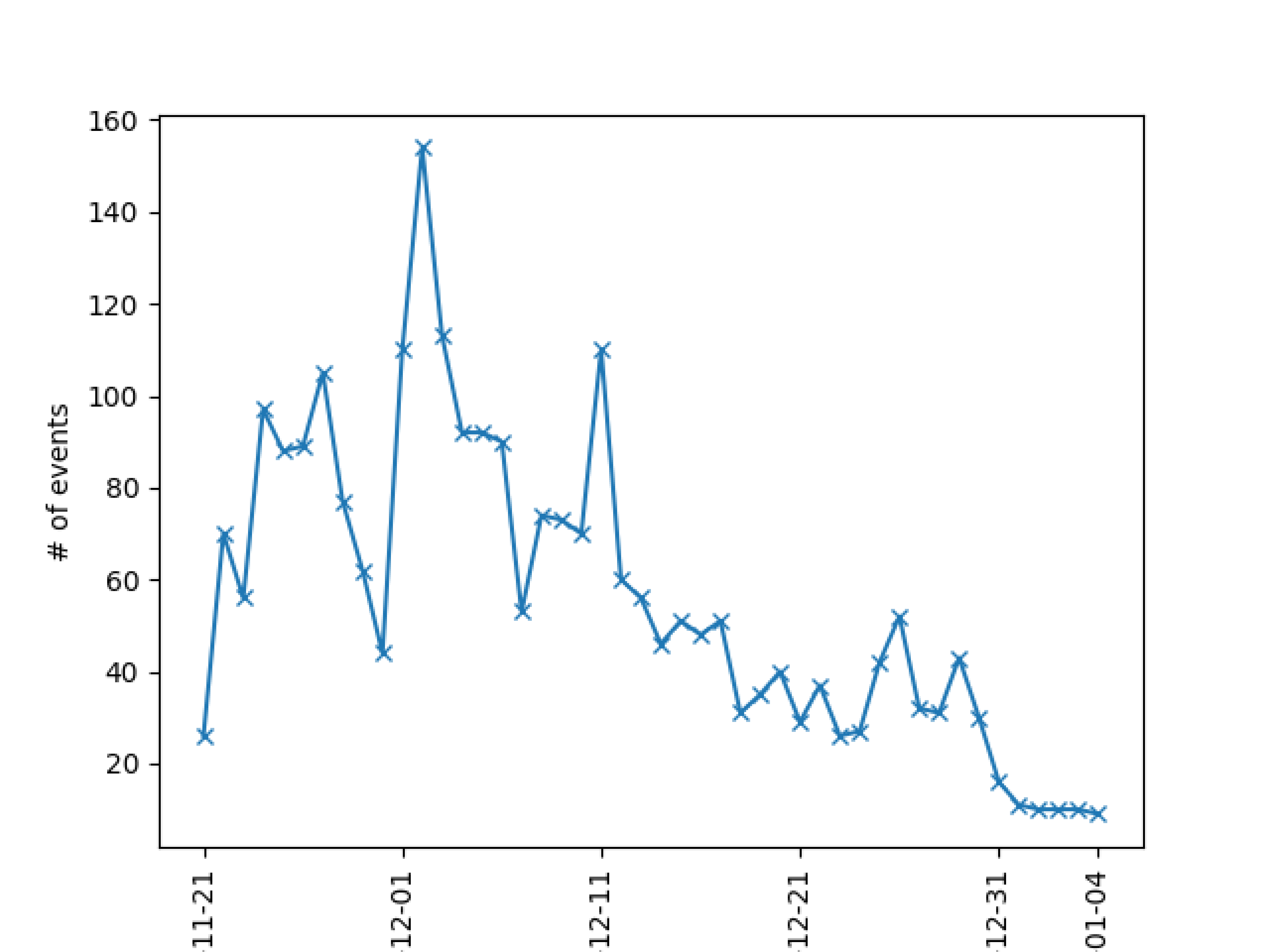}
 \caption{The total number of events per day in Ukraine from November 21, 2013, to January 4, 2014, including protests, rallies, riots, and police crackdowns registered in media under scrutiny.}
 \label{fig:uk1}
\end{figure}

The fact that the data set does not contain events that began in 2014 forces us to focus on modeling only the first half of Euromaidan. Given the fact that there was a clear break in the protests, before the resurgence in mid-January, it is sound to study the first half of Euromaidan as its own contained self-excitation event, which will be further demonstrated in the statistical analysis section. Moreover, we must also model the number of events and not the number of protesters. This can be seen as a limitation given that large protests may have a larger impact than others.  However, as seen in \cite{Bonnasse-Gahot2018}, the magnitude of a protest does not always correlate to the importance of the event. For example, an event that has only a few protesters can cause a chain reaction and lead to more important events in the future. Note also that we do not differentiate between pro and anti-European Union protests, and do not take into account the different political leanings of all groups involved in the different events. The main reason behind this is that all protests do add to the general tension and, as seen from the statistical analysis, lead to more protests. See the statistical analysis section below for details supporting this assertion. As one can see in Figure \ref{fig:uk1} and Figure \ref{fig:oblasttmp}, which respectively illustrate the overall number of events in all of Ukraine and each of its oblast per day from November 21, 2013, to January 4, 2013, there is a lack of activity starting from January 1, 2014. This is because the only events in January and February are those that were initiated in 2013. However, as mentioned above, the protests did increase significantly after the new year and resulted in the ousting of the president of Ukraine. Therefore, the data set is missing most key events that happened during the second part of the protests. 

\subsection*{Statistical analysis}\label{sec:sa}
The data set consisted of one row per protest, with columns containing information on the start date, end date, oblast where the protest occurred, and numbers of protesters, arrests, injuries, and fatalities. Unfortunately, due to missing data issues, most of these numerical fields cannot be used, as remarked above, but we do use the number of injuries in our final model, because of its relevance in a similar model in \cite{Rodriguez-White}. Our primary goal in this section is to build a model to predict the number of protest events on any given day and to show that the time series ``number of events'' exhibits self-excitation behavior. Towards that goal, we first wrangle the given data set into a different form, where the fundamental unit is a day, rather than a protest. Each day in this new data set will have information about how many protests occurred that day, as well as three covariates listed below. 

To create a variable containing the number of events per day, we transformed the data and extracted several time series: 
\begin{itemize}
    \item $p_t$ is the number of events on day $t$ (that is, all events where $t$ is between the start and end date, inclusive)
    \item $nr_t$ is the number of events with a ``negative response" on day $t$
    \item $e_t$ is the number of events associated with Euromaidan on day $t$
    \item $i_t$ is the number of civilians injured on day $t$
\end{itemize}

 In this context, an event $E$ occurring on day $t$ can influence the number of events on day $t+1$ either by extending to last another day, or by spawning subsequent events (e.g., inspired by news coverage regarding $E$). We will show a positive relationship between $p_t$ and $p_{t+1}$, which justifies the slogan ``more events today are associated with even more tomorrow.'' Furthermore, we fit a threshold model to $p_t$, inspired by the spread of epidemics, that illustrates that after a certain tipping point, the relationship between $p_t$ and $p_{t+1}$ is even larger. This provides evidence that a Hawkes process model is appropriate for these data.

The first step is to figure out which of these variables leads/lags the others. Is the number of injuries today associated with the number of events tomorrow, or is it the other way around? We use \textit{cross-correlation analysis} \cite{shumway} after pre-whitening to answer this question.  See Appendix \ref{sec:App3} for details on this methodology. Carrying out this procedure (that is, cross-correlation analysis after suitable pre-whitening) for the time series above demonstrated that $p_t$ is closely and positively associated with lags of the three covariate time series, e.g., most strongly with $i_{t-1}$, $e_{t-3}$, and $nr_{t-5}$. That is, if there were many injuries today, negative response events today, or Euromaidan events today, then one can expect more events overall tomorrow and in the days to come. This suggests that injuries to protesters are associated with more future protests, and provides evidence that Euromaidan events were driving the protests. It also suggests that government negative responses (e.g., police use of force, beatings, etc.) are associated with more future protests, i.e., have an inflammatory rather than suppressing effect on protests. The cross-correlations of the pre-whitened data are useful to be able to say there is a \textit{statistically significant} association between $p_t$ and the lagged variables but lack a meaningful real-world interpretation. So, we instead report the actual correlations (reiterating the caveat that, without pre-whitening, these correlations can be affected by exogenous variables) to illustrate the strength of the relationships between the variables in question.

\[
\begin{tabular}{c|c|c|c|c|c}
     & \mbox{lag 0} & \mbox{lag 1} & \mbox{lag 2} & \mbox{lag 3} & \mbox{lag 4} \\
     \hline 
$e_t$     & 0.91 & 0.79 & 0.75 & 0.72 & 0.7 \\
$nr_t$ & 0.82 & 0.71 & 0.67 & 0.59 & 0.59 \\
$i_t$ & 0.24 & 0.16 & 0.09 & 0.08 & 0.09
\end{tabular}
\]

This table says, for example, that the correlation between $p_t$ and $e_t$ is 0.91, and between $p_t$ and $nr_{t-1}$ is 0.71, etc. Lags on the other side (e.g., between $p_t$ and $nr_{t+1}$) were smaller, and the pre-whitened cross-correlation showed that such lags lack statistical significance. The pattern in the table above is consistent with $p_t$ being affected by each of the three covariates and their lags. As the correlations above are computed without pre-whitening, the lag 2 correlations are likely primarily due to the lag 1 correlation, since $e_{t-2}$ is correlated with $p_{t-1}$ and hence with $p_t$ as we will see below.

We turn now to statistical models. Autoregressive integrated moving average (ARIMA) models are standard statistical models for time series data. They are built of three parts: an AR part, an I part, and an MA part. The methodology for the AR part identifies, for a given time series $y_t$, which lagged time series $y_{t-k}$ have a statistically significant correlation with $y_t$, and then builds a linear model where the response variable is $y_t$, the explanatory variables are the relevant $y_{t-k}$, and the residuals now ideally exhibit independence. In situations of multivariate time series, one can model $y_t$ as a linear function of its past ($y_{t-k}$'s), of another time series $x_t$, and of the lagged explanatory time series $x_{t-k}$. The MA part also allows explanatory variables of the form $\epsilon_{t-k}$, meaning the error term of the model for $y_{t-k}$. The purpose of this is to allow large ``shocks" from the past (e.g., a day on which the number of events was much larger than expected) to affect the present.
The I part involves replacing the starting time series $y_t$ with a differenced time series. For example, the first-order differenced time series $\Delta y_t$ is defined as $y_t - y_{t-1}$, and differencing can be iterated, so $\Delta^2 y_t = \Delta(\Delta y_t) = \Delta(y_t - y_{t-1}) = (y_t - y_{t-1}) - (y_{t-1} - y_{t-2})$. Differencing is used to shift from a non-stationary time series $y_t$ to a stationary time series $\Delta^i y_t$, the $i$-times differenced time series, and then fit a model (using AR and MA terms) to $\Delta^i y_t$. The ARIMA framework assumes the given time series can be made stationary by iterated differencing. To simplify notation, we write $z_t$ for $\Delta^i y_t$ (where $i$ is chosen to make $z_t$ stationary) and we write $x_t$ generically to mean any explanatory variable (including a lagged $z_{t-k}$). Seasonal ARIMA models further allow seasonal AR terms (e.g., $y_{t-7}$, the number of events one week ago), seasonal MA terms (e.g., $\epsilon_{t-7}$), and seasonal differencing. The notation SARIMA(p,d,q)x(P,D,Q)[S] represents a model with $p$ regular lags, $P$ seasonal lags (e.g., $y_{t-S}, y_{t-2S}$, etc.), $d$ regular differencing, $D$ seasonal differencing, $q$ regular MA terms, and $Q$ seasonal MA terms. For us, $S$ will be 7, so our time series will depend on non-seasonal lags (like, yesterday or the day before) as well as seasonal ones (like, what happened seven days ago or 14 days ago).

Once a suitable SARIMA model for $z_t$ has been fit, whose residual vector is independent random white noise that does not depend on time, the model can be used to forecast future values of $z_t$ (and hence of $y_t$). The model coefficients and standard errors can be used to determine which explanatory variables $x_t$ are statistically significant (and the direction of their influence), and the Akaite Information Criterion (AIC, from information theory) can be used to decide between competing models. Using these techniques we see, for example, that $p_t$ is strongly influenced by $nr_t$ and $e_t$, with negative reaction events and Euromaidan events both associated with more total events in the subsequent days.

Time series analysis starts from the assumption that the time series of interest is stationary (or that a differencing operation can make it so), which means, informally, that its mean, variance, and autocorrelation structure do not change over time. In the case of protest data from Ukraine in 2013-2014, the time series does change radically, in November of 2013. Hence, differencing was required to make the time series stationary.

We used both classical time series analysis techniques (e.g., inspection of time plots, unit root tests, differencing, (partial) autocorrelation function graphs, cross-correlation functions, Box-Ljung tests, and inspection/tests of residuals) as well as automated model-fitting techniques. Both approaches resulted in the same model.

\subsubsection*{Statistical analysis results}

Our best models for the ``number of events" from its history required first-order differencing to make $p_t$ stationary. We let $P_t$ denote $\Delta p_t$, and note that $P_t$ passes tests of stationarity. Our best ARIMA model for the autocorrelation structure is an ARIMA(2,1,3) model, meaning that two lags, $P_{t-1}, P_{t-2}$ and three moving-average terms $\epsilon_{t-1}$, $\epsilon_{t-2}$, $\epsilon_{t-3}$ are included in the model. However, the residuals still exhibited autocorrelation, so we allowed seasonal terms, resulting in a SARIMA(2,1,3)(1,0,2)[7] model for $p_t$.  We use the notation {\tt ar} to stand for ``autoregressive'' (e.g., {\tt ar1} refers to the coefficient of $P_{t-1}$), {\tt ma} for ``moving average'' (e.g., {\tt ma3} is the coefficient of $\epsilon_{t-3}$), {\tt sar} for ``seasonal autoregressive'' terms like $P_{t-7}$, and {\tt sma} for ``seasonal moving average'' terms $\epsilon_{t-7}$ and $\epsilon_{t-14}$. Below, we provide the coefficients and standard errors for our best model for the time series of protest events, based on its history.

\begin{verbatim}
         ar1      ar2      ma1     ma2      ma3    sar1     sma1    sma2
coef  1.4594  -0.7899  -1.8331  1.1981  -0.1720  0.9191  -0.9114  0.0902
s.e.  0.0702   0.0688   0.0913  0.1512   0.0742  0.0867   0.1048  0.0576
\end{verbatim}

That is, the model first transforms $p_t$ to $P_t = \Delta p_t$, then determines that $P_t$ depends, in a statistically significant way, on $P_{t-1}, P_{t-2}$, and $P_{t-7}$, as well as `shocks' (e.g., days with an unusual number of events) at times $t-1, t-2, t-3, t-7,$ and $t-14$. One could spell out the model as:

\begin{align}\label{eq:sarima}
 P_t = & 1.4594 * P_{t-1} - 0.7899 * P_{t-2} - 1.8331 * \epsilon_{t-1} + 1.1981 * \epsilon_{t-2} - 0.172 * \epsilon_{t-3} \notag\\
 & + 0.9191 * P_{t-7} - 0.9114 * \epsilon_{t-7} + 0.0902 * \epsilon_{t-14} + \epsilon_t
\end{align}

If one wished to predict the number of events tomorrow, one could use this model to predict $P_{t+1}$ and then use the fact that $P_{t+1} = p_{t+1} - p_t$ to predict $p_{t+1}$ as $p_t + P_{t+1}$, i.e., the number of events today plus the output of this model based on the history of the event time series.

In this model, the residuals exhibit no autocorrelation, all terms are statistically significant, and the AIC is improved relative to the non-seasonal ARIMA model. Because the residuals exhibited mild heteroskedasticity, we investigated fitting GARCH models, but they did not improve on the SARIMA model. The heteroskedasticity was fixed by the final model introduced below, using the covariates. Because the residuals were not normal, non-parametric (bootstrap) methods were used to ascertain statistical significance. This model achieved an AIC of 3162.5.

We turn now to multivariate models. Including $i_t$, $nr_t$, and $e_t$ as explanatory variables for $p_t$, and fitting a SARIMA model to the residuals, resulted in a strong improvement over the model above (AIC = 2680). All three of the ``number of injuries," ``number of negative responses," and ``number of Euromaidan events" have statistically significant predictive power for the number of events (as do all the SARIMA terms in the model). Furthermore, this model no longer exhibits heteroskedasticity, thanks to the inclusion of the Euromaidan terms. 

The best model begins with $i_t, nr_t,$ and $e_t$, then fits a SARIMA(1,1,1)(2,0,0)[7] to the residuals of that model. We list the coefficients and standard errors below:

\begin{verbatim}
         ar1      ma1    sar1    sar2    ts_i   ts_nr    ts_e
      0.2341  -0.9466  0.1829  0.2399  1.0326  1.1902  0.7970
s.e.  0.0536   0.0188  0.0497  0.0485  0.2669  0.1144  0.0279
\end{verbatim}

The model can be spelled out as:
\begin{equation*}
p_t = 1.0326 * i_t + 1.1902 * nr_t + 0.7970 * e_t + x_t
\end{equation*}

where the differenced series $X_t = \Delta x_t$ satisfies:
\begin{equation*}
X_t = 0.2341 * X_{t-1} - 0.9466 * \epsilon_{t-1} + 0.1829 * X_{t-7} + 0.2399 * X_{t-14} + \epsilon_t
\end{equation*}

The dependence on the past is now encapsulated by the residual term $x_t$, which makes the model cleaner but does not change our earlier assertion that $p_t$ depends in a statistically significant way on the lags of $i_t, nr_t$, and $e_t$.

Interpreting the coefficients, we learn that every injury is associated with 1.03 more events (so, if one event led to 100 people getting injured, we would expect 103 more events), and every negative response event is associated with 1.19 more events. We also fit models including lagged terms such as $nr_{t-1}$, but this did not improve the model, as the dependence on the past is already encapsulated by the SARIMA term, $x_t$.


Lastly, we fit a threshold model to $p_t$. Threshold models are the statistical analog of a self-excitation process and are often used to model the spread of epidemics. The idea is to pick a threshold $r$ (e.g., using cross-validation), fit one SARIMA model to all $p_t \leq r$, and fit a different SARIMA model to all $p_t > r$. Threshold models have more parameters than simple SARIMA models, but this allows them to model time series that behave differently during periods of high intensity. While such models can be used for prediction, our goal in fitting this model was to determine if there was a moment when the protest dynamics accelerated and to find the date on which this change occurred. Our best threshold model achieved an AIC of 2107, demonstrating that it was a better fit than the preceding models. Nevertheless, in this case, due to the complexity of the threshold model, we prefer the models above for explaining the relationships between the variables.
When we fit a threshold model to $p_t$, we found that the only time period where $p_t$ was above our threshold (70 events per day) occurred starting on November 22, 2013. Sociologists with expertise in Ukraine selected the date November 29, 2013, as the date on which the protests changed in nature (based on the EU trade agreement and police crackdowns). This perfectly fits our model, which says that the number of events on any given day depends on the days in the preceding week. The two SARIMA models (above/below the threshold, or, equivalently in this case, before/after the cutoff date) were both substantially better fits than using one SARIMA model for the entire 2013 year. We experimented with other cut-off dates in November of 2013 and the choice of cut-off did not significantly change the conclusion.

\subsection*{Dynamic modeling}
\subsubsection*{Hawkes process}
 Initially developed to model seismic events \cite{Ogata1988}, Hawkes processes appear well-suited for characterizing these protests. This category of stochastic point processes functions as a counting process, where the history of events shapes the likelihood of future occurrences. For instance, each event excites the process, heightening the probability of upcoming events. Hawkes processes have found applications in describing financial processes \cite{AZIZPOUR2018154}, mass shootings \cite{Mohler2011}, and the popularity of tweets \cite{ZADEH2022103594}. While some formulations of Hawkes processes adopt a continuous form, the nature of our dataset, presenting a daily sum of events, necessitates the use of a discretized variation.

In a Hawkes process, events are portrayed as temporal points, with each event having the potential to trigger subsequent occurrences. The initiation of new events is influenced by past events, leading to a cascade-like effect. This self-exciting property implies that the occurrence of an event can increase the likelihood of more events happening in its vicinity, forming clusters or bursts of activity.

Mathematically, a Hawkes process is defined by its intensity function, which gauges the rate of event occurrences at any given time. This intensity function is influenced by both the baseline event rate and the impact of past events. As events unfold, they contribute to the intensity function, subsequently affecting the likelihood of future events. To fully comprehend Hawkes processes, one must grasp their two major components.

\textbf{1 - Kernel:}\label{sec:hawkes_ker}
In a Hawkes process, the kernel plays a pivotal role in elucidating how past events influence the occurrence of future events. The kernel, a mathematical function, models the impact of previous events on the process's intensity, reflecting its inherent self-exciting behavior. It quantifies the temporal influence of past events on the intensity at a given time, measuring how much the occurrence of a past event affects the likelihood of a new event happening shortly afterward. The shape of the kernel function determines the form and strength of this influence, capturing the dynamics of the process by indicating the ``memory" of the system and how it responds to past occurrences.

Mathematically, the kernel function is typically non-negative and integrates to a finite value. It is convolved with the history of event occurrences to compute the instantaneous intensity at any given time. As new spikes occur, they contribute to the intensity, influencing the likelihood of subsequent events. Different types of kernel functions, such as exponential, power-law, and Gaussian kernels, can model various temporal patterns and degrees of influence. The choice of the kernel function profoundly impacts the overall behavior of the Hawkes process and its ability to capture real-world dynamics.

 Given that the events occur on a fixed time period, a natural assumption is that the expected number of events $N_{exp}(t)$ and the observed data $Y$ follow a Poisson process. Thus, fitting the model is a matter of minimizing the negative likelihood derived from the said process: 

$$
NLL = \sum_t N_{exp}(t) - Y(t)log(N_{exp}(t)).
$$

\textit{\textbf{1.1 - Classical Exponential Kernel:}}\label{sec:og_ker} The exponential kernel is a fundamental component of the Hawkes process, serving as a common choice to model the temporal influence of past events on the occurrence of future events. This kernel embodies the principle that the influence of a past event on the intensity of the process decreases exponentially over time.
Mathematically, the kernel function takes the form: 

\begin{equation}\label{eq:og_ker}
P(t) = N_{sec}e^{-\left(\frac{t-t_i}{T_{ex}}\right)}.
\end{equation}

Thus, the number of expected events at any time $t$, given self-exciting events happening at the times $\{ t_0,t_1,.....,t_n\}$, is of the form

\begin{equation}\label{eq:og_HWP}
N_{exp}(t, ob) = N_{0,ob} + N_{sec}\sum_{t_i<t} e^{-\left(\frac{t-t_i}{T_{ex}}\right),}
\end{equation}

where $N_{0,ob}$ is the average number of events before self-excitation, which serves as a base that the model will converge to after self-excitation, $N_{sec}$ represents the maximum magnitude of the spike, $T_{ex}$ is a variable influencing the time needed for complete relaxation and return to the steady state after a spike, and $e^{-\left(\frac{t-t_i}{T_{ex}}\right)}$ represents the effect of the spike time on the number of expected events at a time $t$.

\textit{\textbf{1.2 - Susceptible Population Model:}}\label{sec:PS_ker}
The Hawkes process presented in Eq. \eqref{eq:og_HWP} assumes no differences between oblasts. However, the population and their political leaning play a role in the observed number of events. Moreover, the spikes' magnitude changes over time as the tension grows in the system before relaxing. Thus, we modify the Hawkes process to reflect the population in each oblast and the rise and subsequent drop in magnitude.  In addition, we see more activity in the pro-EU regions and consequently make the ansatz that the population that is susceptible to participate in these protests are pro-EU individuals.  This leads to the following model:

\begin{equation}\label{eq:SPS_HW}
    N_{exp}(t, ob) = N_{0,ob} + p_r(ob)vr(ob)N_{sec} e^{-(t-d_e)^2}\sum_{t_i<t}e^{-\left(\frac{t-t_i}{T_{ex}}\right)}, 
\end{equation}

where $p_r(ob)$ is the population ratio of each oblast, $vr(ob)$ is the pro-EU vote in each oblast scaled by the entire countries' vote from 2014, and $e^{-(t-d_e)^2}$ is a term that reduces the magnitude of the spike with $d_e$ being a time delay.
Note that when $N_{sec}$ is multiplied by $p_r(ob)$ and $vr(ob)$, the model will scale the number of expected events for each oblast. We use population ratio as it allows for the population factor to stay between $[0,1]$.

\textit{\textbf{1.3 - Interaction Effect Between Oblasts:}}
While Eq. \eqref{eq:SPS_HW} takes into consideration the specificities of the different regions, it does not reflect their influence on one another. We add a lag term to reflect such such effect to a couple of models, which we discuss below.  

\textit{\textbf{1.3.1 - Geographical Influence:}}
In the study by Bonnasse-Gahot et al., \cite{Bonnasse-Gahot2018}, the dispersion of events was found to be influenced by the geographic distance between different regions in France. In our initial model aimed at accounting for spatial interactions between oblasts, we integrated geographical distance as a factor. We thus have the following model:

\begin{align}\label{eq:Geo_HW}
 N_{exp}(t, ob) = & N_{0,ob} +  \Bigg[ p(ob)vr(ob)N_{sec} e^{-(t-d_e)^2} \sum_{t_i<t} I(t_i, ob)^pe^{-\left(\frac{t-t_i}{T_{ex}}\right)} \notag\\
 & + \sum_{ob_j \neq ob} p(ob)vr(ob) W(ob, ob_j)N_{exp}(t - 1, ob_j) \Bigg].
\end{align}

Where 
\newcommand{\norm}[1]{\left\lVert#1\right\rVert}
$$
W(ob, ob_j) = \frac{d}{(\norm{ob_j-ob}+1)^c},
$$
In this case, $\norm{ob_j-ob}$ is the Euclidean distance of the central point of the two oblasts. 
  
\textit{\textbf{1.3.2 - Political Influence:}}\label{sec:PS_ker}
The political polarization observed in Ukraine motivates the development of a model designed to examine the hypothesis that the contagion of events between oblasts is not primarily determined by geographic proximity but is instead influenced by political alignment.  To investigate this, we employ an alternative factor based on voting ratios to assess such a model:

\begin{align}
 N_{exp}(t, ob) = & N_{0,ob} + p_r(ob)vr(ob)N_{sec} e^{-(t-d_e)^2} \sum_{t_i<t} e^{-\left(\frac{t-t_i}{T_{ex}}\right)} \notag\\
 & + \sum_{ob_j \neq ob} p_r(ob)vr(ob) V(ob, ob_j)N_{exp}(t - 1, ob_j),
\end{align}

where

$$
V(ob,ob_j) = \frac{d}{(|vr(ob) - vr(ob_j)| + 1)^c},
$$

and $vr(ob)$ is the pro-EU vote in the oblast scaled by the entire countries' vote from 2014, and $c$ and $d$ are two variables influencing the effect of this factor. 

\textit{\textbf{1.4 - The Effect of the Number of Injured Individuals:}}\label{sec:IN_ker}
The statistical analysis outlined above and previous research \cite{Rodriguez-White} demonstrates that the count of injured individuals had a discernible impact on forecasting the daily event count. The effect is observed even though a substantial number of injuries during the protests were not accounted for. Consequently, our model evolves to include such events and takes the form: 

\begin{align*}
 N_{exp}(t, ob) = & N_{0,ob} + I(t, ob)^p \Bigg[ p_r(ob)vr(ob)N_{sec} e^{-(t-d_e)^2} \sum_{t_i<t} e^{-\left(\frac{t-t_i}{T_{ex}}\right)} \notag\\
 & + \sum_{ob_j \neq ob} p_r(ob)vr(ob) V(ob, ob_j)N_{exp}(t - 1, ob_j) \Bigg],
\end{align*}

where $I(t, ob)$ is the number of reported injured people per day per oblast.

\textbf{2 - Spike Times:} While self-excitation is observed from the statistical analysis, it is not necessarily the case that all events contribute to self-excitation.  Including all events would lead to an exponential growth of events with no time for relaxation.  Thus, we incorporate thresholding where we consider event time in our model that is above a certain threshold. By implementing a threshold, we can control the events that contribute to the intensity function. This is particularly important in self-exciting processes, as it helps prevent excessive growth in the intensity function due to a cascading effect from numerous closely spaced events. Additionally, thresholding allows us to focus on events that have a significant impact on the process. This enhances the interpretability of the model, as it helps distinguish between events that genuinely contribute to the self-excitation mechanism and those that may be considered background noise. A carefully chosen threshold ensures that only meaningful events are considered in the modeling process. In practice, thresholding can significantly improve the computational efficiency of parameter estimation procedures for Hawkes processes. By excluding events below a certain threshold, the algorithm can concentrate on the most relevant events, reducing the computational burden associated with estimating parameters. Finally, introducing a threshold can contribute to the stability of the modeling process. It helps prevent overfitting and ensures that the model generalizes well to new data. Without thresholding, the model might become overly sensitive to minor fluctuations in the data, potentially leading to poor generalization. We incorporate two thresholding methods: (1) uniform thresholding, where spike times are chosen at regular temporal intervals. While the approach is naive, it does not add much complexity to the system and assumes that the effect of events is delayed. (2) oblast sensitivity, where the threshold deciding whether or not an event is a spike time is decided by the number of events in that particular day relative to the maximum number of events per oblast. While this approach adds a certain level of complexity, it far outperforms the the uniform model from an AIC perspective.

\textbf{3 - Exogenous Effect:}
 In complex social phenomena such as protests, it is important to note the complex interplay between internal processes and external stimuli in shaping protest phenomena. Levels of protest are influenced not solely by internal factors but also by external events. The initial incident that sparks a protest, termed the triggering event, is a crucial internal factor that should be incorporated into the model. However, additional occurrences, like government concessions or their absence, further contribute to the intensification of the situation. For example, the study by Varol et al. in \cite{varol2014evolution} examines the impact of external events on a social media uprising associated with the Gezi Park movement in Turkey.  
In scenarios where political concessions are sought, external political events play a crucial role in shaping the dynamics of protests and should, therefore, be integrated into models.  Mathematically, these external events act as external forcing terms in models \cite{Berestycki2015}.
 Identifying events that influence protests is a nuanced process that requires a multidisciplinary approach and the integration of various data sources. Researchers often rely on comprehensive event databases, such as the Global Database of Events, Language, and Tone, which captures a wide array of events worldwide, including political demonstrations and social movements. Analyzing media reports, social media content, and government records provides valuable insights into the occurrence and context of events leading to protests. Scholars emphasize the importance of triangulating information from diverse sources to enhance the reliability of event identification \cite{earl2004use}. As such we devised a comprehensive list of events that influenced the protests \ref{table:spktb}. This list was devised by checking multiple news sources for events adjacent to the protests that were qualitatively judged to influence the protests. One way in which exogenous events can shape the evolution of protests is exemplified by actions such as a ban on protests, which may decrease their occurrence, whereas confrontations with law enforcement might intensify the tension within the system. In mathematical terms, the impact of such events is represented as a $\delta$ pulse, and we adjust this pulse by the population ratio and vote ratio of each oblast.  This leads to the following model:

\begin{align}
 N_{exp}(t, ob) = & N_{0,ob} + I_{exo}(t,ob)+ I(t, ob)^p \Bigg[ p_r(ob)vr(ob)N_{sec} e^{-(t-d_e)^2} \sum_{t_i<t} e^{-\left(\frac{t-t_i}{T_{ex}}\right)} \notag\\
 & + \sum_{ob_j \neq ob} p_r(ob)vr(ob) V(ob, ob_j)N_{exp}(t - 1, ob_j) \Bigg],
\end{align}

 where 
 $$ I_{exo}(t, ob) = p_r(ob)vr(ob)N_{exo}\delta_{t \in \mathbb{S}},$$
 $N_{exo}$ represents a parameter measuring the impact of exogenous events on the system and $\mathbb{S}$ denotes the set of times when these external occurrences take place.

\begin{table}[!ht]
\centering
\begin{adjustwidth}{-2.25in}{0in} 

\caption{
{2013 exogenous events corresponding to the best fitting spike times to model Euromaidan.}}\label{table:spktb}
\begin{tabular}{|l|l|}
\hline
{\bf Date} & {\bf Exogenous Event}\\ \thickhline
11-22 & Government decree to suspend the signing of the Ukraine-EU agreement. \\ \hline
11-25 & News of the riot police's violent actions spread.\\ \hline
12-08 & President Yanukovych and Russian President Vladimir Putin.\\ \hline
12-11 & Police cut off power to the protesters' headquarters which incited more protests. \\ \hline
12-14 & President Yanukovych suspends multiple Kyiv officials.\\ \hline
12-17 & The proclamation that the former head of police will be put on house arrest.\\ \hline
12-19 & President Yanukovych officially pauses the EU trade agreement.\\ \hline
12-25 & Armed assault against Kharkiv oblast protest organizer was conducted.\\ \hline
12-26 & The brutal assault of a pro-Euromaidan journalist is made known.\\ \hline
12-29/30 & The Government continues to pass laws that target protesters.\\ \hline

\end{tabular}
\end{adjustwidth}
\end{table}



\section*{Results}


Our investigation encompasses six primary variations of the Hawkes process framework, each building on the previous to address specific characteristics of protest events, including self-excitation, external influences, effects of different regions, and the number of injured individuals. Under the final model, $N_{sec}$, $d_e$, $T_{ex}$, $c$, $d$, and $p$ are the parameters to be determined by minimizing the negative log-likelihood. By systematically comparing these models and assessing their goodness-of-fit, we gain valuable insights into not only the suitability of Hawkes processes for characterizing the complex temporal dynamics inherent in protest data but also the different protest drivers that governed the Euromaidan revolution.

\begin{figure}[!h]
    \centering
    \subcaptionbox{\label{fig:obpd}}{\includegraphics[width=.31\textwidth]{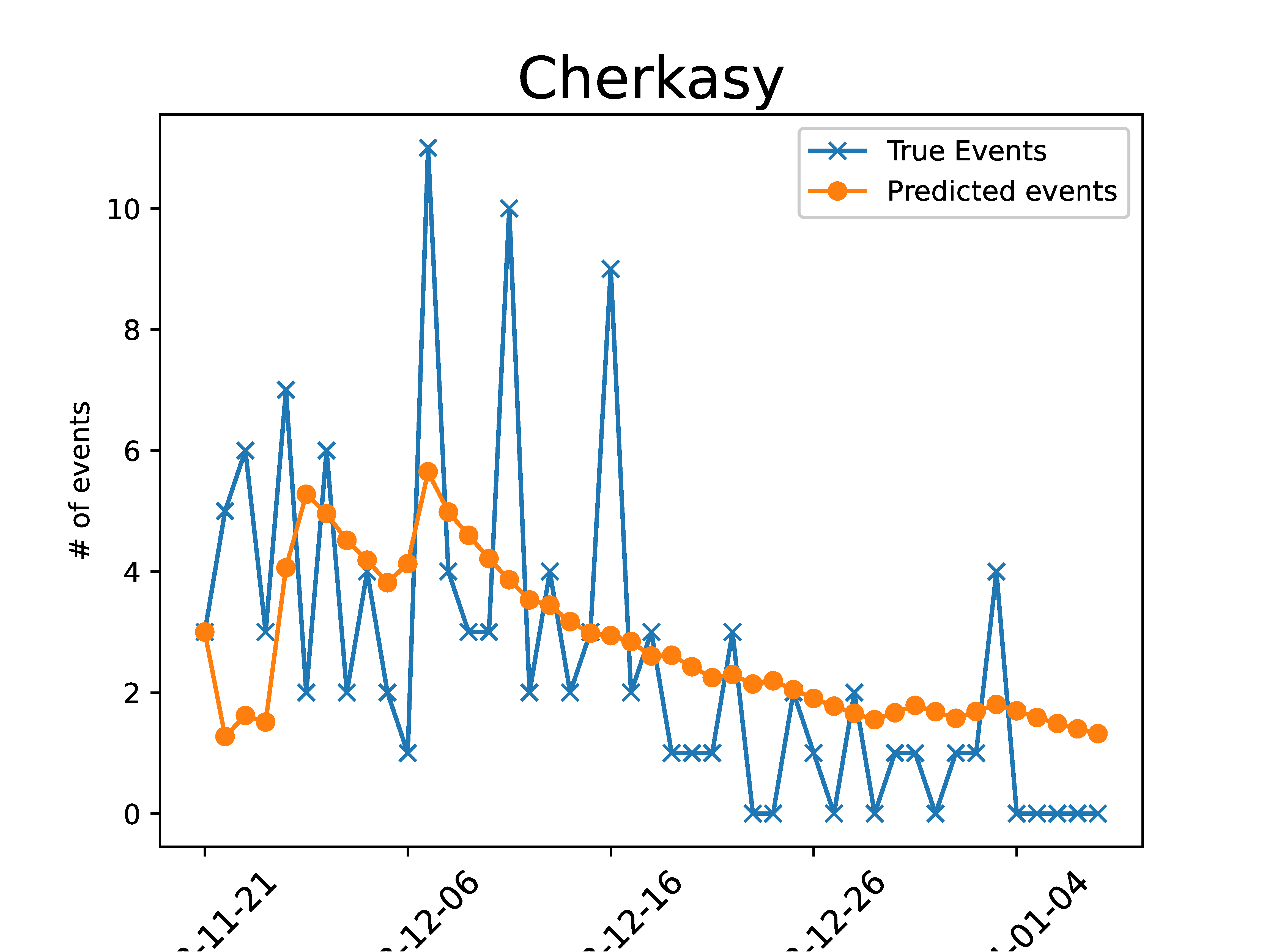}}\hspace{1em}%
    \subcaptionbox{\label{fig:obpd}}{\includegraphics[width=.31\textwidth]{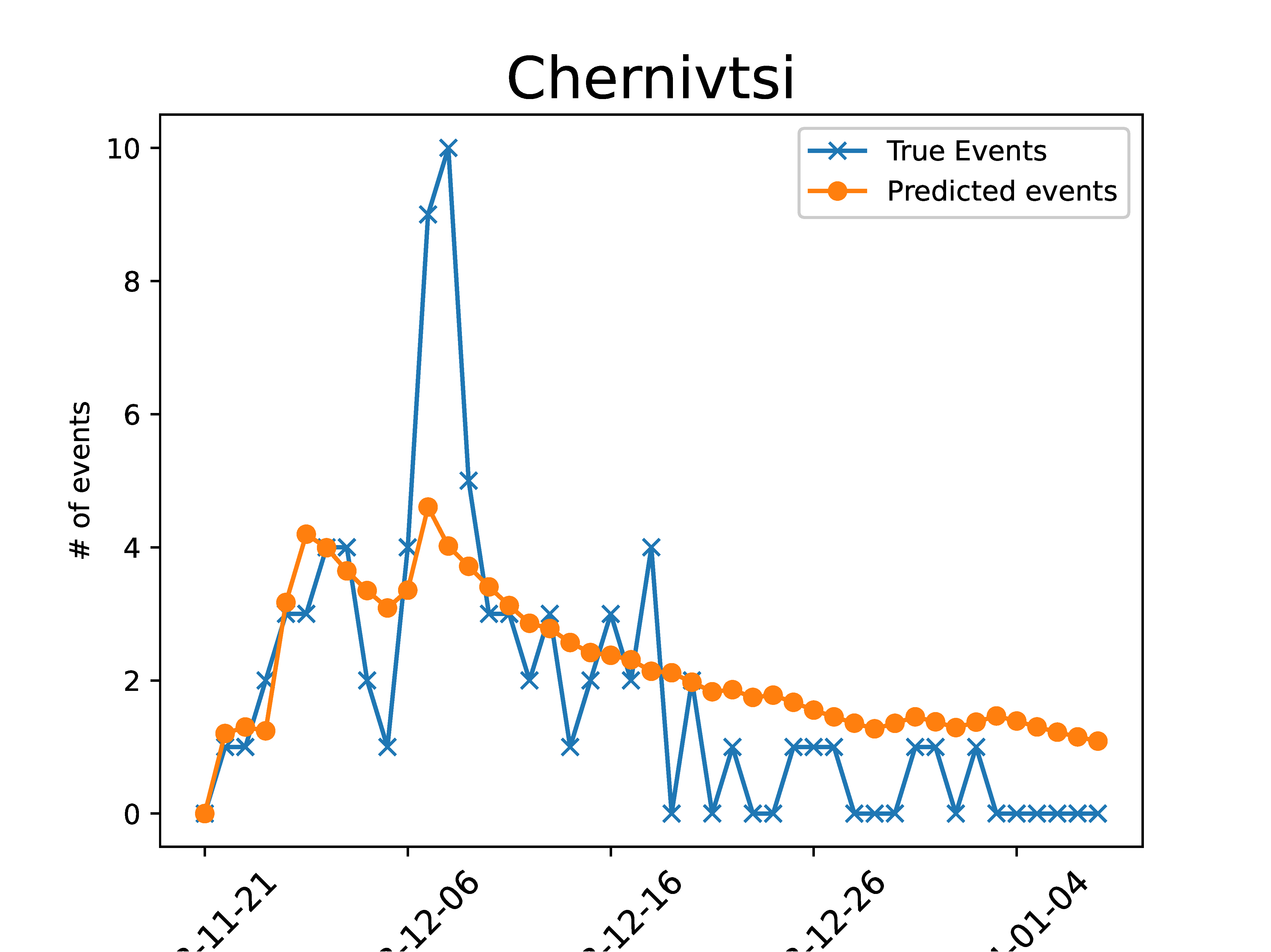}}\hspace{1em}%
    \subcaptionbox{\label{fig:obpd}}{\includegraphics[width=.31\textwidth]{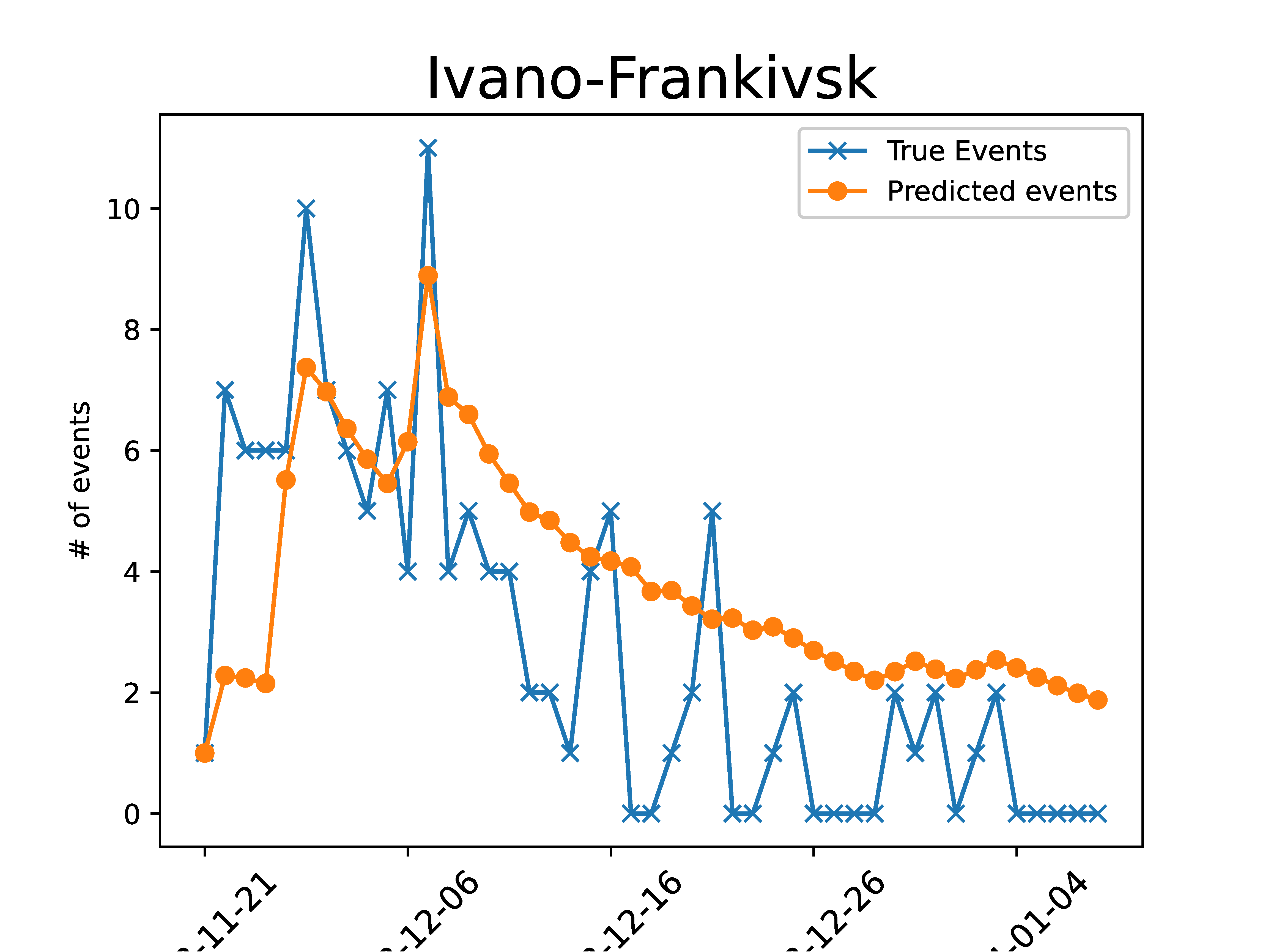}}\hspace{1em}%
    \\
    \subcaptionbox{\label{fig:obpd}}{\includegraphics[width=.31\textwidth]{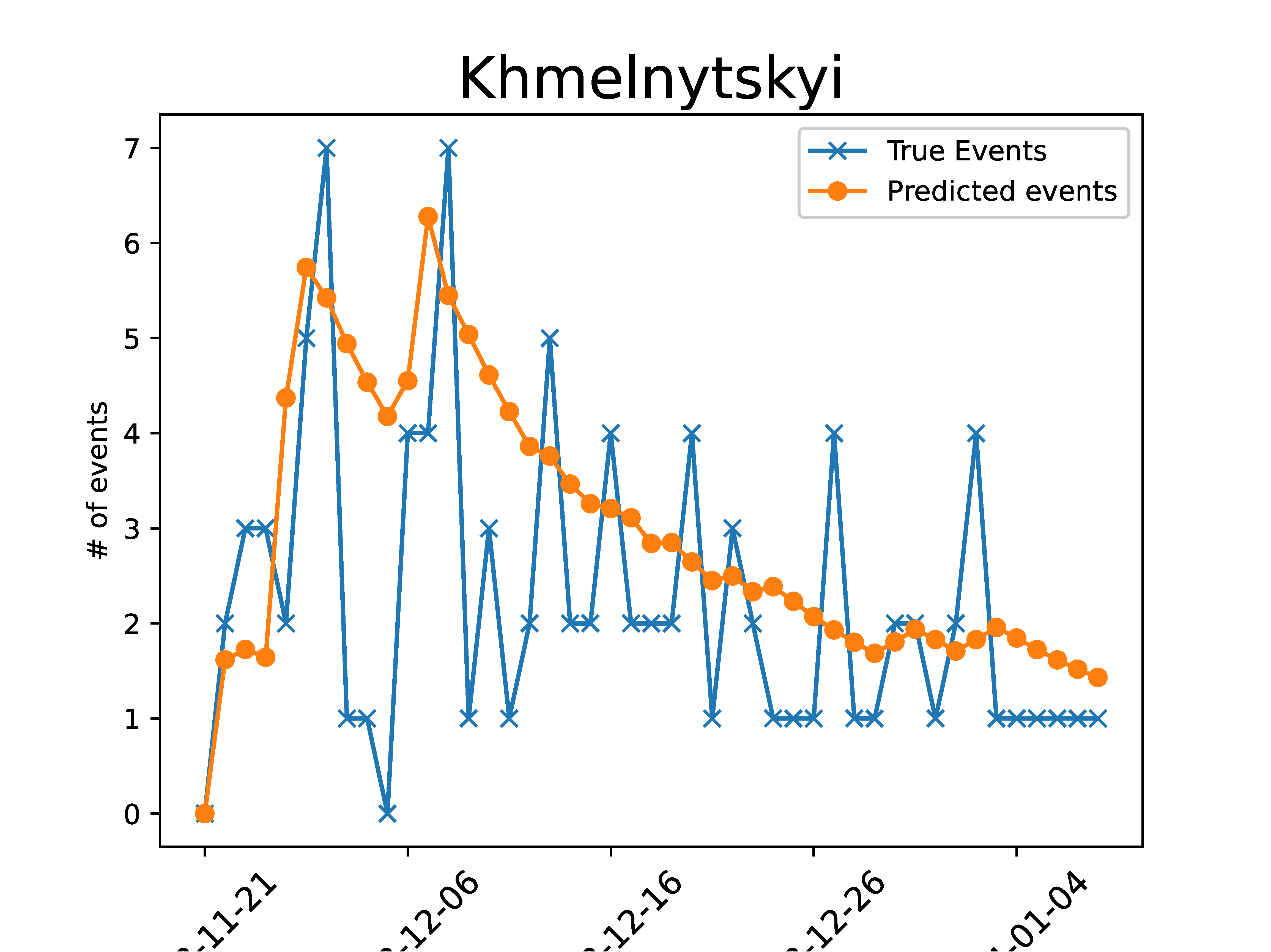}}\hspace{1em}%
    \subcaptionbox{\label{fig:obpd}}{\includegraphics[width=.31\textwidth]{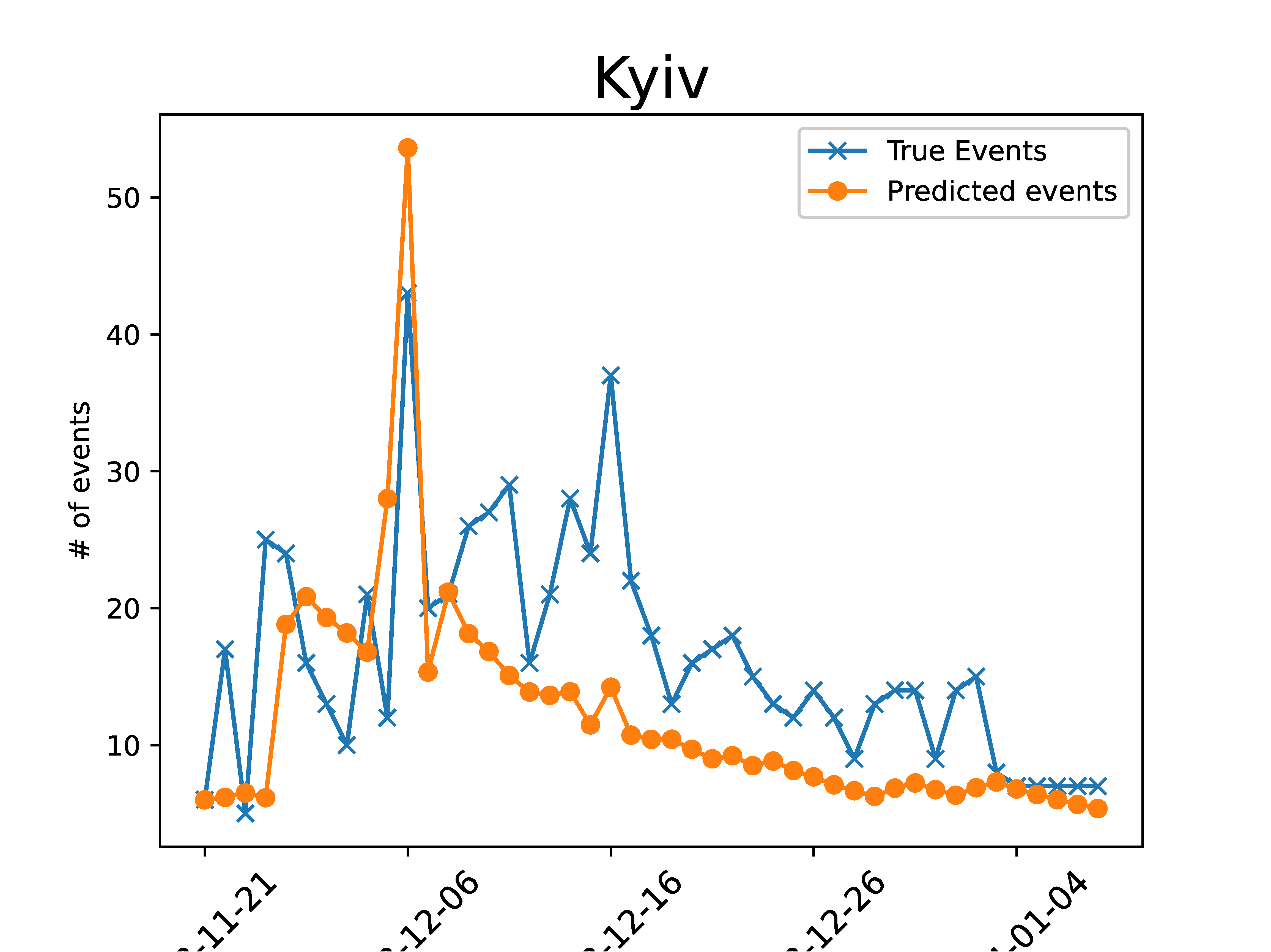}}\hspace{1em}%
    \subcaptionbox{\label{fig:obpd}}{\includegraphics[width=.31\textwidth]{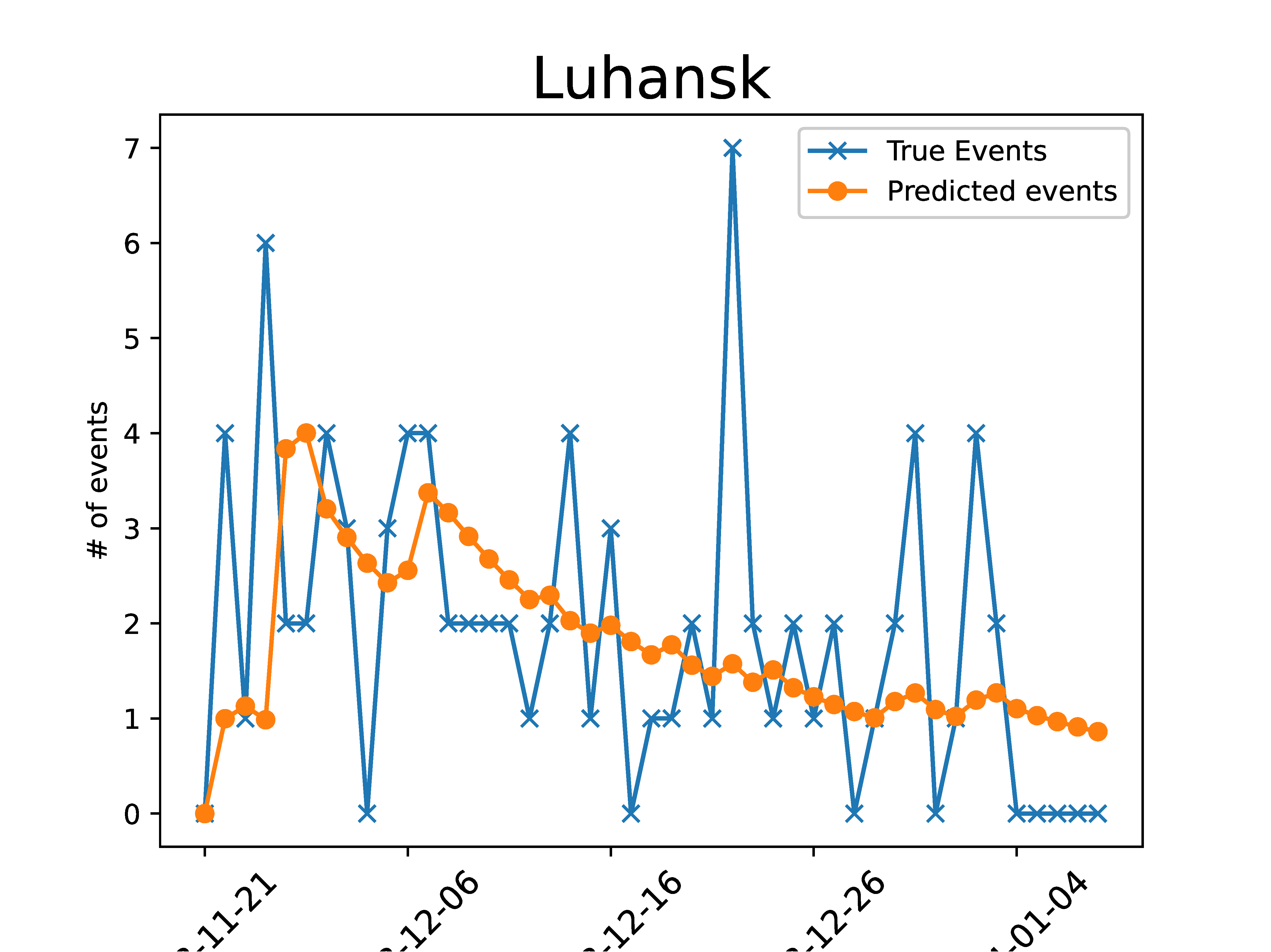}}\hspace{1em}%
    \\
    \subcaptionbox{\label{fig:obpd}}{\includegraphics[width=.31\textwidth]{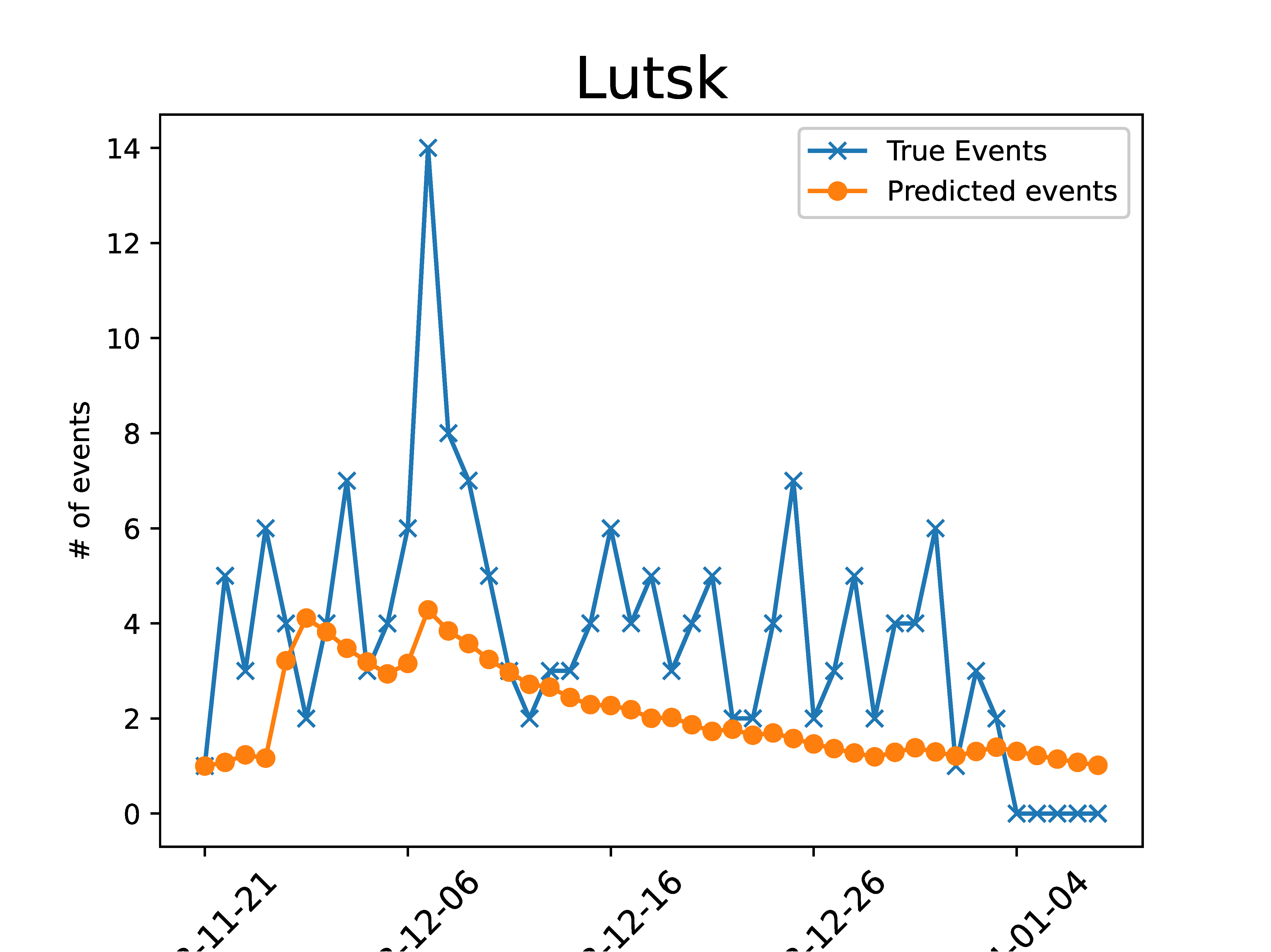}}\hspace{1em}%
    \subcaptionbox{\label{fig:obpd}}{\includegraphics[width=.31\textwidth]{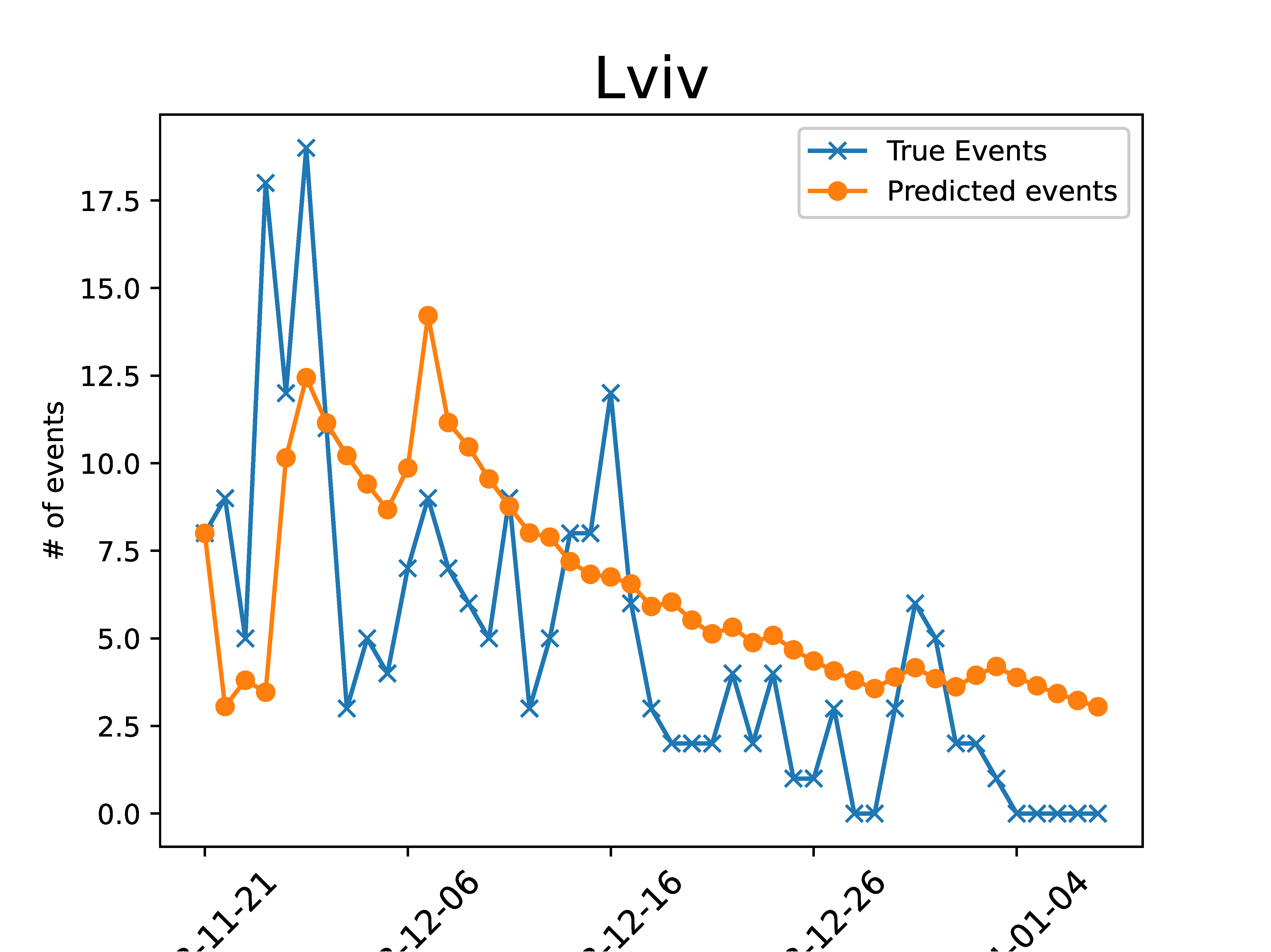}}\hspace{1em}%
    \subcaptionbox{\label{fig:obpd}}{\includegraphics[width=.31\textwidth]{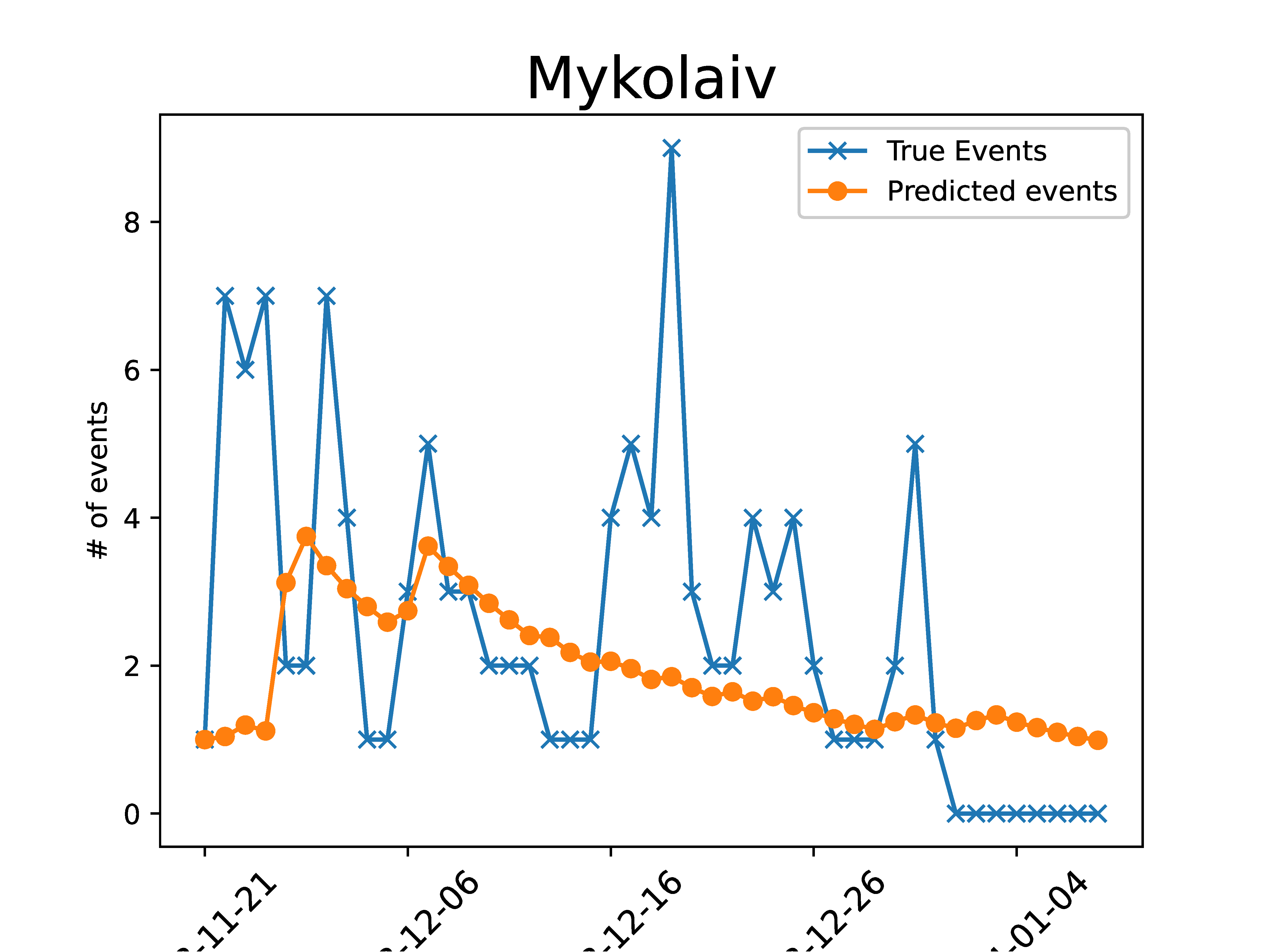}}\hspace{1em}%
    \\
    \subcaptionbox{\label{fig:obpd}}{\includegraphics[width=.31\textwidth]{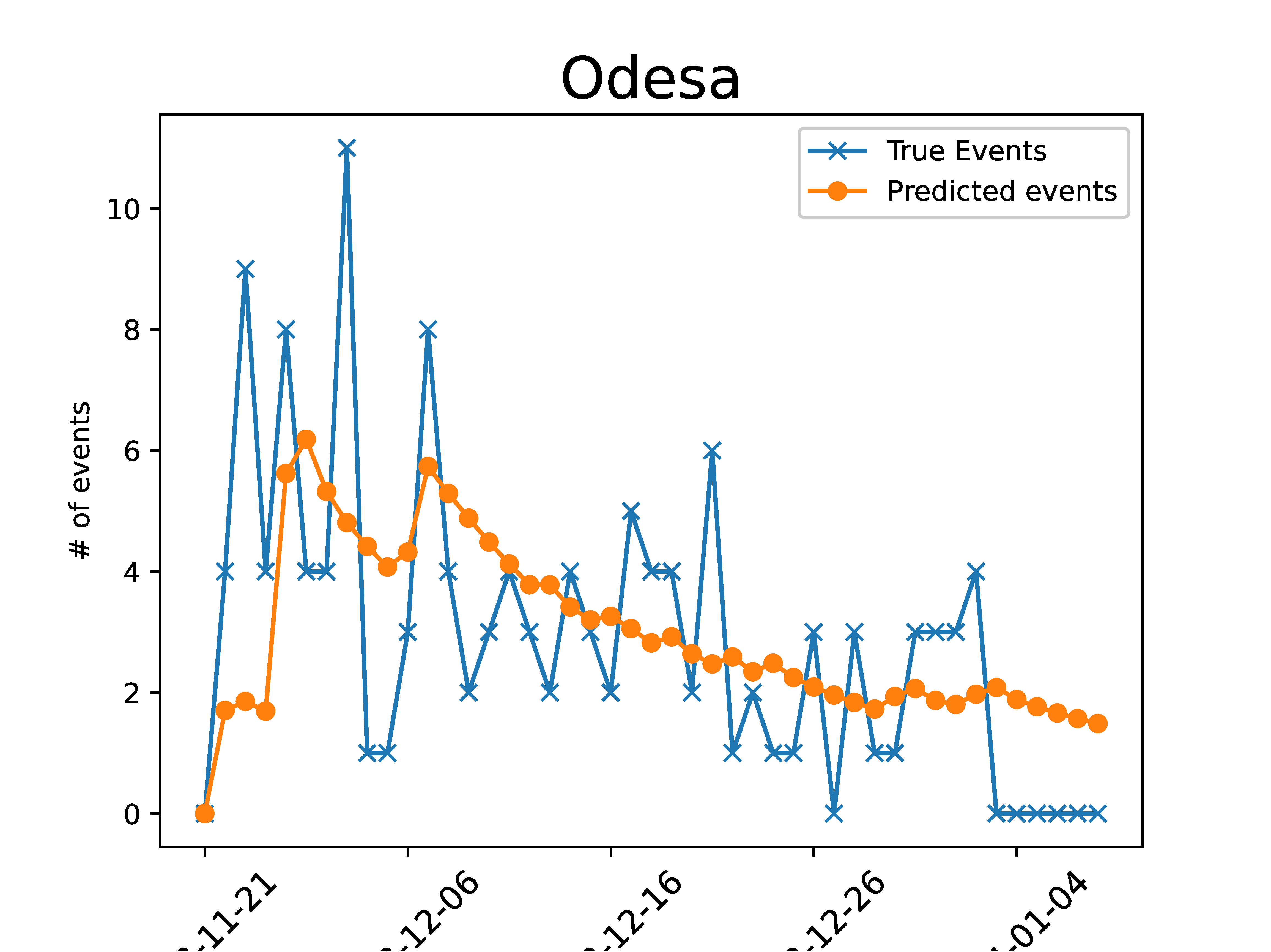}}\hspace{1em}%
    \subcaptionbox{\label{fig:obpd}}{\includegraphics[width=.31\textwidth]{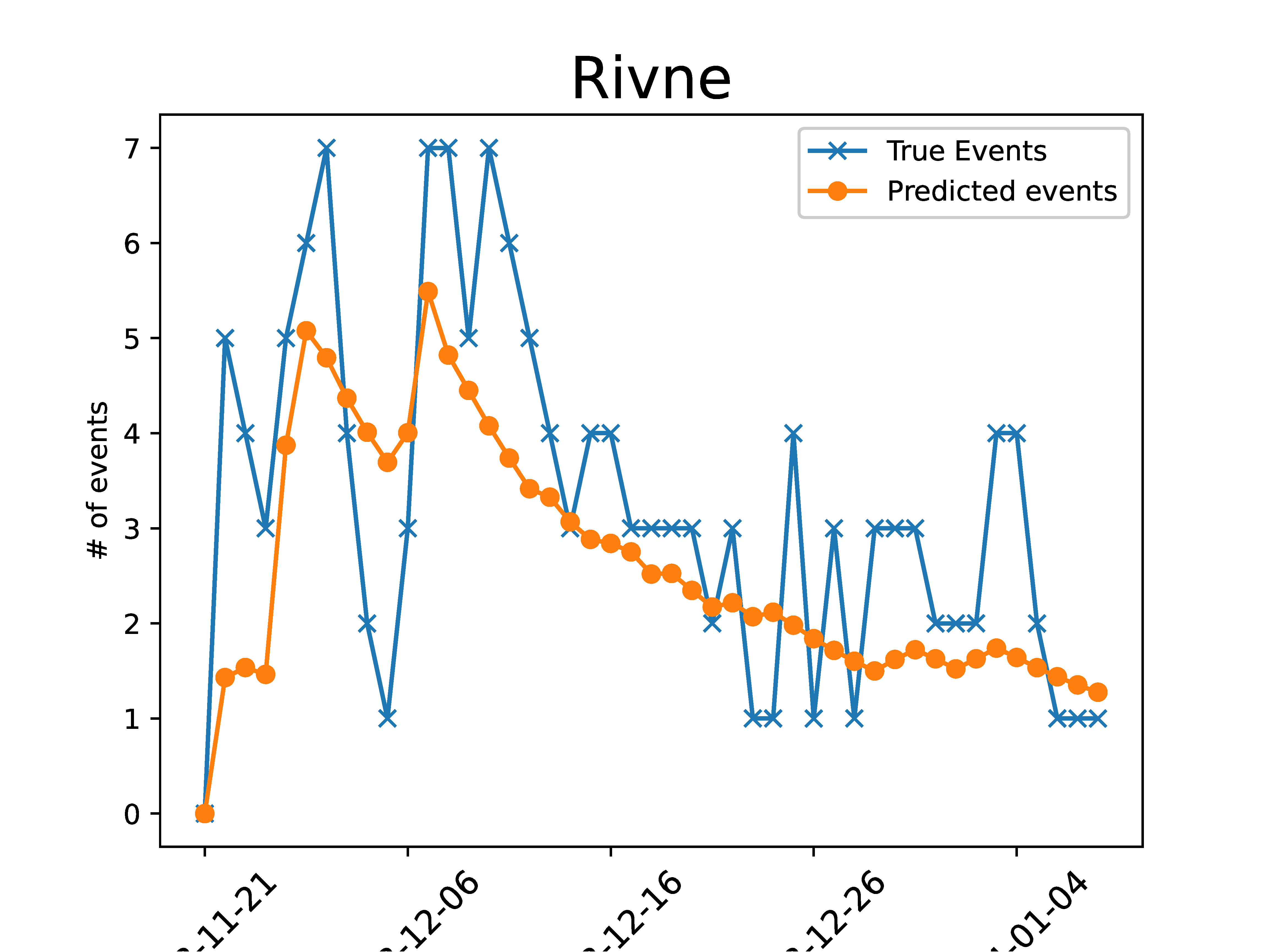}}\hspace{1em}%
    \subcaptionbox{\label{fig:obpd}}{\includegraphics[width=.31\textwidth]{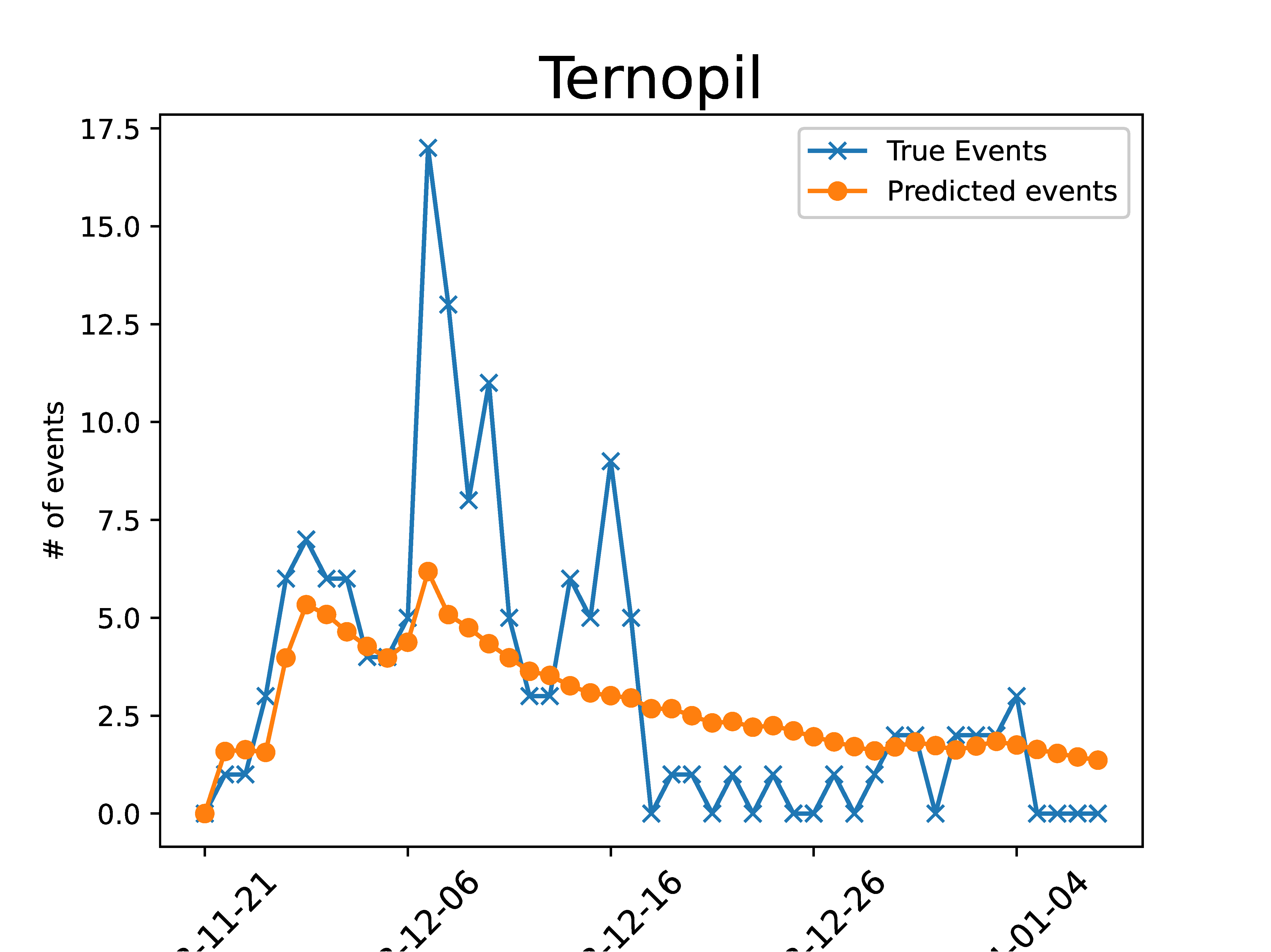}}\hspace{1em}%
    \\
    \subcaptionbox{\label{fig:obpd}}{\includegraphics[width=.31\textwidth]{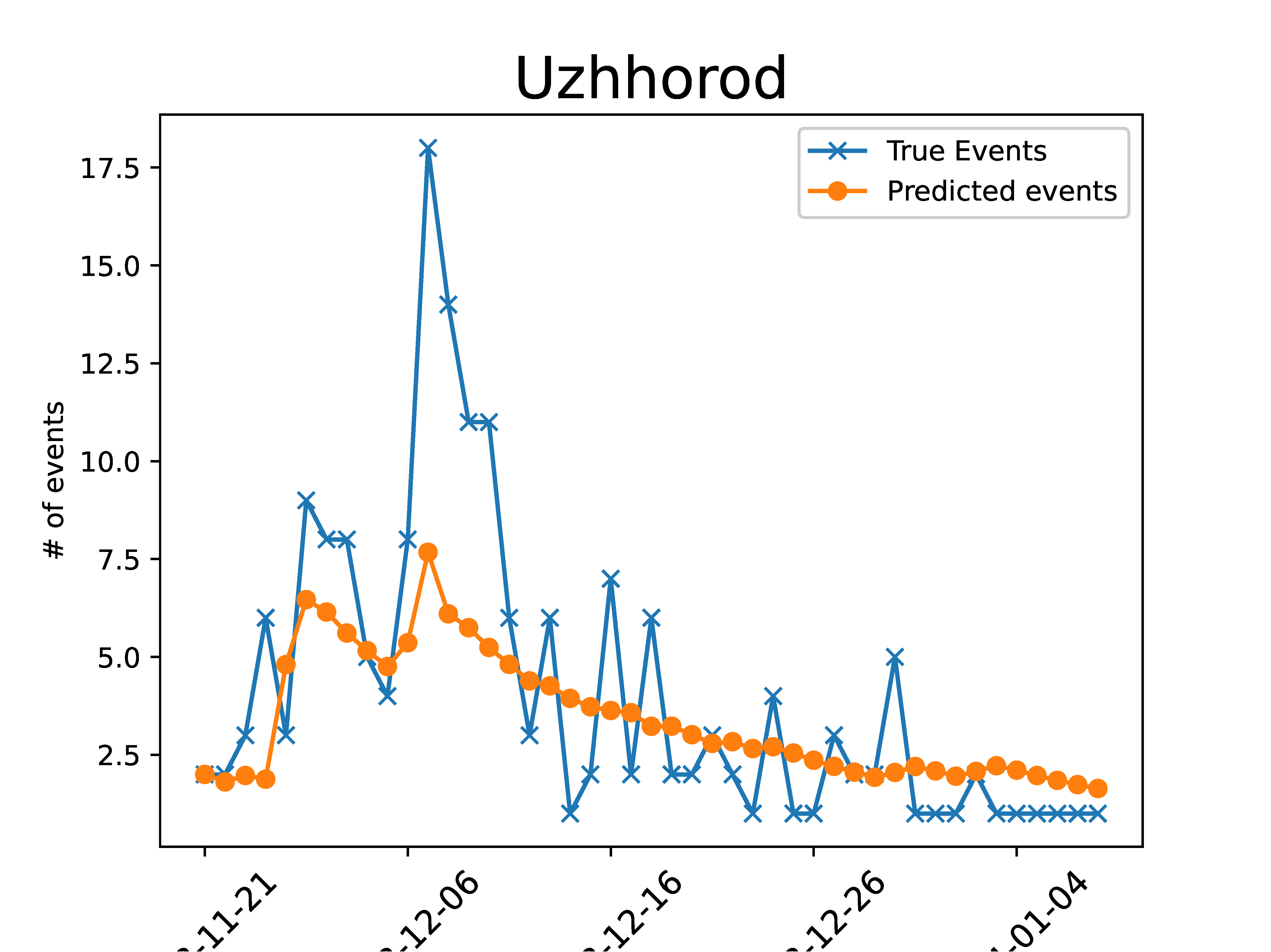}}\hspace{1em}%
    \subcaptionbox{\label{fig:obpd}}{\includegraphics[width=.31\textwidth]{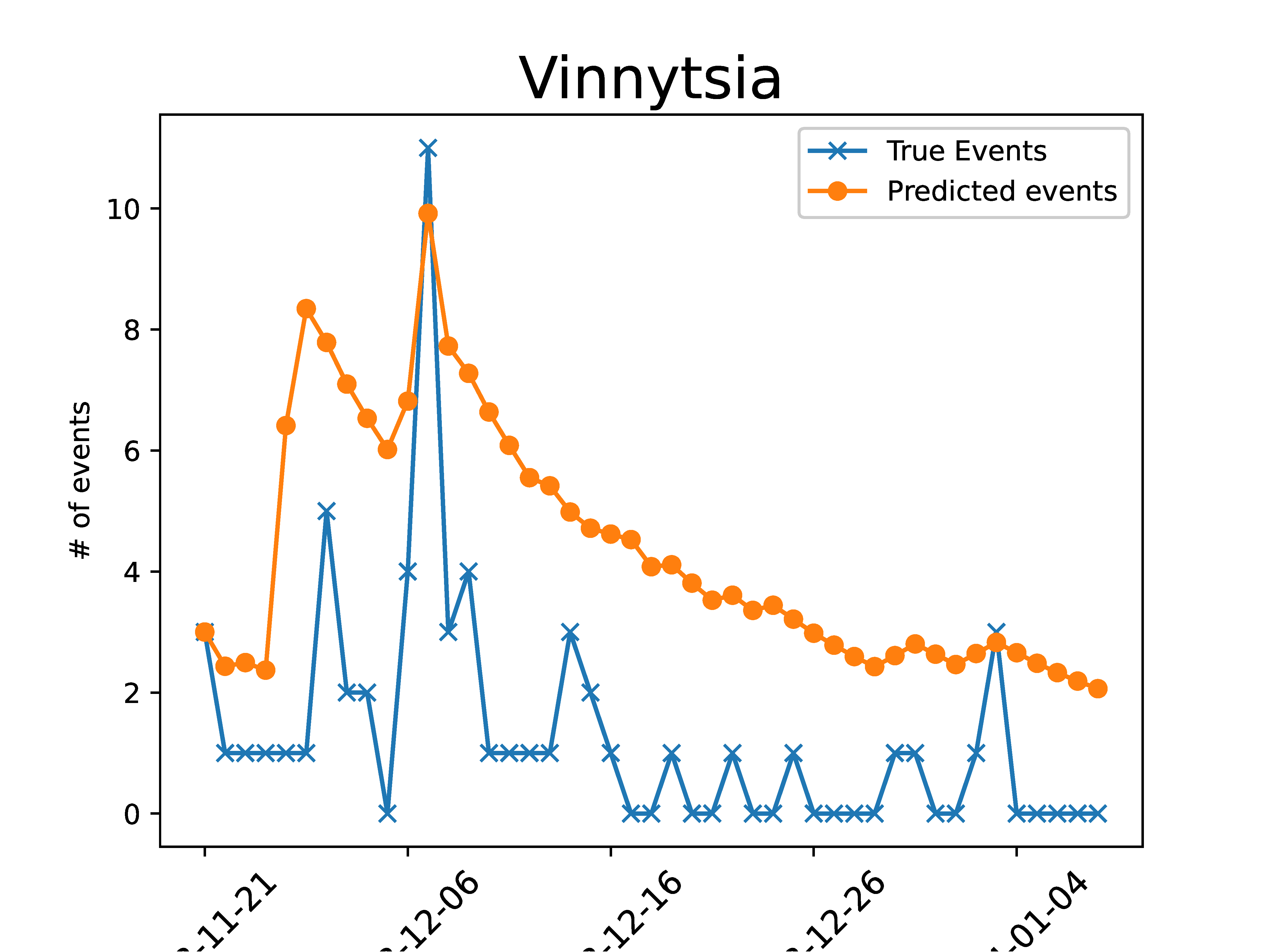}}\hspace{1em}%
    \subcaptionbox{\label{fig:obpd}}{\includegraphics[width=.31\textwidth]{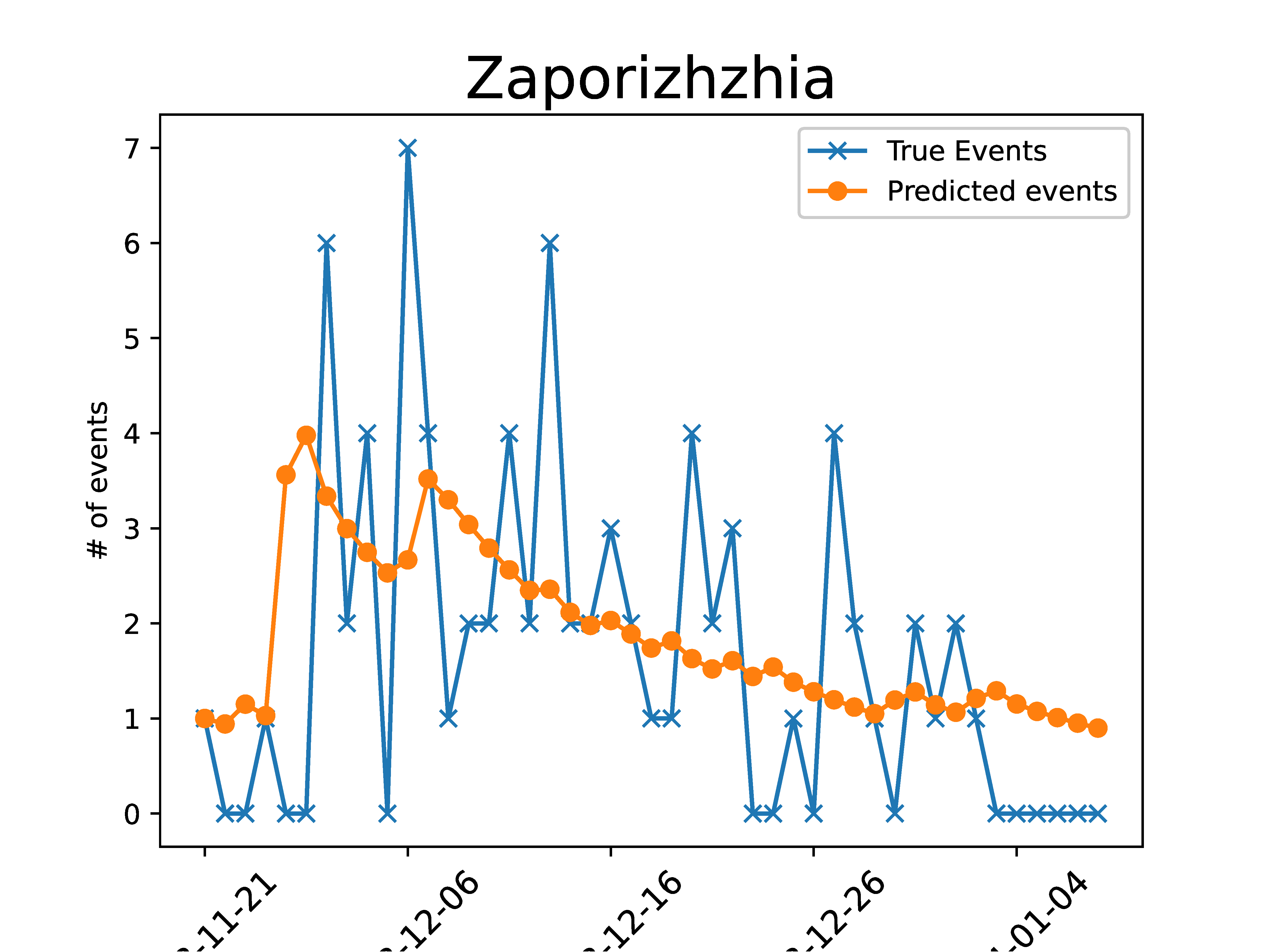}}\hspace{1em}%
  \caption{The predicted number of events per day in each oblast weighted by the susceptible population. $N_{sec} = 102.6.$, $d_e = 4.18$, $T_{ex} = 5.8$, $c = 2.6$, $d =2.4$, $p = 2$, and $Nexo = 7.22$}
  \label{fig:oblastpred}
\end{figure}

The classical kernel performs poorly, which is predictable given that it fails to capture the specificities of each oblast. Only including the susceptible population does not lead to a model that better predicts the dynamics of the protests. It is only when accounting for the effect of different oblasts that we see a significant increase in performance in both the negative log-likelihood and Akaike information criterion (AIC). With regards to spatial influence, while the distance-based model presented in Eq. \eqref{eq:Geo_HW} performs well, the voting-based interaction model is far more effective at predicting the spread of events (Appendix \ref{sec:App1}). 
The fit of the Hawkes process experiences a remarkable enhancement when spike times are adjusted to exceed a two-day threshold and surpass $40\%$ frequency. This improvement is consistent across various configurations, primarily because the model encounters difficulties in accurately capturing the self-excitation process when the threshold is set too low. By raising the threshold to a minimum of $40\%$ of the maximum potential events in each oblast, the model aligns more closely with the expected behavior, resulting in a more precise fit. This threshold setting ensures that the model accounts for the underlying dynamics and optimally represents the observed data, thereby enhancing its overall performance. These results highlight the key fact that self-excitation depends on the sensitivity of the different regions and that $40\%$ constitutes a global threshold for events to cause self-excitation. Once we add the effect of identified exogenous events presented in Table \ref{table:spktb}, the model using the 2014 voting data excels at predicting the protest dynamics as seen in Fig. \ref{fig:oblastpred} achieving a negative log-likelihood of $-1590.08$ which stays far better than any configuration as presented in Table. \ref{table:com_table}.

Fitting the parameters for the best-fitting model gives an insight into the behavior of the protests. Given that $T_{ex} = 5.8$, the effect of each spike decreases by a factor of $e^{-\frac{1}{5.8}}$. The magnitude of spikes that influence the number of expected events at reaches its highest value $d_e = 4$ days into the protests, the following major spikes are all due to the self-excitation from the previous event. The voting affinity between different oblasts plays a major role in the spread of protests. The number of injured individuals does affect the number of events as the number of injured to the power of $p = 2$ is proportional to the number of events. 

\begin{figure}[!h]
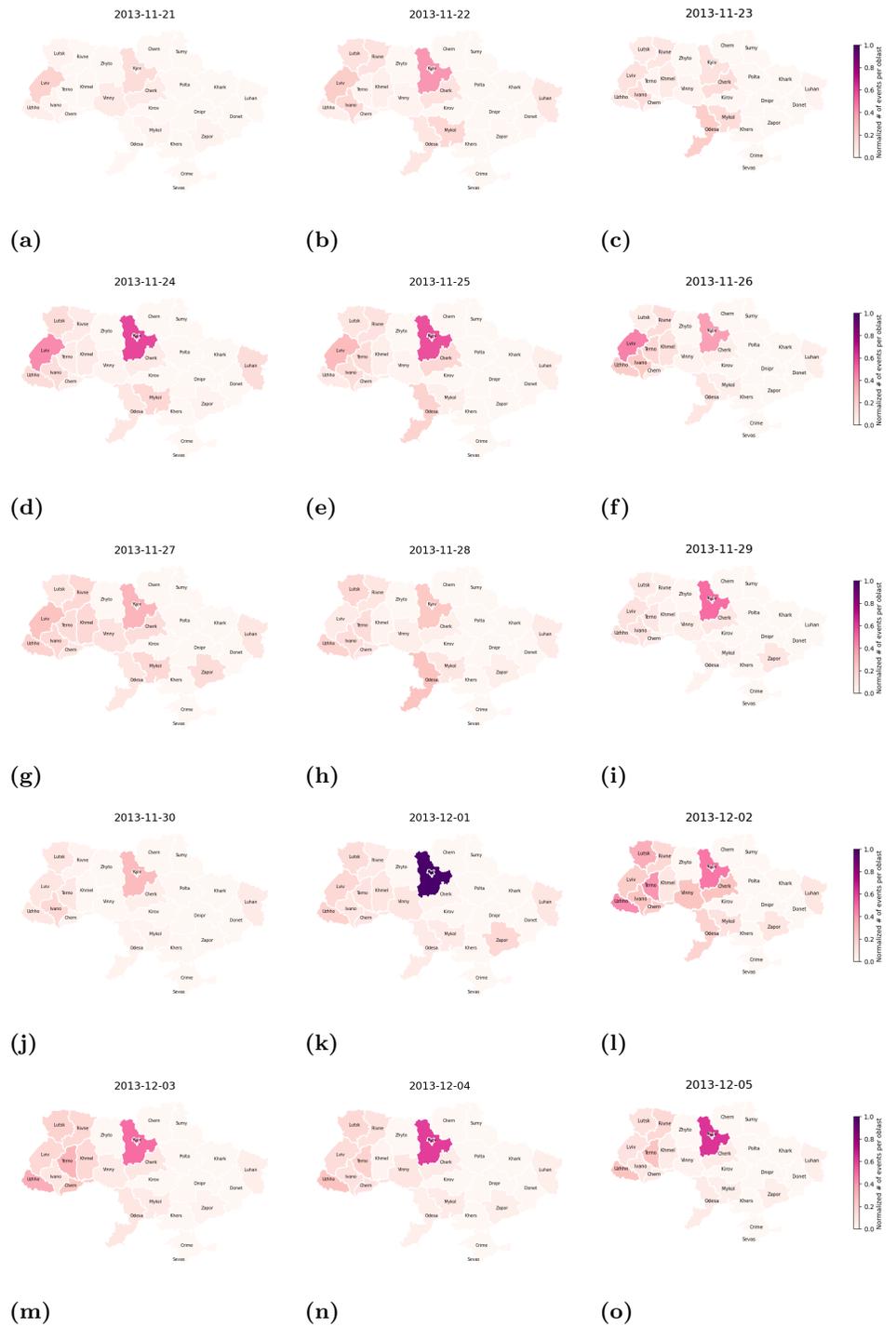

    \centering
    \subcaptionbox{\label{fig:spt0}}{\includegraphics[width=.29\textwidth]{predsptemp0.png}}\hspace{1em}%
    \subcaptionbox{\label{fig:spt1}}{\includegraphics[width=.29\textwidth]{predsptemp1.png}}\hspace{1em}%
    \subcaptionbox{\label{fig:spt2}}{\includegraphics[width=.32\textwidth]{predsptemp2.png}}\hspace{1em}%
    \\
    \subcaptionbox{\label{fig:spt3}}{\includegraphics[width=.29\textwidth]{predsptemp3.png}}\hspace{1em}%
    \subcaptionbox{\label{fig:spt4}}{\includegraphics[width=.29\textwidth]{predsptemp4.png}}\hspace{1em}%
    \subcaptionbox{\label{fig:spt5}}{\includegraphics[width=.32\textwidth]{predsptemp5.png}}\hspace{1em}%
    \\
    \subcaptionbox{\label{fig:spt6}}{\includegraphics[width=.29\textwidth]{predsptemp6.png}}\hspace{1em}%
    \subcaptionbox{\label{fig:spt7}}{\includegraphics[width=.29\textwidth]{predsptemp7.png}}\hspace{1em}%
    \subcaptionbox{\label{fig:spt8}}{\includegraphics[width=.32\textwidth]{predsptemp8.png}}\hspace{1em}%
    \\
    \subcaptionbox{\label{fig:spt9}}{\includegraphics[width=.29\textwidth]{predsptemp9.png}}\hspace{1em}%
    \subcaptionbox{\label{fig:spt10}}{\includegraphics[width=.29\textwidth]{predsptemp10.png}}\hspace{1em}%
    \subcaptionbox{\label{fig:spt11}}{\includegraphics[width=.32\textwidth]{predsptemp11.png}}\hspace{1em}%
    \\
    \subcaptionbox{\label{fig:spt12}}{\includegraphics[width=.29\textwidth]{predsptemp12.png}}\hspace{1em}%
    \subcaptionbox{\label{fig:spt13}}{\includegraphics[width=.29\textwidth]{predsptemp13.png}}\hspace{1em}%
    \subcaptionbox{\label{fig:spt14}}{\includegraphics[width=.32\textwidth]{predsptemp14.png}}\hspace{1em}%

    \caption{A spatiotemporal illustration of the normalized predicted number of total events per oblast from November 21, 2013, to December 05, 2013, where spike times are determined from the average rate of change per oblast between two days.  These events include protests, rallies, riots, and police crackdowns.}
    \label{fig:sptmppred}
\end{figure}

The model succeeds in capturing the general behavior of protests. It excels at predicting the spatio-temporal spread of events. More specifically, the model captures the massive spike that happened on December 1, 2013, in Kyiv, thanks to the addition of the influence of the number of injured individuals, and its following spread through the western oblasts (Fig. \ref{fig:sptmppred}). The model captures the peaks and following self-excitation behavior in Ivano-Frankivsk, Khmelnytskyi, Kyiv, Luhansk, Lviv, Rivne, Odesa, and Vinnytsia. However, it struggles to capture the spikes in Cherkasy, Chernivtsi, Lutsk, and Ternopil. 


%

\begin{table*}[!ht]
\begin{adjustwidth}{-2.25in}{0in} 
\centering
\caption{
{\bf Summary of the negative log-likelihood and AIC scores }} 
\resizebox{\textwidth}{!}{\begin{tabular}{|c|c|c|c|c|c|c|c|c|}
\hline
\multirow{3}*{Spike time configurations} &\multicolumn{4}{|c|}{\bf Classical Hawkes Process} & \multicolumn{4}{|c|}{\bf Susceptible Population Model}\\ \cline{2-9}
& \multicolumn{2}{|c|}{Vote $2010$} & \multicolumn{2}{|c|}{Vote $2014$}  & \multicolumn{2}{|c|}{Vote $2010$} & \multicolumn{2}{|c|}{Vote $2014$} \\ \cline{2-9}
& NLL & AIC & NLL & AIC & NLL & AIC & NLL & AIC \\ \hline
global fixed threshold 0 events &	5817.41	&	5819.41	&	5817.41	&	5819.41	&	5454.79	&	5456.79	&	5455.69	&	5457.69	\\ \hline
global fixed threshold 1 events &	5817.41	&	5819.41	&	5817.41	&	5819.41	&	5682.34	&	5684.34	&	5686.78	&	5688.78	\\ \hline
global fixed threshold 2 events &	5817.41	&	5819.41	&	5817.41	&	5819.41	&	5682.34	&	5684.34	&	5686.78	&	5688.78	\\ \hline
global fixed threshold 3 events &	-744.12	&	-742.12	&	-744.12	&	-742.12	&	5710.11	&	5712.11	&	5712.75	&	5714.75	\\ \hline
global fixed threshold 4 events &	-382.85	&	-380.85	&	-382.85	&	-380.85	&	4726.1	&	4728.1	&	4697.93	&	4699.93	\\ \hline
global fixed threshold 5 events &	424.341	&	426.341	&	424.341	&	426.341	&	5817.41	&	5819.41	&	5817.41	&	5819.41	\\ \hline
global fixed threshold 6 events &	1799.03	&	1801.03	&	1799.03	&	1801.03	&	5817.41	&	5819.41	&	5817.41	&	5819.41	\\ \hline
global fixed threshold 7 events &	1799.03	&	1801.03	&	1799.03	&	1801.03	&	5817.41	&	5819.41	&	5817.41	&	5819.41	\\ \hline
global fixed threshold 8 events &	1799.03	&	1801.03	&	1799.03	&	1801.03	&	5817.41	&	5819.41	&	5817.41	&	5819.41	\\ \hline
global fixed threshold 9 events &	1799.03	&	1801.03	&	1799.03	&	1801.03	&	5817.41	&	5819.41	&	5817.41	&	5819.41	\\ \hline
0\% of maximum activity threshold &	5817.41	&	5819.41	&	5817.41	&	5819.41	&	5454.79	&	5456.79	&	5455.69	&	5457.69	\\ \hline
10\% of maximum activity threshold &	5817.41	&	5819.41	&	5817.41	&	5819.41	&	5682.34	&	5684.34	&	5686.78	&	5688.78	\\ \hline
20\% of maximum activity threshold &	5817.41	&	5819.41	&	5817.41	&	5819.41	&	5682.34	&	5684.34	&	5686.78	&	5688.78	\\ \hline
30\% of maximum activity threshold &	-744.12	&	-742.12	&	-744.12	&	-742.12	&	5710.11	&	5712.11	&	5712.75	&	5714.75	\\ \hline
40\% of maximum activity threshold &	-382.85	&	-380.85	&	-382.85	&	-380.85	&	4726.1	&	4728.1	&	4697.93	&	4699.93	\\ \hline
50\% of maximum activity threshold &	424.341	&	426.341	&	424.341	&	426.341	&	5817.41	&	5819.41	&	5817.41	&	5819.41	\\ \hline
60\% of maximum activity threshold &	1799.03	&	1801.03	&	1799.03	&	1801.03	&	5817.41	&	5819.41	&	5817.41	&	5819.41	\\ \hline
70\% of maximum activity threshold &	1799.03	&	1801.03	&	1799.03	&	1801.03	&	5817.41	&	5819.41	&	5817.41	&	5819.41	\\ \hline
80\% of maximum activity threshold &	1799.03	&	1801.03	&	1799.03	&	1801.03	&	5817.41	&	5819.41	&	5817.41	&	5819.41	\\ \hline
90\% of maximum activity threshold &	1799.03	&	1801.03	&	1799.03	&	1801.03	&	5817.41	&	5819.41	&	5817.41	&	5819.41	\\ \hline
\multirow{3}*{Spike time configurations} &\multicolumn{4}{|c|}{\bf Distance Based Model} & \multicolumn{4}{|c|}{\bf Voting Based Model}\\ \cline{2-9}
& \multicolumn{2}{|c|}{Vote $2010$} & \multicolumn{2}{|c|}{Vote $2014$}  & \multicolumn{2}{|c|}{Vote $2010$} & \multicolumn{2}{|c|}{Vote $2014$} \\ \cline{2-9}
& NLL & AIC & NLL & AIC & NLL & AIC & NLL & AIC \\ \hline
global fixed threshold 0 events & 5128.48	&	5134.48	&	5172.68	&	5178.68	&	5454.72	&	5460.72	&	5455.62	&	5461.62	\\ \hline
global fixed threshold 1 events & 5128.48	&	5134.48	&	5172.68	&	5178.68	&	5454.72	&	5460.72	&	5455.62	&	5461.62	\\ \hline
global fixed threshold 2 events & 5419.41	&	5425.41	&	5459.65	&	5465.65	&	5682.28	&	5688.28	&	5686.73	&	5692.73	\\ \hline
global fixed threshold 3 events & 5419.41	&	5425.41	&	5459.65	&	5465.65	&	5682.28	&	5688.28	&	5686.73	&	5692.73	\\ \hline
global fixed threshold 4 events & -1442.61	&	-1436.61	&	-1414.08	&	-1408.08	&	5710.07	&	5716.07	&	5712.71	&	5718.71	\\ \hline
global fixed threshold 5 events & -1031.78	&	-1025.78	&	-1190.28	&	-1184.28	&	-1133.38	&	-1127.38	&	-1324.21	&	-1318.21	\\ \hline
global fixed threshold 6 events & -1031.78	&	-1025.78	&	-1190.28	&	-1184.28	&	-1133.38	&	-1127.38	&	-1324.21	&	-1318.21	\\ \hline
global fixed threshold 7 events & -1030.70	&	-1024.7	&	-1190.28	&	-1184.28	&	-1133.38	&	-1127.38	&	-1324.21	&	-1318.21	\\ \hline
global fixed threshold 8 events & -1030.70	&	-1024.7	&	-1190.28	&	-1184.28	&	-1133.38	&	-1127.38	&	-1324.21	&	-1318.21	\\ \hline
global fixed threshold 9 events & -1030.70	&	-1024.7	&	-1190.28	&	-1184.28	&	-1133.38	&	-1127.38	&	-1324.21	&	-1318.21	\\ \hline
0\% of maximum activity threshold & 5128.48	&	5134.48	&	5172.68	&	5178.68	&	5454.72	&	5460.72	&	5455.62	&	5461.62	\\ \hline  
10\% of maximum activity threshold & 5419.41	&	5425.41	&	5459.65	&	5465.65	&	5682.28	&	5688.28	&	5686.73	&	5692.73	\\ \hline  
20\% of maximum activity threshold & 5419.41	&	5425.41	&	5459.65	&	5465.65	&	5682.28	&	5688.28	&	5686.73	&	5692.73	\\ \hline  
30\% of maximum activity threshold & 5440.73	&	5446.73	&	-1621.33	&	-1615.33	&	5710.07	&	5716.07	&	5712.71	&	5718.71	\\ \hline  
40\% of maximum activity threshold & -1374.18	&	-1368.18	&	-1190.33	&	-1184.33	&	-996.14	&	-990.14	&	-1562.77	&	-1556.77	\\ \hline  
50\% of maximum activity threshold & -1031.78	&	-1025.78	&	-1190.28	&	-1184.28	&	-996.09	&	-990.09	&	-1189.64	&	-1183.64	\\ \hline  
60\% of maximum activity threshold & -1030.7	&	-1024.7	&	-1190.28	&	-1184.28	&	-996.09	&	-990.09	&	-1189.64	&	-1183.64	\\ \hline  
70\% of maximum activity threshold & -1030.7	&	-1024.7	&	-1190.28	&	-1184.28	&	-996.09	&	-990.09	&	-1189.64	&	-1183.64	\\ \hline  
80\% of maximum activity threshold & -1030.7	&	-1024.7	&	-1190.28	&	-1184.28	&	-996.09	&	-990.09	&	-1189.64	&	-1183.64	\\ \hline  
90\% of maximum activity threshold & -1030.7	&	-1024.7	&	-1190.28	&	-1184.28	&	-996.09	&	-990.09	&	-1189.64	&	-1183.64	\\ \hline  
40\% of max threshold w. exogenous events  & -	&	-	&	-	&	-	&	-	&	-	&	-1590.08	&	-1583.08  \\
\hline
\hline
\end{tabular}}
\label{table:com_table}
\end{adjustwidth}
\end{table*}

\section*{Discussion and Conclusions}


Euromaidan was a significant political and social movement that unfolded in Ukraine from November 2013 to February 2014. The protests, which were primarily peaceful at first, grew in scale and intensity as they continued. Protesters demanded President Yanukovych's resignation and a shift towards closer ties with the EU. The protests took a violent turn in January 2014, after the Ukrainian government passed legislation restricting the right to protest. This led to clashes between protesters and law enforcement officials, resulting in several fatalities. Despite the violence, protesters remained steadfast in their efforts, and their resolve ultimately led to President Yanukovych's removal from power in February 2014. The events of Euromaidan had significant implications, both within Ukraine and beyond. The movement had a profound impact on the country's political landscape, leading to the formation of a new government and a pivot towards the EU. However, it also contributed to the annexation of Crimea by Russia and a prolonged conflict in eastern Ukraine. As such, the Euromaidan movement is a significant case study from a sociological and mathematical standpoint. Its unfolding provides insight into the dynamics of political movements and the role of civil society in shaping the direction of a country.

We use data from the Center for Social and Labor Research to study the spatiotemporal dynamics of the protest events.  While the data set is missing important activity that occurred during the first two months of 2014 and also shows some discrepancies, such as the missing number of injured and deceased individuals, it remains useful for modeling the first half of Euromaidan. We choose to model the number of events and not the number of protesters, as the magnitude of an event does not always correlate with the importance of the event. As pointed out above, all protest events add to the general tension and lead to more events. Additionally, our 
statistical analysis shows the number of protest events per day was closely associated with the number of injuries, negative response events, and Euromaidan events on the previous day. We found that the number of events per day was strongly influenced by the number of events with a negative response and the number of events associated with Euromaidan, with negative reaction events and Euromaidan events both associated with more total events in the subsequent days. One can derive two main results. First, there is indeed self-excitation in the data, which justifies the use of the Hawkes process. Second, injuries and negative police responses play an important role in the process.  The kernel used in our model takes into account the interactions between oblasts and the influence they have on each other due to political affinity.  

The model accurately predicts the spatiotemporal dynamics of events during the Euromaidan protests in Ukraine. While the model struggles to accurately predict the magnitude of events in certain regions, it excels in predicting the spread of events. The political affinity between regions was found to be a more significant factor in determining protest spread than geographic distance, highlighting the importance of social and cultural factors.  This is in contrast to other studies, such as that of the 2005 French riots where geographic distance was seen to play a key role in the spread of activity \cite{Bonnasse-Gahot2018}.  Thus, our study highlights the shift in protest contagion as being less dependent on geography (perhaps due to the immediate spread of information in current days) to being more influenced by political affinity. 
We note that the under-reporting of police violence negatively affects the model's accuracy. The best spike times are those that take into account the specific reactions of each region. The differences between the chosen spike models indicate that certain regions are significantly affected by the national upheaval, while others benefit from considering them individually. 

\section*{Appendix}\label{sec:s_i}

\subsection{Missing data}\label{sec:App2}

 In this section, we discuss missing data in more detail. The data set consists of 6627 events that began during 2013. Each event has a start date, though only 6591 have an end date recorded. The latest end date is 2014-02-13. Because the events missing an end date appear to be missing at random, we omit them from our analysis. Of the 6627 events, 3044 are missing data on the number of protesters present, and even in cases where this data is not missing, it is frequently an estimate. For this reason, we do not include the ``number of protesters" as a variable in any of our models. Similarly, data on the number of arrests, injuries, and deaths is missing from most of the events. For example, the data set contains only 8 deaths total in 2013, a vast undercount. The data set also shows that for the eight most serious events (including one with 800,000 protesters) there were no arrests and no injuries, which is impossible. In general, the ``number of civilians injured" variable had much less missing data than the variables for the number of civilians arrested/killed. This variable is coded `NA' in only 65 of the events, but it takes the value zero more often than it should, as noted above. It is only positive in 212 events but takes a range of values, the highest of which is 122. Of the events with the ``number of civilians injured" recorded as zero, we do not know how many actually had zero injuries. 

 The lack of data on the number of protesters is a clear limitation in our model, as is the extent of missing data in the ``number of civilians injured" variable. The latter variable has been shown to have a statistically significant relationship to the number of protests in the USA \cite{Rodriguez-White}, and for this reason, we do include this variable in our model. To our knowledge, no research paper has ever included the former variable, because of the difficulty in reliably measuring the number of protesters. We hope that future researchers will be able to study the relationship between the number of protesters, the extent of media coverage, and the effect on protest dynamics, if technological advances allow for these data to be correctly measured.

In our final model, the ``number of civilians injured" variable is a statistically significant explanatory variable, whose inclusion slightly improves the model.  Due to the missing data issues, we urge the reader to use caution interpreting this variable as it relates to protests in Ukraine and to consider the model in \cite{Rodriguez-White}, on a data set that exhibits less cause for concern regarding missing data for this variable.
Even without reliable data on the number of protesters, our analysis still shows self-excitation in the number of protests, analogously to the situation of \cite{Rodriguez-White}.

\subsection{Cross-correlation analysis}\label{sec:App3}

In this section, we discuss the idea behind \textit{cross-correlation analysis} \cite{shumway}. First, we pick either of the two-time series, e.g., $i_t$, and compute shifted versions of it. For example, $i_{t-1}$ means we shift the entire column down by one, so there's no data at time 1, and at time 2 the data recorded is $i_1$, etc. Similarly, $i_{t+1}$ means we shift the column up by one so that at time 1 we record $i_2$, etc. Secondly, we write down all the correlations, first between the column $p_t$ and the column $i_t$, then between the column $p_t$ and the column $i_{t-1}$ (hence, the number $p_1$, of events on the first day, is dropped, as it would be paired with $i_0$, which doesn't exist), then between $p_t$ and $i_{t-2}$, etc., and also between $p_t$ and $i_{t+1}, i_{t+2}$, etc. For each shift, $h$, we record the correlation between $p_t$ and $i_{t-h}$. Finally, we look for a spike in this collection of correlations, e.g., if the strongest correlation is between $p_t$ and $i_{t-1}$ then this means the number of civilians injured on any given day (time $t-1$) is associated with the number of events on the next day (time $t$).

Carrying out this plan is slightly more complicated in practice because there could be some exogenous variables that affect both time series. To solve that issue, the standard technique, \textit{pre-whitening}, is to first model one of the time series (say $p_t$) as a function of its own history, then write down a new time series of residuals $r_t$ of that model, which should be white noise (i.e., a random time series with no dependence on its own history, and no dependence on any exogenous variable). This process provides a \textit{filter} that transforms $p_t$ into $r_t$. We apply that filter to $i_t$ to transform it into $r_t'$, and compute the correlation between $r_t$ and $r_t'$. This correlation shows how related $p_t$ and $i_t$ are once we remove any dependence on anything else. We repeat this process for each lag, e.g., $i_{t-h}$ becomes $r_{t-h}'$, and again write down these correlations for every $h$.

\subsection{Voting Based Hawkes Configuration with no Exogenous Events:} \label{sec:App1}
\phantom{\cite{fig:oblastprednoexo}}
\begin{figure}[H]
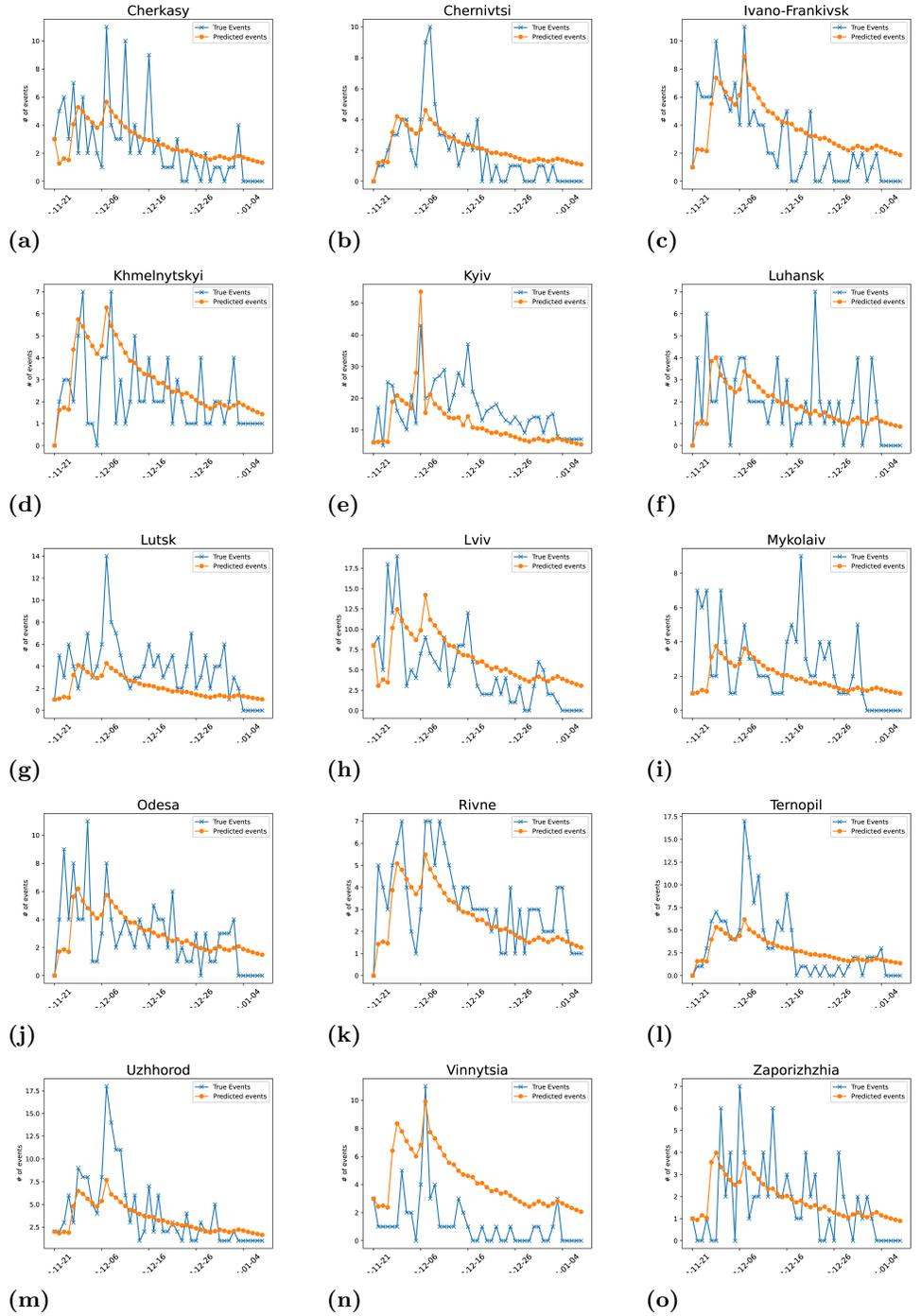

    \centering
    \subcaptionbox{\label{fig:obpd}}{\includegraphics[width=.31\textwidth]{predictionCherkasy.png}}\hspace{1em}%
    \subcaptionbox{\label{fig:obpd}}{\includegraphics[width=.31\textwidth]{predictionChernivtsi.png}}\hspace{1em}%
    \subcaptionbox{\label{fig:obpd}}{\includegraphics[width=.31\textwidth]{predictionIvano-Frankivsk.png}}\hspace{1em}%
    \\
    \subcaptionbox{\label{fig:obpd}}{\includegraphics[width=.31\textwidth]{predictionKhmelnytskyi.png}}\hspace{1em}%
    \subcaptionbox{\label{fig:obpd}}{\includegraphics[width=.31\textwidth]{predictionKyiv.png}}\hspace{1em}%
    \subcaptionbox{\label{fig:obpd}}{\includegraphics[width=.31\textwidth]{predictionLuhansk.png}}\hspace{1em}%
    \\
    \subcaptionbox{\label{fig:obpd}}{\includegraphics[width=.31\textwidth]{predictionLutsk.png}}\hspace{1em}%
    \subcaptionbox{\label{fig:obpd}}{\includegraphics[width=.31\textwidth]{predictionLviv.png}}\hspace{1em}%
    \subcaptionbox{\label{fig:obpd}}{\includegraphics[width=.31\textwidth]{predictionMykolaiv.png}}\hspace{1em}%
    \\
    \subcaptionbox{\label{fig:obpd}}{\includegraphics[width=.31\textwidth]{predictionOdesa.png}}\hspace{1em}%
    \subcaptionbox{\label{fig:obpd}}{\includegraphics[width=.31\textwidth]{predictionRivne.png}}\hspace{1em}%
    \subcaptionbox{\label{fig:obpd}}{\includegraphics[width=.31\textwidth]{predictionTernopil.png}}\hspace{1em}%
    \\
    \subcaptionbox{\label{fig:obpd}}{\includegraphics[width=.31\textwidth]{predictionUzhhorod.png}}\hspace{1em}%
    \subcaptionbox{\label{fig:obpd}}{\includegraphics[width=.31\textwidth]{predictionVinnytsia.png}}\hspace{1em}%
    \subcaptionbox{\label{fig:obpd}}{\includegraphics[width=.31\textwidth]{predictionZaporizhzhia.png}}\hspace{1em}%
  \caption{The predicted number of events per day in each oblast weighted by the susceptible population without exogenous events. $N_{sec} = 100$, $d_e = 3.84$, $T_{ex} = 15.8$, $c = 2.3$, $d =2.6$, $p = 2$.}
  \label{fig:oblastprednoexo}
\end{figure}
\subsection*{Dedication}
An unfortunate consequence of the Euromaidan revolution has been the current war between Ukraine and Russia.  We would like to dedicate this work to the people of Ukraine.

\subsection*{Acknowledgements}
We are indebted to the anonymous referees whose careful review significantly improved this paper. We would like to express our deepest appreciation to Leonid Berlyand for his help with this project in the fall of 2021, and we thank the Center for Social and Labor Research for providing the data used in this project.  

\subsection*{Software archive:}
The code used to generate all graphs and models presented in this paper can be found at: \url{https://github.com/bahidyassin/Protest-Modeling}

%
%
%




\bibliographystyle{plainurl}

\begin{thebibliography}{10}



\bibitem{varol2014evolution}
Varol O, Ferrara E, Ogan CL, Menczer F, Flammini A.
\newblock {Evolution of online user
behavior during a social upheaval.} 
\newblock {In: Proceedings of the 2014 ACM conference
on Web science; 2014. p. 81–90.}

\bibitem{agreement1994}
Ukraine Partnership and Cooperation Agreement.
\newblock \url{https://cordis.europa.eu/article/id/2672-euukraine-partnership-and-cooperation-agreement},
\newblock{EU/Ukraine Partnership and Cooperation Agreement - Cordis}

\bibitem{data}
{The Center for Social and Labor Research.}
\newblock{Databases of protest events and reports. 2014. \url{https://www.cslr.org.ua/en/databases-of-protest-events-and-reports/}}

\bibitem{codebook}
{The Center for Social and Labor Research}
\newblock {Ukrainian Protest And Coercion Data Codebook. 2015}
\newblock \url{http://cslr.org.ua/wp-content/uploads/2015/01/UPCD_Codebook_31-Dec-2012.pdf}

\bibitem{earl2004use}
Earl J, Martin A, McCarthy JD, Soule SA.
\newblock {The Use of Newspaper Data in the
Study of Collective Action.}
\newblock {Annu Rev Sociol. 2004;30:65–80.}


\bibitem{data-prov}
The Center for Social and Labor Research.
\newblock {Ukrainian Protest and Coercion
Data Project; 2023.}
\newblock \url{https://www.cslr.org.ua/en/ukrainian-protest-and-coercion-data-project/}

\bibitem{Smith_Spark_Brumfield_Krever_2014}
Smith-Spark L, Brumfield B, Krever M.
\newblock {Ukraine Signs EU Deal that Sparked
Months of Upheaval, Extends Cease-Fire; CNN; 2014.}
\newblock \url{https://www.cnn.com/2014/06/27/world/europe/ukraine-crisis/index.html}

\bibitem{BBCNews_2014}
{Ukraine Elections: Runners and Risks.}
\newblock {BBC; 2014}
\newblock \url{https://www.bbc.com/news/world-europe-27518989}

\bibitem{alsulami2022dynamical}
Alsulami A, Glukhov A, Shishlenin M, Petrovskii S.
\newblock {Dynamical Modelling of
Street Protests Using the Yellow Vest Movement and Khabarovsk as Case
Studies.}
\newblock{Scientific Reports. 2022;12(1):20447.}

\bibitem{loeffler2018gun}
Loeffler C, Flaxman S.
\newblock{Is Gun Violence Contagious? A spatiotemporal Test.
Journal of quantitative criminology.}
\newblock {2018;34:999–1017.}

\bibitem{brantingham2021gang}
Brantingham JP, Yuan B, Herz D.
\newblock {Is Gang Violent Crime More Contagious than
Non-Gang Violent Crime?}
\newblock {Journal of quantitative criminology. 2021;37:953–977.}

\bibitem{tench2016spatio}
Tench S, Fry H, Gill P.
\newblock {Spatio-temporal Patterns of IED Usage by the
Provisional Irish Republican Army.}
\newblock {European Journal of Applied Mathematics. 2016;27(3):377–402.}

\bibitem{quek2009evolutionary}
Quek HY, Tan KC, Abbass HA.
\newblock {Evolutionary Game Theoretic Approach for
Modeling Civil Violence.}
\newblock {IEEE Transactions on Evolutionary Computation.
2009;13(4):780–800.}

\bibitem{campedelli2021temporal}
Campedelli GM, D’Orsogna MR.
\newblock {Temporal Clustering of Disorder Events During the COVID-19 Pandemic.}
\newblock {PLOS One. 2021;16(4):e0250433.}

\bibitem{EUtrade}
European Neighbourhood Policy and Enlargement Negotiations (DG NEAR).
\newblock \url{https://neighbourhood-enlargement.ec.europa.eu/enlargement-policy/glossary/association-agreement_en}

\bibitem{Lapatina}
Lapatina A.
\newblock {Berkut officers who attacked protesters during Euromaidan given first ever prison sentences.}
\newblock \url{https://www.kyivpost.com/post/8065.}

\bibitem{dimarco2021kinetic}
Dimarco G, Perthame B, Toscani G, Zanella M.
\newblock {Kinetic Models for Epidemic
Dynamics with Social Heterogeneity.}
\newblock {Journal of Mathematical Biology. 2021;83(1):4.}

\bibitem{caroca2020anatomy}
Caroca Soto P, Cartes C, Davies TP, Olivari J, Rica S, Vogt-Geisse K.
\newblock {The Anatomy of the 2019 Chilean social Unrest.}
\newblock {Chaos: An Interdisciplinary Journal of Nonlinear Science. 2020;30(7).}


\bibitem{young2015}
Young T.
\newblock {10 maps that Explain Ukraine’s Struggle for Independence; 2015.}
\newblock \url{"https://www.brookings.edu/blog/brookings-now/2015/05/
21/10-maps-that-explain-ukraines-struggle-for-independence/".}

\bibitem{kyivpost2014}
Interns.
\newblock {EuroMaidan Rallies in Ukraine (Jan. 19 updates); Kyiv Post; 2014.}

\bibitem{ZADEH2022103594}
Zadeh A, Sharda R.
\newblock { How Can Our Tweets Go Viral? Point-Process Modelling of Brand Content.}
\newblock {Information Management. 2022;59(2):103594.
doi:https://doi.org/10.1016/j.im.2022.103594.}

\bibitem{Ogata1988}
Ogata Y.
\newblock {Statistical Models for Earthquake Occurrences and Residual Analysis for Point Processes.}
\newblock {Journal of the American Statistical Association.
1988;83(401):9–27.}

\bibitem{AZIZPOUR2018154}
Azizpour S, Giesecke K, Schwenkler G.
\newblock {Exploring the sources of default clustering. Journal of Financial Economics.}
\newblock {2018;129(1):154–183. doi:https://doi.org/10.1016/j.jfineco.2018.04.008.}

\bibitem{Mohler2011}
Mohler GO, Short MB, Brantingham PJ, Schoenberg FP, Tita GE.
\newblock {Self-Exciting Point Process Modeling of Crime.}
\newblock {Journal of the American Statistical Association. 2011;106(493):100–108. doi:10.1198/jasa.2011.ap09546.}

\bibitem{Lemos2013}
Lemos CM, Coelho H, Lopes RJ.
\newblock {Agent-based Modeling of Social Conflict, Civil Violence and Revolution: State-of-the-art Review and Further Prospects.}
\newblock {In:EUMAS; 2013.}

\bibitem{Epstein2012}
Epstein JM.
\newblock{Modeling civil violence: An Agent-based Computational Approach.}
\newblock{Proceedings of the National Academy of Sciences. 2002;99(suppl 3):7243–7250.}


\bibitem{Reznik2016-vb}
Reznik O.
\newblock {From the Orange Revolution to the Revolution of Dignity}
\newblock {East Eur Polit Soc. 2016;30(4):750–765.}

\bibitem{Andreev1997}
Andreev A, Borodkin L, Levandovskii M.
\newblock {Using Methods of Non-linear Dynamics in Historical Social Research: Application of Chaos Theory in the Analysis of the Worker’s Wovement in Pre Revolutionary Russia.}
\newblock {Hist Soc Res. 1997;22(3/4 (83)):64–83.}

\bibitem{Gavrilets2015}
Gavrilets S.
\newblock {Collective Action Problem in Heterogeneous Groups.}
\newblock {The Royal Society. Philos Trans R Soc Lond B Biol Sci. 2015;370(1683):20150016.}

\bibitem{Rodriguez-White}
Rodr{\'\i}guez N, White D.
\newblock {An Analysis of Protesting Activity and Trauma through Mathematical and Statistical Models.}
\newblock {Crime Science. 2023;12(1):17.}

\bibitem{Bonnasse-Gahot2018}
Bonnasse-Gahot L, Berestycki H, Depuiset M, Gordon MB, Roché S, Rodr{\'\i}guez N, et al.
\newblock {Epidemiological Modelling of the 2005 French riots: a Spreading Wave and the Role of Contagion.}
\newblock {n. Scientific Reports. 2018;8(1):107. doi:10.1038/s41598-017-18093-4.}

\bibitem{Berestycki2015}
Berestycki H, Nadal J, Rodr{\'\i}guez N.
\newblock {A Model Of Riots Dynamics: Shocks,
Diffusion And Thresholds.}
\newblock {Networks and Heterogeneous Media.  2015;10. doi:10.3934/nhm.2015.10.443.}

\bibitem{Khosaeva2015}
Khosaeva ZK.
\newblock {The mathematics Model of Protests.}
\newblock {Computer Research and Modeling. 2009;13(4):780–800.}

\bibitem{EaP}
Official European Union Site
\newblock {European Union Eastern Partnership. 2009}
\newblock \url{https://www.eeas.europa.eu/eeas/eastern-partnership_en}

\bibitem{EuAssAgree}
Association Agreement between the European Union and Ukraine. 2017.
\newblock \url{https://www.kmu.gov.ua/en/yevropejska-integraciya/ugoda-pro-asociacyu}

\bibitem{Aljazeera13}
Ukraine Drops EU Plans and Looks to Russia.
\newblock {Aljazeera. 2013. \url{https://www.aljazeera.com/news/2013/11/21/ukraine-drops-eu-plans-and-looks-to-russia/}}

\bibitem{NYT2013}
Kotsyuba O.
\newblock {Ukraine's Battle for Europe.}
\newblock {The New York Times. 2013. \url{https://www.nytimes.com/2013/11/30/opinion/ukraines-battle-for-europe.html}}

\bibitem{WikiDec1}
1 December 2013 Euromaidan riots.
\newblock {Wikipedia. 2013.}
\newblock \url{https://en.wikipedia.org/wiki/1_December_2013_Euromaidan_riots}

\bibitem{WikiDec11}
11 December 2013 Euromaidan assault
\newblock {Wikipedia. 2013.}
\newblock \url{https://en.wikipedia.org/wiki/11_December_2013_Euromaidan_assault}

\bibitem{Marples15}
Marples D.
\newblock {Ukraine’s Euromaidan: Analyses of a Civil Revolution.}
\newblock {Ibidem Verlag; 2015.}

\bibitem{shumway}
Shumway RH, Stoffer DS.
\newblock{Time Series Analysis and Its Applications: With R Examples.}
\newblock{ 4th ed. Springer; 2017.}

\bibitem{Guard14}
\newblock {Ukrainian President Approves Strict Anti-Protest Laws.}
\newblock{The Guardian; 2014. \url{https://www.theguardian.com/world/2014/jan/17/ukrainian-president-anti-protest-laws}.}

\bibitem{NST14}
200,000 Mass in Ukraine in Defiance of Protest Curbs
\newblock{The Guardian; 2014. \url{https://www.theguardian.com/world/2014/jan/17/ukrainian-president-anti-protest-laws}.}

\bibitem{Econ14}
Europe’s New Battlefield
\newblock {The Economist; 2014. \url{https://www.economist.com/briefing/2014/02/20/europes-new-battlefield}.}

\bibitem{Unian14}
Understanding Ukraine’s Euromaidan Protests
\newblock {Open Society Foundations; 2019.}
\newblock \url{https://www.opensocietyfoundations.org/explainers/understanding-ukraines-euromaidan-protests}

\bibitem{BBC14}
Hewitt G.
\newblock {Ukraine Conflict: Tymoshenko Speech Ends Historic Day of Revolution}
\newblock {The BBC News; 2024. \url{https://www.bbc.com/news/av/world-europe-26302277}}

\bibitem{ConFR}
Masters J.
\newblock {Ukraine: Conflict at the Crossroads of Europe and Russia}
\newblock {Council of Foreign Relations; 2022}
\newblock \url{https://www.cfr.org/backgrounder/ukraine-conflict-crossroads-europe-and-russia}

\bibitem{Ind10}
Polityuk P, Balmforth R.
\newblock {Yanukovich declared Winner in Ukraine Pol.}
\newblock {The Independent; 2010.}
\newblock \url{https://www.independent.co.uk/news/world/europe/yanukovich-declared-winner-in-ukraine-poll-1899552.html}

\bibitem{BritHist}
Ukraine Summary.
\newblock{The Encyclopedia Britannica; 2010. \url{https://www.britannica.com/summary/Ukraine}}

\bibitem{Kuzio2010}
Kutzio T.
\newblock {Nationalism, Identity and Civil Society in Ukraine.}
\newblock {Communist and Post Communist Studies; 2010.}
\newblock \url{https://www.britannica.com/summary/Ukraine}

\bibitem{khpg2013}
EuroMaidan protester allegedly beaten by Berkut dies.
\newblock {2013.}
\newblock {https://khpg.org/en/1387893053}

\bibitem{fisher_2021}
Fisher M.
\newblock {The Three Big Reasons that Protests Reignited in Ukraine}
\newblock {The Washington Post; 2021.}
\newblock \url{https://www.washingtonpost.com/news/worldviews/wp/2014/02/18/the-three-big-reasons-that-protests-reignited-in-ukraine/}

\bibitem{UA_2014}
Fisher M.
\newblock{War on the Streets of Kyiv.}
\newblock {Ukrayinska Pravda; 2024.}
\newblock \url{https://www.pravda.com.ua/articles/2014/02/18/7014151/}
\end{thebibliography}

\end{document}